# What should we understand by the four-momentum of physical system?

Sergey G. Fedosin

PO box 614088, Sviazeva str. 22-79, Perm, Perm Krai, Russia

E-mail: fedosin@hotmail.com

It is shown that, in general, in curved spacetime, none of the known definitions of four-momentum correspond to the definition, in which all the system's particles and fields, including fields outside matter, make an explicit contribution to the four-momentum. This drawback can be eliminated under the assumption that the primary representation of four-momentum is the sum of two nonlocal four-vectors of the integral type with covariant indices. The first of these four-vectors is the generalized four-momentum, found with the help of Lagrangian density. The time component of the generalized four-momentum, in theory of vector fields, is proportional to the particles' energy in scalar field potentials, and the space component is related to vector field potentials. The second four-vector is the four-momentum of fields themselves, and its time component is related to the energy given by tensor invariants. As a result, the system's four-momentum is defined as a four-vector with a covariant index. The standard approach makes it possible to find the four-momentum in covariant form only for a free point particle. In contrast, the obtained formulas for calculating the four-momentum components are applied to a stationary and moving relativistic uniform system, consisting of many particles. In this case, the main fields of the system under consideration are taken into account, including the electromagnetic and gravitational fields, the acceleration field and the pressure field. All these fields are considered vector fields, which makes it possible to unambiguously determine the equations of motion of the fields themselves and the equations of motion of matter in these fields. The formalism used includes the principle of least action, charged and neutral four-currents, corresponding four-potentials and field tensors, which ensures unification and the possibility of combining fields into a single interaction. Within the framework of the special theory of relativity, it is shown that due to motion, the four-momentum of the system increases in proportion to the Lorentz factor of the system's center of momentum, while in the matter of the system the sum of the energies of all fields is equal to zero. The calculation of the integral vector's components in the relativistic uniform system shows that the so-called integral vector is not equal to four-momentum and is not a four-vector at all, although it is conserved in a closed system. Thus, in the theory of relativistic vector fields, the four-momentum cannot be found with the help of an integral vector and components of the system's stress-energy tensor, in contrast to how it is assumed in the general theory of relativity.

## 1. Introduction

By the standard definition adopted in the theory of relativity, the four-momentum of a physical system is a four-vector of the following form:

$$P^\mu = \left( \frac{\mathbb{E}}{c}, \mathbf{P} \right), \tag{1}$$

where $\mathbb{E}$ is energy, $c$ is speed of light, and $\mathbf{P}$ is three-dimensional relativistic momentum of the system.

Definition (1) is widely used in mechanics for equations of motion, where the derivative of four-momentum with respect to proper time defines four-forces acting on a system. Both $\mathbb{E}$ and $\mathbf{P}$ are additive quantities, thus, the energy is obtained by summing the energies contained in all the volume elements of a system. The system momentum should be determined by vector summation of the momenta of all volume elements, including those in which there is no matter where there is only a field.

The energy and momentum of a system, which contains $N$ particles or individual elements of continuously distributed matter, can be derived via Lagrangian formalism [1], for which Lagrangian $L$ is used as an integral over an infinite moving volume:

$$L = \int \mathcal{L} \sqrt{-g}\, dx^1 dx^2 dx^3, \tag{2}$$

$$\mathbb{E} = \sum_{n=1}^{N} \left( \mathbf{v}_n \cdot \frac{\partial L}{\partial \mathbf{v}_n} \right) - L, \tag{3}$$

$$\mathbf{P} = \sum_{n=1}^{N} \frac{\partial L}{\partial \mathbf{v}_n} = \sum_{n=1}^{N} \mathbf{P}_n, \tag{4}$$

here $\mathcal{L}$ is Lagrangian density as volumetric density of Lagrange function, $dx^1 dx^2 dx^3$ is product of differentials of space coordinates, $g$ is determinant of metric tensor $g^{\mu\nu}$, $\mathbf{v}_n$ is velocity of matter particle with the current number $n$, and $\mathbf{P}_n$ is three-dimensional momentum of one volume element of the system.

Substituting (3) and (4) in (1) allows us to find $P^\mu$. However, such a definition of four-momentum is unsatisfactory in the sense that it is not a direct consequence of Lagrangian four-dimensional formalism for four-vectors and four-tensors. As far as we know, $P^\mu$ as a four-vector is not expressed in a covariant form either with the help of Lagrangian $L$ or with the help of Lagrangian density $\mathcal{L}$, the four-momentum $P^\mu$ is simply constructed manually using formula (1). Instead, in [2], we can find a covariant expression of four-momentum but only for one free particle, on which no external forces are acting. In this case, the definition of the action function $S$ with a variable upper limit of integration is used:

$$S = S(t_2) = \int_{t_1}^{t_2} L \, dt . \tag{5}$$

$$P_\mu = -\frac{\partial S}{\partial x^\mu} = -\partial_\mu S = \left( \frac{E}{c}, -\mathbf{P} \right). \tag{6}$$

The characteristic feature of (5) is that the upper limit in the time integral for the action function is not fixed, in contrast to the lower limit. In addition, a particle must move with zero four-acceleration along a certain true trajectory according to its equation of motion. Under such conditions, the variation in the particle's location at the initial time point is equal to zero, $\delta x^\mu(t_1) = 0$; however, the variation $\delta x^\mu(t_2)$ at time point $t_2$ is not equal to zero. According to (6), the four-momentum $P_\mu$ of a single particle turns out to be a four-vector with a covariant index, in contrast to (1), where the four-momentum $P^\mu$ of a system of particles is presented as a four-vector with a contravariant index.

In the flat Minkowski spacetime, the difference between a four-vector with a covariant index and the same four-vector with a contravariant index consists only of the fact that their space components have opposite signs. In curved spacetime, the difference is more significant since to turn to a contravariant form, the four-vector with a covariant index must be multiplied by the metric tensor. In this case, it is more convenient to write equations for the particle through $P_\mu$ (6) rather than through $P^\mu$, since then the metric tensor is not required. This can significantly simplify the solution of the equation of motion, since the metric tensor may not be known in advance.

In a system, consisting of many closely interacting particles, different forces are acting and the particles acquire four-accelerations. This violates conditions, under which definition (6) is valid, so that the summation of four-momenta of individual particles may not provide the total four-momentum $P_\mu$ of the system. Thus, for continuously distributed matter, another covariant expression is required for both the four-momentum of a single particle and the four-momentum of the entire system.

According to [3], the covariant four-dimensional Euler–Lagrange equation should have the following form:

$$\frac{d}{d\tau}\left(\frac{\partial L}{\partial u^\mu}\right) - \frac{\partial L}{\partial x^\mu} = 0, \tag{7}$$

where $u^\mu = \dfrac{dx^\mu}{d\tau}$ is four-velocity, $\tau$ is proper time of a particle, and $x^\mu$ is four-position that determines location of the particle.

The equation (7) is the result of variation of the Lagrange function $L$ in the principle of least action and represents the equation of particle motion. To apply (7), it is necessary to know the dependence $L$ on the values $u^\mu$ and $x^\mu$ of each individual particle of the system, which turns out to be difficult with a large number of particles.

The quantity $P_\mu = -\dfrac{\partial L}{\partial u^\mu}$ in (7) can be interpreted as the four-momentum of an arbitrary particle of a physical system, the quantity $F_\mu = -\dfrac{\partial L}{\partial x^\mu}$ can be considered as a four-force acting on the particle, and the total four-momentum must be obtained by summing individual quantities $P_\mu$ over all the system's particles. However, here, a difficulty arises with respect to contributing to four-momentum from fields outside the matter's limits, which are characterized in Lagrangian form with the help of volume integrals of tensor invariants. The expression of volume integrals of the tensor invariants in terms of the four-velocity of particles is by itself a rather complicated and nontrivial task.

There is a completely different approach to determining the system's four-momentum, associated with the general theory of relativity. Therefore, in [2] we can find the following expression:

$$\mathcal{J}^\mu = \frac{1}{c}\int (-g)\left(T^{\mu 0} + t^{\mu 0}\right)dx^1 dx^2 dx^3, \qquad (8)$$

where the time components of the stress-energy tensor of matter and nongravitational fields are denoted by $T^{\mu 0}$, and the time components of the gravitational field pseudotensor are denoted by $t^{\mu 0}$.

It is argued that quantity $\mathcal{J}^\mu$ is simply the four-momentum $P^\mu$ of a physical system, taking into account the contribution from gravitational field, while in a closed system $\mathcal{J}^\mu$ is conserved. In this regard, we should note that to obtain $\mathcal{J}^\mu$, it is necessary to proceed from the equation of motion in the form $\nabla_\nu T^{\mu\nu} = 0$. Then, such a gravitational field pseudotensor $t^{\mu\nu}$ is introduced into this equation to transform from a covariant derivative to a partial derivative. The equation of motion takes the form $\partial_\nu\left[\left(T^{\mu\nu} + t^{\mu\nu}\right)(-g)\right] = 0$, after which it is integrated over volume, resulting in (8).

The drawback of this approach is the lack of evidence that $\mathcal{J}^\mu$ is truly the system's four-momentum. Expressions (3) and (4) for the energy and momentum do not automatically follow from (8), and it is difficult to determine whether (8) is associated with (6) or with $P_\mu = -\dfrac{\partial L}{\partial u^\mu}$ in (7). Moreover, according to [4], there are many different gravitational field pseudotensors, which give different expressions for $\mathcal{J}^\mu$; therefore, there is no guarantee that at least in one case the equality $\mathcal{J}^\mu = P^\mu$ would hold. In addition, in [5], the correspondence principle was not fulfilled in the general theory of relativity, and the inertial mass in general case within the limits of weak fields and low velocities was not included in the corresponding expression in Newton's theory. According to [6], the same holds true for the energy, momentum and angular momentum of a system.

In [7], a comparison was made between the vector $\mathcal{J}^{\mu}$ and the integral vector, obtained by the following formula:

$$\mathcal{J}_{\alpha} = \int T_{\alpha}{}^{0} dx^1 dx^2 dx^3 . \tag{9}$$

Expression (9) for the integral vector $\mathcal{J}_{\alpha}$ is valid for the case of weak fields and low velocities, when the covariant derivative $\nabla_{\nu}$ in the equation of motion $\nabla_{\nu} T^{\mu\nu} = 0$ can be replaced with the partial derivative $\partial_{\nu}$ with a small error. The quantity $T_{\alpha}{}^{0}$ with mixed indexes in (9) defines the time components of the system's stress-energy tensor, while the gravitational field is considered a vector field within the framework of the covariant theory of gravitation [8]. Analysis of the vector $\mathcal{J}_{\alpha}$ shows that its time component is related to the sum of energies of all the system's fields, and the space component is related to the vector sum of field energy flux vectors. If the four-momentum $P^{\mu}$ defines the energy and momentum of the system's particles and fields, then the vector $\mathcal{J}_{\alpha}$ defines only the energies and fields' energy fluxes.

According to the method of its construction, the integral vector $\mathcal{J}_{\alpha}$ is not a four-vector and can be considered a four-dimensional pseudovector. For the vector $\mathcal{J}^{\mu}$ in (8), this vector has the property that does not allow us to simultaneously fulfill two conditions for a closed system [9]:

1) Conservation over time of the sum of all types of energy, including gravitational energy given by the pseudotensor; 2) independence of the sum of all types of energy at a given time point from the choice of reference frame. As a result, in [10] vectors such as $\mathcal{J}^{\mu}$ in the general theory of relativity are considered not as four-vectors but rather as pseudovectors that cannot define four-momentum $P^{\mu}$.

The purpose of this work is to derive a covariant formula for the system's four-momentum, which is valid for curved spacetime and for continuously distributed matter. Considering the latter circumstance leads to the fact that instead of Lagrangian $L$, the Lagrangian density $\mathcal{L}$ takes the first place in calculations. Our analysis includes the four most common fields, electromagnetic and gravitational fields, acceleration field [11], and pressure field [12], which are considered vector fields. The Lagrangian formalism we use allows us to consider all these fields as components of a single general field [13-15], while the forces acting in the system from each field have the same form, similar to the Lorentz force.

As we will show below, the derived formula for four-momentum will differ from the well-known standard definitions. In addition, by direct calculation of the integral vector $\mathcal{J}_\alpha$, we will show its difference from the four-momentum of the moving physical system.

Appendix A briefly describes two ways of representing the four-momentum $P_\mu$ of a physical system. In the first of these methods, it is necessary to calculate the energy and momentum of the system, and in the second method, the four-momentum $P_\mu$ is represented as the sum of two nonlocal integral vectors in the form $P_\mu = p_\mu + K_\mu$. Appendix B provides details of calculations in relations (119-122). Appendix C provides a list of symbols used.

## 2. Methods

Before considering the four-momentum of a physical system, it is necessary to define the generalized four-momentum, which is the main part of the four-momentum.

To calculate the generalized four-momentum, we will proceed from the Lagrangian formalism for continuously distributed matter in four-dimensional form. In the general case, the Lagrangian density depends on coordinate time $t$, on charge four-currents $j^\mu$ and mass four-current $J^\mu$, on four-potentials and field tensors at each point in the field, including inside the particles, as well as on the metric tensor $g^{\mu\nu}$ and the scalar curvature $R$:

$$\mathcal{L} = \mathcal{L}\left[t, j^\mu\left(x_n^\mu, u_n^\mu\right), J^\mu\left(x_n^\mu, u_n^\mu\right), A_\mu, D_\mu, U_\mu, \pi_\mu, F_{\mu\nu}, \Phi_{\mu\nu}, u_{\mu\nu}, f_{\mu\nu}, g^{\mu\nu}, R(g^{\mu\nu})\right], \quad (10)$$

where $x_n^\mu$ specifies the observation point at which a typical particle with number $n$ is located at a given moment in time, and $u_n^\mu$ is a four-velocity of the typical particle at this point.

In (10), $A_\mu, D_\mu, U_\mu, \pi_\mu$ denote the four-potentials of electromagnetic and gravitational fields, acceleration field and pressure field, respectively, and $F_{\mu\nu}, \Phi_{\mu\nu}, u_{\mu\nu}, f_{\mu\nu}$ are tensors of these fields. Considering (2), the action function within the time interval $t_2 - t_1$ with fixed integration limits is equal to:

$$S = \int_{t_1}^{t_2} L \, dt = \int_{t_1}^{t_2} \left( \int \mathcal{L} \sqrt{-g} \, dx^1 dx^2 dx^3 \right) dt. \quad (11)$$

After substituting (10) into (11), we can vary the action function and obtain equations for each field, equation for metric and equation of motion of particles of matter [11], [16]. In addition, we obtain the four-dimensional Euler–Lagrange equation for each typical particle [17]:

$$\frac{\partial \mathcal{L}_p}{\partial x_n^\mu} - \frac{d}{d\tau_n}\left(\frac{\partial \mathcal{L}_p}{\partial u_n^\mu}\right) = 0. \qquad (12)$$

The quantity $\mathcal{L}_p$ in (12) represents the part of the Lagrangian density $\mathcal{L}$, which contains mass four-current $J^\mu$ and charge four-current $j^\mu$ since only four-currents can depend directly on the four-position $x_n^\mu$ and on the four-velocity $u_n^\mu$ of particles. All the other quantities in the Lagrangian, including four-potentials, field tensors and metric tensor, become functions of $x_n^\mu$ and $u_n^\mu$ only after solving corresponding equations; therefore, they are not differentiated in (12), behaving as constants.

Indeed, the equation of any field is obtained only after varying the Lagrangian in the principle of least action on the four-potential of the corresponding field. This equation gives a relation between the four-current generating the field and the field tensor. Considering the expression of the field tensor in terms of the four-potential, the field equation can also be represented as a relation between the four-current and four-potential. When varying, it is assumed that the field tensor directly depends only on the four-potential and its derivatives. As an example, we can consider Maxwell's equations for the electromagnetic field, the solutions of which give either the electromagnetic tensor or the four-potential as a function of the four-current with a known dependence on time, coordinates and velocities.

On the other hand, in (10) all quantities in the large bracket are assumed to be independent of each other from the point of view of the procedure for varying these quantities. At the same time, $x_n^\mu$ and $u_n^\mu$ appear in the Lagrangian not directly, but indirectly, since the four-currents $j^\mu$ and $J^\mu$ depend on them. As a result, variations of four-currents in the Lagrangian are reduced to variations from $x_n^\mu$ [8], [11], [18].

The characteristic feature of (12) is that it is valid for a small interval $t_2 - t_1$ when the time components $u_n^0 = c\frac{dt}{d\tau_n}$ of the particles' four-velocities can be assumed to be constant. If the interval $t_2 - t_1$ cannot be considered small, it should be divided into small time intervals, and at each of these intervals, we should perform synchronous variation of the action function and specify averaged constant time components $u_n^0$ of the particles' four-velocities. On the other hand, equation (12) can be understood as an equation for typical particles of a system; in this case, the constancy $u_n^0$ for each of the particles is obtained automatically as a result of averaging the parameters of the particles at each point of the system.

In contrast to the equation of motion (7), in which the four-force $F_\mu = -\dfrac{\partial L}{\partial x^\mu}$ appears, the equation of motion (12) is written for the rate of change of the density of the four-momentum and for the density of the four-force $\left(f_\mu\right)_n = -\dfrac{\partial \mathcal{L}_p}{\partial x_n^\mu}$ acting in a unit element of the volume in which a typical particle is located.

### 3. Results
#### 3.1. Generalized four-momentum

With the help of (12) in [17], the generalized four-momentum was determined:

$$p_\mu = -\frac{1}{c}\int_{V_m} \frac{\partial \mathcal{L}_p}{\partial u^\mu} u^0 \sqrt{-g}\, dx^1 dx^2 dx^3 = \frac{1}{c}\int_{V_m} \mathcal{P}_\mu u^0 \sqrt{-g}\, dx^1 dx^2 dx^3 = \int_{V_{m0}} \mathcal{P}_\mu dV_0 \,. \tag{13}$$

In (13) $\mathcal{P}_\mu = -\dfrac{\partial \mathcal{L}_p}{\partial u^\mu}$ is the density of the generalized four-momentum, which is presented in (12), $u^0$ defines the time component of the particles' four-velocity at each integration point over the volume $V_m$, occupied by matter. In addition, a relation from [2] is used:

$$\frac{dt}{d\tau}\sqrt{-g}\, dx^1 dx^2 dx^3 = \frac{u^0}{c}\sqrt{-g}\, dx^1 dx^2 dx^3 = dV_0, \tag{14}$$

where $dV_0$ is the differential of invariant volume of any of the particles of continuously distributed matter, calculated in the particle's comoving reference frame.

In a closed system, the four-vector $p_\mu$ is conserved and represents the generalized four-momentum of all the system's particles.

In addition to $p_\mu$ in (13), the following four-dimensional quantity can be determined:

$$\Im_\mu = -\frac{1}{c} \int_{V_m} \frac{\partial(\mathcal{L}_p/u^0)}{\partial(u^\mu/u^0)} u^0 \sqrt{-g}\, dx^1 dx^2 dx^3 \,. \tag{15}$$

The quantity $\Im_\mu$ is not a four-vector, but under the condition of constancy of $u^0$, expression (15) becomes expression (13) for $p_\mu$. As shown in [17], for purely vector fields, part of the Lagrangian density $\mathcal{L}_p$ is such that the space components of $p_\mu$ and $\Im_\mu$ coincide with each other and, up to a sign, give the particles' relativistic momentum $\mathbf{P}_p$, which is part of (4). To calculate $\mathbf{P}_p$, instead of entire Lagrangian $L$, we need to substitute into (4) its part $L_p = \int_{V_m} \mathcal{L}_p \sqrt{-g}\, dx^1 dx^2 dx^3$, associated with four-currents.

### 3.2. Field energy in matter

By solving equations for fields and metric inside matter, we can express four-potentials, field tensors, metric tensor and scalar curvature in terms of four-positions $x_n^\mu$ and four-velocities $u_n^\mu$ of system particles. In this case, inside the matter the Lagrangian density (10) takes the form $\mathcal{L}_m = \mathcal{L}_m(t, x_1^\mu, x_2^\mu, \ldots x_N^\mu, u_1^\mu, u_2^\mu, \ldots u_N^\mu)$. Using (2) and (14), we find:

$$L_m = L_m(t, x_1^\mu, x_2^\mu, \ldots x_N^\mu, u_1^\mu, u_2^\mu, \ldots u_N^\mu) = \int_{V_m} \mathcal{L}_m \sqrt{-g}\, dx^1 dx^2 dx^3 = c \int_{V_{m0}} \frac{\mathcal{L}_m}{u^0} dV_0 \,. \tag{16}$$

In addition, we can write:

$$\frac{dL_m}{dt} = \frac{\partial L_m}{\partial t} + \sum_{n=1}^{N} \left( \frac{\partial L_m}{\partial x_n^\mu} \frac{dx_n^\mu}{dt} + \frac{\partial L_m}{\partial u_n^\mu} \frac{du_n^\mu}{dt} \right). \tag{17}$$

Expression (17) represents the derivative of the Lagrange function $L_m$ with respect to coordinate time $t$. This derivative is written as a derivative of a complex function under the assumption that $x_n^\mu$ and $u_n^\mu$ are functions of time $t$.

The isochronous variation $\delta L_m$ of the Lagrange function $L_m = L_m(t, x_1^\mu, x_2^\mu, ... x_N^\mu, u_1^\mu, u_2^\mu, ... u_N^\mu)$, taking into account standard equality to zero of variation of coordinate time $\delta t = 0$, is expressed in terms of variations $\delta x_n^\mu$ and $\delta u_n^\mu$:

$$\delta L_m = \sum_{n=1}^{N} \left( \frac{\partial L_m}{\partial x_n^\mu} \delta x_n^\mu + \frac{\partial L_m}{\partial u_n^\mu} \delta u_n^\mu \right). \tag{18}$$

Since $u_n^\mu = \frac{d x_n^\mu}{d\tau_n}$, $u_n^0 = c \frac{dt}{d\tau_n}$, we obtain:

$$\frac{\partial L_m}{\partial u_n^\mu} \delta u_n^\mu = \frac{1}{c} \frac{\partial L_m}{\partial u_n^\mu} \delta \left( u_n^0 \frac{d x_n^\mu}{dt} \right) = \frac{u_n^0}{c} \frac{\partial L_m}{\partial u_n^\mu} \frac{d(\delta x_n^\mu)}{dt} + \frac{1}{c} \frac{\partial L_m}{\partial u_n^\mu} \frac{d x_n^\mu}{dt} \delta u_n^0 =$$
$$= \frac{u_n^0}{c} \frac{d}{dt} \left( \frac{\partial L_m}{\partial u_n^\mu} \delta x_n^\mu \right) - \frac{u_n^0}{c} \frac{d}{dt} \left( \frac{\partial L_m}{\partial u_n^\mu} \right) \delta x_n^\mu + \frac{1}{c} \frac{\partial L_m}{\partial u_n^\mu} \frac{d x_n^\mu}{dt} \delta u_n^0. \tag{19}$$

In (19), variation of the product of two functions was used in the form:

$$\delta \left( u_n^0 \frac{d x_n^\mu}{dt} \right) = u_n^0 \delta \left( \frac{d x_n^\mu}{dt} \right) + \frac{d x_n^\mu}{dt} \delta u_n^0 = u_n^0 \frac{d(\delta x_n^\mu)}{dt} + \frac{d x_n^\mu}{dt} \delta u_n^0.$$

Substitution $\frac{\partial L_m}{\partial u_n^\mu} \delta u_n^\mu$ from (19) to (18) gives the following:

$$\delta L_m = \sum_{n=1}^{N} \left[ \frac{\partial L_m}{\partial x_n^\mu} \delta x_n^\mu - \frac{u_n^0}{c} \frac{d}{dt} \left( \frac{\partial L_m}{\partial u_n^\mu} \right) \delta x_n^\mu + \frac{u_n^0}{c} \frac{d}{dt} \left( \frac{\partial L_m}{\partial u_n^\mu} \delta x_n^\mu \right) + \frac{1}{c} \frac{\partial L_m}{\partial u_n^\mu} \frac{d x_n^\mu}{dt} \delta u_n^0 \right]. \tag{20}$$

We substitute $L_m$ into (11) instead of $L$, find the action variation and, in view of (20), equate it to zero:

$$\delta S = \int_{t_1}^{t_2} \delta L_m \, dt =$$

$$= \int_{t_1}^{t_2} \sum_{n=1}^{N} \left[ \frac{\partial L_m}{\partial x_n^\mu} \delta x_n^\mu - \frac{u_n^0}{c} \frac{d}{dt}\left(\frac{\partial L_m}{\partial u_n^\mu}\right) \delta x_n^\mu + \frac{u_n^0}{c} \frac{d}{dt}\left(\frac{\partial L_m}{\partial u_n^\mu} \delta x_n^\mu\right) + \frac{1}{c} \frac{\partial L_m}{\partial u_n^\mu} \frac{d x_n^\mu}{dt} \delta u_n^0 \right] dt = 0.$$

(21)

As in [17], we assume that in the volume of each particle the time components $u_n^0$ of the particles' four-velocity are constant during the action variation, so that $\delta u_n^0 = 0$ and the last term in (21) is equal to zero. Then for the next-to-last term in (21), we can write the following:

$$\int_{t_1}^{t_2} \sum_{n=1}^{N} \frac{u_n^0}{c} \frac{d}{dt}\left(\frac{\partial L_m}{\partial u_n^\mu} \delta x_n^\mu\right) dt \approx \frac{1}{c} \sum_{n=1}^{N} \int_{t_1}^{t_2} \frac{d}{dt}\left(u_n^0 \frac{\partial L_m}{\partial u_n^\mu} \delta x_n^\mu\right) dt = \frac{1}{c} \sum_{n=1}^{N} \left(u_n^0 \frac{\partial L_m}{\partial u_n^\mu} \delta x_n^\mu\right)\bigg|_{t_1}^{t_2} = 0. \quad (22))$$

The equality to zero in (22) follows from the fact that variations $\delta x_n^\mu$ at the time points $t_1$ and $t_2$ vanish according to the conditions of variation. In (21) the first two terms remain, the difference between which must be equal to zero, as a consequence of $\delta S = 0$ in the principle of least action. This gives the following:

$$\frac{\partial L_m}{\partial x_n^\mu} - \frac{u_n^0}{c} \frac{d}{dt}\left(\frac{\partial L_m}{\partial u_n^\mu}\right) = 0. \quad (23)$$

The equation (23) corresponds to (7) with the difference that the Lagrangian $L_m$ inside matter is used instead of $L$.

Let us express $\dfrac{\partial L_m}{\partial x_n^\mu}$ from (23) and substitute it into (17), taking into account the relation $u_n^0 = c \dfrac{dt}{d\tau_n}$:

$$\frac{dL_m}{dt} = \frac{\partial L_m}{\partial t} + \sum_{n=1}^{N}\left[\frac{u_n^0}{c}\frac{d}{dt}\left(\frac{\partial L_m}{\partial u_n^\mu}\right)\frac{d x_n^\mu}{dt} + \frac{\partial L_m}{\partial u_n^\mu}\frac{du_n^\mu}{dt}\right] =$$

$$= \frac{\partial L_m}{\partial t} + \sum_{n=1}^{N}\left[u_n^\mu \frac{d}{dt}\left(\frac{\partial L_m}{\partial u_n^\mu}\right) + \frac{\partial L_m}{\partial u_n^\mu}\frac{du_n^\mu}{dt}\right] = \frac{\partial L_m}{\partial t} + \sum_{n=1}^{N}\frac{d}{dt}\left(u_n^\mu \frac{\partial L_m}{\partial u_n^\mu}\right).$$

(24)

Relation (24) can be written as follows: $\dfrac{dZ}{dt} = -\dfrac{\partial L_m}{\partial t}$, where

$$Z = \sum_{n=1}^{N}\left(u_n^\mu \frac{\partial L_m}{\partial u_n^\mu}\right) - L_m. \tag{25}$$

If the Lagrangian $L_m$ inside matter does not depend on time, then $\dfrac{\partial L_m}{\partial t} = 0$ and will be $\dfrac{dZ}{dt} = 0$, $Z = const$; that is, the quantity $Z$ will be constant in time and will not depend on coordinates.

From the sum over particles in (25) we can pass on to the integral over volume of continuously distributed matter, expressing the Lagrangian $L_m$ through the Lagrangian density $\mathcal{L}_m$ using (16):

$$Z = \sum_{n=1}^{N}\left(u_n^\mu \frac{\partial L_m}{\partial u_n^\mu}\right) - L_m = \sum_{n=1}^{N}\left(u_n^\mu \frac{\partial\left(c\int_{V_{m0}} \frac{\mathcal{L}_m}{u^0} dV_0\right)}{\partial u_n^\mu}\right) - c\int_{V_{m0}} \frac{\mathcal{L}_m}{u^0} dV_0 =$$

$$= \sum_{n=1}^{N}\left(u_n^\mu \frac{\partial\left(c\int_{V_{n0}} \frac{\mathcal{L}_m}{u^0} dV_{n0}\right)}{\partial u_n^\mu}\right) - c\int_{V_{m0}} \frac{\mathcal{L}_m}{u^0} dV_0 =$$

$$= c\sum_{n=1}^{N}\left(\int_{V_{n0}} u_n^\mu \frac{\partial}{\partial u_n^\mu}\left(\frac{\mathcal{L}_m}{u^0}\right) dV_{n0}\right) - c\int_{V_{m0}} \frac{\mathcal{L}_m}{u^0} dV_0 =$$

$$= c\int_{V_{m0}} u^\mu \frac{\partial}{\partial u^\mu}\left(\frac{\mathcal{L}_m}{u^0}\right) dV_0 - c\int_{V_{m0}} \frac{\mathcal{L}_m}{u^0} dV_0.$$

In the relation presented above, the sum $\sum_{n=1}^{N}\left( u_n^\mu \dfrac{\partial\left( c\int_{V_{m0}} \dfrac{\mathcal{L}_m}{u^0} dV_0 \right)}{\partial u_n^\mu} \right)$ was replaced by a sum

$\sum_{n=1}^{N}\left( u_n^\mu \dfrac{\partial\left( c\int_{V_{n0}} \dfrac{\mathcal{L}_m}{u^0} dV_{n0} \right)}{\partial u_n^\mu} \right)$ in which the integral $\int_{V_{n0}} \dfrac{\mathcal{L}_m}{u^0} dV_{n0}$ is taken only over the volume of one particle with number $n$. This is possible because the derivative $\dfrac{\partial\left( c\int_{V_{n0}} \dfrac{\mathcal{L}_m}{u^0} dV_{n0} \right)}{\partial u_n^\mu}$ is taken only with respect to the four-velocity $u_n^\mu$ of the particle with number $n$. Therefore, the integral $\int_{V_{m0}} \dfrac{\mathcal{L}_m}{u^0} dV_0$ over the volume of all other particles of the system does not depend on the four-velocity $u_n^\mu$ of the particle with number $n$ and the derivative $\dfrac{\partial\left( c\int_{V_{n0}} \dfrac{\mathcal{L}_m}{u^0} dV_{n0} \right)}{\partial u_n^\mu}$ for all other particles becomes equal to zero.

After this, the four-velocity $u_n^\mu$ is entered under the sign of the integral $\int_{V_{n0}} \dfrac{\mathcal{L}_m}{u^0} dV_{n0}$ over the invariant volume $V_{n0}$ of one particle, taking into account that $u_n^\mu$ is constant within the volume of this particle. In the integral $\int_{V_{n0}} \dfrac{\mathcal{L}_m}{u^0} dV_{n0}$, the volume element $dV_{n0}$ does not depend on the four-velocity $u_n^\mu$, so the partial derivative $\dfrac{\partial}{\partial u_n^\mu}$ is also introduced inside the integral and acts there on $\dfrac{\mathcal{L}_m}{u^0}$.

Next, the sum $\sum_{n=1}^{N}\left( \int_{V_{n0}} u_n^\mu \dfrac{\partial}{\partial u_n^\mu}\left(\dfrac{\mathcal{L}_m}{u^0}\right) dV_{n0} \right)$ in the expression for $Z$ is converted into an integral $\int_{V_{m0}} u^\mu \dfrac{\partial}{\partial u^\mu}\left(\dfrac{\mathcal{L}_m}{u^0}\right) dV_0$ over the volume of all particles. Subsequent use of expression (14) for the volume element gives:

$$Z = c \int_{V_{m0}} u^\mu \frac{\partial}{\partial u^\mu} \left( \frac{\mathcal{L}_m}{u^0} \right) dV_0 - c \int_{V_{m0}} \frac{\mathcal{L}_m}{u^0} dV_0 =$$

$$= \int_{V_m} u^\mu \frac{\partial}{\partial u^\mu} \left( \frac{\mathcal{L}_m}{u^0} \right) u^0 \sqrt{-g} \, dx^1 dx^2 dx^3 - \int_{V_m} \mathcal{L}_m \sqrt{-g} \, dx^1 dx^2 dx^3 = \qquad (26)$$

$$= \int_{V_m} u^\mu \frac{\partial \mathcal{L}_m}{\partial u^\mu} \sqrt{-g} \, dx^1 dx^2 dx^3 - \int_{V_m} \mathcal{L}_m \sqrt{-g} \, dx^1 dx^2 dx^3.$$

In (21), we assumed that the time components $u_n^0$ of the four-velocity of particles are constant during the action variation. In this regard, the time component $u^0$ of the four-velocity in (26) is also considered as a constant value when calculating the derivative $\frac{\partial}{\partial u^\mu}$.

By comparing (25) and (3), we can see that the quantity $Z$ has dimension of energy. To better understand the meaning of $Z$, we use the Lagrangian density expression for four vector fields [11], [19], which consists of two parts:

$$\mathcal{L} = \mathcal{L}_p + \mathcal{L}_f, \qquad \mathcal{L}_p = -A_\mu j^\mu - D_\mu J^\mu - U_\mu J^\mu - \pi_\mu J^\mu. \qquad (27)$$

$$\mathcal{L}_f = -\frac{1}{4\mu_0} F_{\mu\nu} F^{\mu\nu} + \frac{c^2}{16\pi G} \Phi_{\mu\nu} \Phi^{\mu\nu} - \frac{c^2}{16\pi \eta} u_{\mu\nu} u^{\mu\nu} - \frac{c^2}{16\pi \sigma} f_{\mu\nu} f^{\mu\nu} + ckR - 2ck\Lambda,$$

(28)

where $A_\mu = \left( \frac{\varphi}{c}, -\mathbf{A} \right)$ is four-potential of electromagnetic field, defined by scalar potential $\varphi$ and vector potential $\mathbf{A}$ of this field,

$j^\mu = \rho_{0q} u^\mu$ is charge four-current,

$\rho_{0q}$ is charge density in particle's comoving reference frame,

$u^\mu$ is four-velocity of a point particle,

$D_\mu = \left( \frac{\psi}{c}, -\mathbf{D} \right)$ is four-potential of gravitational field, described with the help of scalar potential $\psi$ and vector potential $\mathbf{D}$ within the framework of covariant theory of gravitation,

$J^\mu = \rho_0 u^\mu$ is mass four-current,

$\rho_0$ is mass density in particle's comoving reference frame,

$U_\mu = \left( \dfrac{\vartheta}{c}, -\mathbf{U} \right)$ is four-potential of acceleration field, where $\vartheta$ and $\mathbf{U}$ denote scalar and vector potentials, respectively,

$\pi_\mu = \left( \dfrac{\wp}{c}, -\mathbf{\Pi} \right)$ is four-potential of pressure field, consisting of scalar potential $\wp$ and vector potential $\mathbf{\Pi}$;

$\mu_0$ is magnetic constant,

$F_{\mu\nu} = \nabla_\mu A_\nu - \nabla_\nu A_\mu = \partial_\mu A_\nu - \partial_\nu A_\mu$ is electromagnetic tensor,

$G$ is gravitational constant,

$\Phi_{\mu\nu} = \nabla_\mu D_\nu - \nabla_\nu D_\mu = \partial_\mu D_\nu - \partial_\nu D_\mu$ is gravitational tensor,

$\eta$ is acceleration field coefficient,

$u_{\mu\nu} = \nabla_\mu U_\nu - \nabla_\nu U_\mu = \partial_\mu U_\nu - \partial_\nu U_\mu$ is acceleration tensor, calculated as four-curl of four-potential of acceleration field,

$\sigma$ is pressure field coefficient,

$f_{\mu\nu} = \nabla_\mu \pi_\nu - \nabla_\nu \pi_\mu = \partial_\mu \pi_\nu - \partial_\nu \pi_\mu$ is pressure field tensor,

$k = -\dfrac{c^3}{16\pi G \beta}$, where $\beta$ is a certain coefficient of the order of unity to be determined,

$R$ is scalar curvature,

$\Lambda$ is cosmological constant.

The components $\mathcal{L}_p$ and $\mathcal{L}_f$ of Lagrangian density $\mathcal{L}$ in (27-28) have the remarkable feature that all fields, be it an electromagnetic field or a pressure field, are expressed in the same form, that is, through their own four-potentials and through the corresponding tensors. It is well known that the electromagnetic field in this form completely describes electromagnetic phenomena, including all phenomena in curved spacetime with known metric, and taking into account quantization it describes phenomena in the microworld within the framework of quantum electrodynamics with very high accuracy. The same should be expected for other fields in (27-28).

For example, in [19] it was shown that the covariant equation of particle motion in electromagnetic and gravitational fields, in the acceleration field, in the pressure field and in the dissipation field, after simplification within the framework of the special theory of relativity, exactly transforms into the phenomenological Navier-Stokes equation in hydrodynamics.

As another example, let us take the pressure field, which is still, even in models of stars at high pressures, treated as a scalar field. However, considering the pressure field as a vector field significantly increases the accuracy of the results obtained, since in this case a new degree of freedom appears in the form of a vector potential of the pressure field, which is responsible for vector effects depending on the particle velocity. Thus, we can consider our choice of the Lagrangian in (27-28) to be completely justified.

Outside the matter, part of the Lagrangian density $\mathcal{L}_p$ vanishes since four-currents $j^\mu$ and $J^\mu$ are equal to zero, and in $\mathcal{L}_f$, tensors of acceleration field and pressure field, which are present only inside the matter, vanish. When calculating energy and momentum in the matter, we can neglect the last two terms in $\mathcal{L}_f$ for two reasons. First, the scalar curvature $R$ is a function of the metric tensor and its derivatives, and it does not directly contain four-velocity; thus, $\frac{\partial R}{\partial u^\mu} = 0$. Second, we use such energy gauge and equation for metric, that difference $ckR - 2ck\Lambda$ in (28) vanishes [11], [20].

The Lagrangian density $\mathcal{L}_m$ is similar to the Lagrangian density $\mathcal{L}$ in (27), but is calculated only inside the matter. We take into account that four-velocity $u^\mu$ is present only in $\mathcal{L}_p$ (27), where it is part of four-currents. Consequently,

$$\frac{\partial \mathcal{L}_m}{\partial u^\mu} = \frac{\partial \mathcal{L}_p}{\partial u^\mu} = -\rho_{0q} A_\mu - \rho_0 D_\mu - \rho_0 U_\mu - \rho_0 \pi_\mu. \tag{29}$$

In view of (27-29), we find the quantity $Z$ in (26):

$$Z = \int_{V_m} \left( \frac{1}{4\mu_0} F_{\mu\nu} F^{\mu\nu} - \frac{c^2}{16\pi G} \Phi_{\mu\nu} \Phi^{\mu\nu} + \frac{c^2}{16\pi \eta} u_{\mu\nu} u^{\mu\nu} + \frac{c^2}{16\pi \sigma} f_{\mu\nu} f^{\mu\nu} \right) \sqrt{-g}\, dx^1 dx^2 dx^3. \tag{30}$$

In (30), integration is performed over the volume occupied by the matter. Hence, the energy $Z$ is expressed exclusively in terms of field tensors and is conserved if the Lagrangian inside the matter does not directly depend on time. The last condition is satisfied for Lagrangian (27) so that in a closed system the field energy, associated with tensor invariants, must be conserved.

### 3.3. Energy and momentum of a system

In (27-28), the Lagrangian density is presented in the form $\mathcal{L} = \mathcal{L}_p + \mathcal{L}_f$, where $\mathcal{L}_p$ depends on four-potentials and four-currents, and $\mathcal{L}_f$ contains fields' tensor invariants. Additionally, Lagrangian $L$ is divided into two parts, one of which $L_p$ is associated with particles, and the other $L_f$ is associated with fields.

In view of (2), we can write: $L = L_p + L_f = \int (\mathcal{L}_p + \mathcal{L}_f) \sqrt{-g}\, dx^1 dx^2 dx^3$. To calculate derivative $\dfrac{\partial L}{\partial \mathbf{v}_n}$ in expression for energy (3), it is necessary to express $L_p$ of the Lagrangian in terms of the integral over invariant volumes of particles. In view of (14), we find:

$$\sqrt{-g}\, dx^1 dx^2 dx^3 = \frac{c}{u^0}\, dV_0, \qquad L_p = c \int \frac{\mathcal{L}_p}{u^0}\, dV_0,$$

$$\sum_{n=1}^{N}\left(\mathbf{v}_n \cdot \frac{\partial L}{\partial \mathbf{v}_n}\right) = \sum_{n=1}^{N}\left[\mathbf{v}_n \cdot \frac{\partial}{\partial \mathbf{v}_n}\left(c\int \frac{\mathcal{L}_p}{u^0}\, dV_0\right)\right] + \sum_{n=1}^{N}\left(\mathbf{v}_n \cdot \frac{\partial L_f}{\partial \mathbf{v}_n}\right) =$$

$$= \sum_{n=1}^{N}\left[\mathbf{v}_n \cdot \frac{\partial}{\partial \mathbf{v}_n}\left(c\int_{V_n} \frac{\mathcal{L}_p}{u^0}\, dV_{n0}\right)\right] + \sum_{n=1}^{N}\left(\mathbf{v}_n \cdot \frac{\partial L_f}{\partial \mathbf{v}_n}\right) =$$

$$= \sum_{n=1}^{N}\left[c\int_{V_n} \mathbf{v}_n \cdot \frac{\partial}{\partial \mathbf{v}_n}\left(\frac{\mathcal{L}_p}{u^0}\right) dV_{n0}\right] + \sum_{n=1}^{N}\left(\mathbf{v}_n \cdot \frac{\partial L_f}{\partial \mathbf{v}_n}\right) =$$

$$= c\int \mathbf{v} \cdot \frac{\partial}{\partial \mathbf{v}}\left(\frac{\mathcal{L}_p}{u^0}\right) dV_0 + \sum_{n=1}^{N}\left(\mathbf{v}_n \cdot \frac{\partial L_f}{\partial \mathbf{v}_n}\right) =$$

$$= \int \mathbf{v} \cdot \frac{\partial}{\partial \mathbf{v}}\left(\frac{\mathcal{L}_p}{u^0}\right) u^0 \sqrt{-g}\, dx^1 dx^2 dx^3 + \sum_{n=1}^{N}\left(\mathbf{v}_n \cdot \frac{\partial L_f}{\partial \mathbf{v}_n}\right).$$

In the sum presented above, the integral $\int \dfrac{\mathcal{L}_p}{u^0} dV_0$ over the volume of all particles was replaced by the integral $\int_{V_n} \dfrac{\mathcal{L}_p}{u^0} dV_{n0}$ over the volume of a particle with respect to which the partial derivative $\dfrac{\partial}{\partial \mathbf{v}_n}$ is taken, the result does not change. After this, $\mathbf{v}_n$ and $\dfrac{\partial}{\partial \mathbf{v}_n}$ are entered under the integral sign of $\int_{V_n} \dfrac{\mathcal{L}_p}{u^0} dV_{n0}$, then the sum $\sum_{n=1}^{N}$ of the integrals over all particles turns into an integral over the volume of all particles, giving $\int \mathbf{v} \cdot \dfrac{\partial}{\partial \mathbf{v}} \left( \dfrac{\mathcal{L}_p}{u^0} \right) dV_0$. Taking this into account, from (3) we find:

$$\mathbb{E} = \sum_{n=1}^{N} \left( \mathbf{v}_n \cdot \frac{\partial L}{\partial \mathbf{v}_n} \right) - L =$$

$$= \int \mathbf{v} \cdot \frac{\partial}{\partial \mathbf{v}} \left( \frac{\mathcal{L}_p}{u^0} \right) u^0 \sqrt{-g}\, dx^1 dx^2 dx^3 + \sum_{n=1}^{N} \left( \mathbf{v}_n \cdot \frac{\partial L_f}{\partial \mathbf{v}_n} \right) - \int (\mathcal{L}_p + \mathcal{L}_f) \sqrt{-g}\, dx^1 dx^2 dx^3 =$$

$$= \int \left[ \mathbf{v} \cdot \frac{\partial}{\partial \mathbf{v}} \left( \frac{\mathcal{L}_p}{u^0} \right) u^0 - \mathcal{L}_p \right] \sqrt{-g}\, dx^1 dx^2 dx^3 - \int \mathcal{L}_f \sqrt{-g}\, dx^1 dx^2 dx^3 + \sum_{n=1}^{N} \left( \mathbf{v}_n \cdot \frac{\partial L_f}{\partial \mathbf{v}_n} \right).$$

(31)

Taking into account the definitions $u^0 = c \dfrac{dt}{d\tau}$ and $\mathbf{v} = \dfrac{d\mathbf{r}}{dt}$, we express both the charge four-current $j^\mu$ and the mass four-current $J^\mu$ in the following form:

$$j^\mu = \rho_{0q} u^\mu = \rho_{0q} \frac{dx^\mu}{d\tau} = \rho_{0q} \left( c \frac{dt}{d\tau}, \frac{d\mathbf{r}}{d\tau} \right) = \rho_{0q} \left( u^0, \frac{u^0}{c} \frac{d\mathbf{r}}{dt} \right) = \rho_{0q} \left( u^0, \frac{u^0}{c} \mathbf{v} \right).$$

$$J^\mu = \rho_0 u^\mu = \rho_0 \left( u^0, \frac{u^0}{c} \mathbf{v} \right).$$

(32)

The products of the electromagnetic four-potential by charge four-current and of the gravitational four-potential by mass four-current in view of (32) can be represented as follows:

$$A_\mu j^\mu = \frac{1}{c} \rho_{0q} u^0 \varphi - \frac{1}{c} \rho_{0q} u^0 \mathbf{A} \cdot \mathbf{v}, \qquad D_\mu J^\mu = \frac{1}{c} \rho_0 u^0 \psi - \frac{1}{c} \rho_0 u^0 \mathbf{D} \cdot \mathbf{v}.$$

(33)

Similarly, we can write for acceleration field and for pressure field:

$$U_\mu J^\mu = \frac{1}{c} \rho_0 u^0 \vartheta - \frac{1}{c} \rho_0 u^0 \mathbf{U} \cdot \mathbf{v}, \qquad \pi_\mu J^\mu = \frac{1}{c} \rho_0 u^0 \wp - \frac{1}{c} \rho_0 u^0 \mathbf{\Pi} \cdot \mathbf{v}.$$

(34)

Using expressions (33-34), in (27) $\mathcal{L}_p$ is expressed in terms of velocity $\mathbf{v}$ of motion of a matter element or a typical particle:

$$\mathcal{L}_p = \frac{u^0}{c}\left(-\rho_{0q}\varphi + \rho_{0q}\mathbf{A}\cdot\mathbf{v} - \rho_0\psi + \rho_0\mathbf{D}\cdot\mathbf{v} - \rho_0\vartheta + \rho_0\mathbf{U}\cdot\mathbf{v} - \rho_0\wp + \rho_0\mathbf{\Pi}\cdot\mathbf{v}\right). \quad (35)$$

We substitute $\mathcal{L}_p$ from (35) and $\mathcal{L}_f$ from (28) into (31) and obtain the following expression for the energy of the system:

$$E = \frac{1}{c}\int\left[\begin{array}{c}\rho_{0q}\varphi + \rho_0\psi + \rho_0\vartheta + \rho_0\wp - \mathbf{v}\cdot\dfrac{\partial}{\partial\mathbf{v}}\left(\rho_{0q}\varphi + \rho_0\psi + \rho_0\vartheta + \rho_0\wp\right) + \\ +v^2\dfrac{\partial}{\partial\mathbf{v}}\left(\rho_{0q}\mathbf{A} + \rho_0\mathbf{D} + \rho_0\mathbf{U} + \rho_0\mathbf{\Pi}\right)\end{array}\right]u^0\sqrt{-g}\,dx^1 dx^2 dx^3 +$$

$$+\int\left(\frac{1}{4\mu_0}F_{\mu\nu}F^{\mu\nu} - \frac{c^2}{16\pi G}\Phi_{\mu\nu}\Phi^{\mu\nu} + \frac{c^2}{16\pi\eta}u_{\mu\nu}u^{\mu\nu} + \frac{c^2}{16\pi\sigma}f_{\mu\nu}f^{\mu\nu}\right)\sqrt{-g}\,dx^1 dx^2 dx^3 +$$

$$+\sum_{n=1}^{N}\left(\mathbf{v}_n\cdot\frac{\partial L_f}{\partial\mathbf{v}_n}\right).$$

(36)

It was assumed in (36) that, in the general case, the average field potentials in the particles' volume, mass density $\rho_0$ and charge density $\rho_{0q}$ of the particles can depend on velocity $\mathbf{v}$ of these particles. When substituting $\mathcal{L}_f$, the energy gauge condition was used, according to which the difference $ckR - 2ck\Lambda$ in (28) was taken to be equal to zero [11], [20].

For momentum (4), in view of (2), (35) and expression $\sqrt{-g}\,dx^1 dx^2 dx^3 = \dfrac{c}{u^0}dV_0$, we can write:

$$\begin{aligned}
\mathbf{P} &= \sum_{n=1}^{N}\left(\frac{\partial L}{\partial \mathbf{v}_n}\right) = \sum_{n=1}^{N}\left(\frac{\partial L_p}{\partial \mathbf{v}_n}\right) + \sum_{n=1}^{N}\left(\frac{\partial L_f}{\partial \mathbf{v}_n}\right) = \\
&= \sum_{n=1}^{N}\left(\frac{\partial}{\partial \mathbf{v}_n}\int \mathcal{L}_p \sqrt{-g}\, dx^1 dx^2 dx^3\right) + \sum_{n=1}^{N}\left(\frac{\partial L_f}{\partial \mathbf{v}_n}\right) = \\
&= c\sum_{n=1}^{N}\left(\frac{\partial}{\partial \mathbf{v}_n}\int \frac{\mathcal{L}_p}{u^0}\, dV_0\right) + \sum_{n=1}^{N}\left(\frac{\partial L_f}{\partial \mathbf{v}_n}\right) = c\sum_{n=1}^{N}\left(\frac{\partial}{\partial \mathbf{v}_n}\int_{V_n} \frac{\mathcal{L}_p}{u^0}\, dV_{n0}\right) + \sum_{n=1}^{N}\left(\frac{\partial L_f}{\partial \mathbf{v}_n}\right) = \\
&= c\sum_{n=1}^{N}\left[\int_{V_n}\frac{\partial}{\partial \mathbf{v}_n}\left(\frac{\mathcal{L}_p}{u^0}\right) dV_{n0}\right] + \sum_{n=1}^{N}\left(\frac{\partial L_f}{\partial \mathbf{v}_n}\right) = c\int \frac{\partial}{\partial \mathbf{v}}\left(\frac{\mathcal{L}_p}{u^0}\right) dV_0 + \sum_{n=1}^{N}\left(\frac{\partial L_f}{\partial \mathbf{v}_n}\right) = \\
&= \frac{1}{c}\int\left[\begin{array}{l}\rho_{0q}\mathbf{A}+\rho_0\mathbf{D}+\rho_0\mathbf{U}+\rho_0\mathbf{\Pi}-\dfrac{\partial}{\partial \mathbf{v}}\left(\rho_{0q}\varphi+\rho_0\psi+\rho_0\vartheta+\rho_0\wp\right)+ \\ +\mathbf{v}\cdot\dfrac{\partial}{\partial \mathbf{v}}\left(\rho_{0q}\mathbf{A}+\rho_0\mathbf{D}+\rho_0\mathbf{U}+\rho_0\mathbf{\Pi}\right)\end{array}\right] u^0\sqrt{-g}\, dx^1 dx^2 dx^3 + \\
&\quad +\sum_{n=1}^{N}\left(\frac{\partial L_f}{\partial \mathbf{v}_n}\right).
\end{aligned}$$

(37)

In (37), the derivative $\dfrac{\partial}{\partial \mathbf{v}_n}$ of the integral $\int \dfrac{\mathcal{L}_p}{u^0} dV_0$ over the volume of all particles was replaced by the derivative $\dfrac{\partial}{\partial \mathbf{v}_n}$ of the integral $\int_{V_n} \dfrac{\mathcal{L}_p}{u^0} dV_{n0}$ over the invariant volume of the particle with number $n$, which has velocity $\mathbf{v}_n$. After this, the derivative $\dfrac{\partial}{\partial \mathbf{v}_n}$ was introduced under the sign of this integral and the indices $n$ inside the integral were removed.

### 3.4. Components of four-momentum with covariant index

For the system's volume occupied by matter, we found in (26) and in (30) the fields' energy $Z$ associated with the fields. In addition, in this volume the energy is associated with the generalized four-momentum $p_\mu$ in (13). Both of these energies are conserved in a closed stationary physical system. By adding the energy of fields outside matter to these energies, we obtain relativistic energy, which is also conserved in a closed system. This approach implies conservation of each energy component separately as a consequence of energy distribution invariance for systems in equilibrium state.

We use the part of Lagrangian density $\mathcal{L}_p$ from (27) and express with the help of $\mathcal{L}_p$ the time and space components of $p_\mu$ in (13). If we present a generalized four-momentum in the form $p_\mu = (p_0, -\mathbf{p})$ and take into account expressions for the fields' four-potentials, we will obtain the following:

$$p_\mu = \frac{1}{c} \int_{V_m} \left( \rho_{0q} A_\mu + \rho_0 D_\mu + \rho_0 U_\mu + \rho_0 \pi_\mu \right) u^0 \sqrt{-g}\, dx^1 dx^2 dx^3 . \tag{38}$$

$$p_0 = \frac{1}{c^2} \int_{V_m} \left( \rho_{0q} \varphi + \rho_0 \psi + \rho_0 \vartheta + \rho_0 \wp \right) u^0 \sqrt{-g}\, dx^1 dx^2 dx^3 . \tag{39}$$

$$\mathbf{p} = \frac{1}{c} \int_{V_m} \left( \rho_{0q} \mathbf{A} + \rho_0 \mathbf{D} + \rho_0 \mathbf{U} + \rho_0 \mathbf{\Pi} \right) u^0 \sqrt{-g}\, dx^1 dx^2 dx^3 . \tag{40}$$

The time component $p_0$ of the generalized four-momentum depends on scalar field potentials in the matter, and the total generalized momentum $\mathbf{p}$ of matter particles depends on vector field potentials.

Furthermore, in addition to the generalized four-momentum $p_\mu$ with a covariant index, we need another form of it with a contravariant index:

$$p^\nu = \frac{1}{c} \int_{V_m} \left( \rho_{0q} A^\nu + \rho_0 D^\nu + \rho_0 U^\nu + \rho_0 \pi^\nu \right) u^0 \sqrt{-g}\, dx^1 dx^2 dx^3 . \tag{41}$$

To obtain (41), in (38) for each matter element we need to multiply the fields' four-potentials by the metric tensor in this matter element to write four-potentials with a contravariant index. Having an integral form, the generalized four-momenta $p_\mu$ and $p^\mu$ differ from standard four-vectors by their nonlocality. As a result, expressions of the form $p_\mu = g_{\mu\nu} p^\nu$ for generalized four-momenta in curved spacetime are not valid.

Indeed, when events occur locally, in a small volume, as in a point particle, we can well assume an expression for the four-velocity of the particle in the form $u_\mu = g_{\mu\nu} u^\nu$, in which the covariant components $u_\mu$ of the four-velocity are related to the contravariant components $u^\nu$ through the metric tensor $g_{\mu\nu}$. However, the volume $V_m$ of integration in integrals (38-41) includes the entire volume in which all typical particles of the system are located, and this volume greatly exceeds the volume of one particle. Therefore, within the volume $V_m$, the values of the metric tensor $g_{\mu\nu}$ can vary significantly. Expression (38) can be written as follows:

$$p_\mu = \frac{1}{c} \int_{V_m} g_{\mu\nu} \left( \rho_{0q} A^\nu + \rho_0 D^\nu + \rho_0 U^\nu + \rho_0 \pi^\nu \right) u^0 \sqrt{-g}\, dx^1 dx^2 dx^3. \tag{42}$$

If the metric tensor $g_{\mu\nu}$ in (42) could be taken out of the integral sign, then, taking into account expression (41), the relation $p_\mu = g_{\mu\nu} p^\nu$ would be obtained. However, this is only possible in the case when $g_{\mu\nu} = const$, that is, within the framework of the special theory of relativity, but not in curved spacetime.

The time components of the generalized four-momentum in (41-42) can be written as follows:

$$p^0 = \frac{1}{c} \int_{V_m} \left( \rho_{0q} A^0 + \rho_0 D^0 + \rho_0 U^0 + \rho_0 \pi^0 \right) u^0 \sqrt{-g}\, dx^1 dx^2 dx^3. \tag{43}$$

$$p_0 = \frac{1}{c} \int_{V_m} g_{0\nu} \left( \rho_{0q} A^\nu + \rho_0 D^\nu + \rho_0 U^\nu + \rho_0 \pi^\nu \right) u^0 \sqrt{-g}\, dx^1 dx^2 dx^3 =$$

$$= \frac{1}{c} \int_{V_m} \begin{bmatrix} g_{00} \left( \rho_{0q} A^0 + \rho_0 D^0 + \rho_0 U^0 + \rho_0 \pi^0 \right) + \\ + g_{01} \left( \rho_{0q} A^1 + \rho_0 D^1 + \rho_0 U^1 + \rho_0 \pi^1 \right) + \\ + g_{02} \left( \rho_{0q} A^2 + \rho_0 D^2 + \rho_0 U^2 + \rho_0 \pi^2 \right) + \\ + g_{03} \left( \rho_{0q} A^3 + \rho_0 D^3 + \rho_0 U^3 + \rho_0 \pi^3 \right) \end{bmatrix} u^0 \sqrt{-g}\, dx^1 dx^2 dx^3. \tag{44}$$

Comparison of (43) and (44) shows that in the general case of curved spacetime, the contravariant time component $p^0$ of the generalized four-momentum does not coincide with the covariant time component $p_0$. Moreover, it is clear that the product $cp_0$ (39) is present in (36) as one of the energy components, and the generalized momentum $\mathbf{p}$ (40) is part of the system's momentum $\mathbf{P}$ in (37). Since $p_0$ and $\mathbf{p}$ are the components of the generalized four-momentum $p_\mu$ in (38), the system's four-momentum, which contains energy $\mathbb{E}$ and momentum $\mathbf{P}$ in its components, must be a four-vector with a covariant index. Thus, the primary generalized four-momentum is one in the form $p_\mu$, and not in the form $p^\mu$, and the same applies to the system's four-momentum.

In this regard, we define the components with the covariant index of the four-momentum of the system similarly to (6) as follows:

$$P_\mu = (P_0, P_1, P_2, P_3) = \left(\frac{E}{c}, -\mathbf{P}\right), \qquad (45)$$

where $\mathbf{P}$ is a three-dimensional relativistic momentum of the system, which in Cartesian coordinates has components $\mathbf{P} = (P_x, P_y, P_z)$.

In (45), we assume that the energy $E$ (36) is part of the time component $P_\mu$, that is $P_0 = \frac{E}{c}$, in contrast to the standard definition (1) for $P^\mu$, where it is implied that $P^0 = \frac{E}{c}$.

The components of the four-momentum $P_\mu$ can be related to corresponding components of the generalized four-momenta of particles and fields. To express $P_\mu$ in terms of these components we need to

1) Take from (13) or from (38) the generalized four-momentum with a covariant index and write it by components: $p_\mu = (p_0, p_1, p_2, p_3) = (p_0, -\mathbf{p})$. In Cartesian coordinates it turns out $p_\mu = (p_0, -p_x, -p_y, -p_z)$, so that $p_1 = -p_x$, $p_2 = -p_y$, $p_3 = -p_z$, where $\mathbf{p} = (p_x, p_y, p_z)$ is the total generalized momentum of the particles of matter (40).

2) Add to $p_\mu$ another four-vector with a covariant index $K_\mu = (K_0, K_1, K_2, K_3) = (K_0, -\mathbf{K})$. In Cartesian coordinates there will be $K_\mu = (K_0, -K_x, -K_y, -K_z)$, where $\mathbf{K} = (K_x, K_y, K_z)$ is a three-dimensional impulse associated with the fields acting in the system.

As a result, we obtain the following:

$$P_\mu = p_\mu + K_\mu, \qquad E = cp_0 + cK_0, \qquad \mathbf{P} = \mathbf{p} + \mathbf{K}. \qquad (46)$$

Taking into account (36) and (39) for $K_0$ in (46), we can write:

$$K_0 = \frac{E}{c} - p_0 = \frac{1}{c^2} \int \begin{bmatrix} -\mathbf{v} \cdot \dfrac{\partial}{\partial \mathbf{v}}\left(\rho_{0q}\varphi + \rho_0 \psi + \rho_0 \vartheta + \rho_0 \wp\right) + \\ +v^2 \dfrac{\partial}{\partial \mathbf{v}}\left(\rho_{0q}\mathbf{A} + \rho_0 \mathbf{D} + \rho_0 \mathbf{U} + \rho_0 \mathbf{\Pi}\right) \end{bmatrix} u^0 \sqrt{-g}\, dx^1 dx^2 dx^3 +$$

$$+ \frac{1}{c} \int \left( \frac{1}{4\mu_0} F_{\mu\nu} F^{\mu\nu} - \frac{c^2}{16\pi G} \Phi_{\mu\nu} \Phi^{\mu\nu} + \frac{c^2}{16\pi \eta} u_{\mu\nu} u^{\mu\nu} + \frac{c^2}{16\pi \sigma} f_{\mu\nu} f^{\mu\nu} \right) \sqrt{-g}\, dx^1 dx^2 dx^3 +$$

$$+ \frac{1}{c} \sum_{n=1}^{N} \left( \mathbf{v}_n \cdot \frac{\partial L_f}{\partial \mathbf{v}_n} \right).$$

(47)

To determine the vector $\mathbf{K}$, it is necessary to take into account (37), (40) and (46):

$$\mathbf{K} = \mathbf{P} - \mathbf{p} = \frac{1}{c} \int \begin{bmatrix} -\dfrac{\partial}{\partial \mathbf{v}}\left(\rho_{0q}\varphi + \rho_0 \psi + \rho_0 \vartheta + \rho_0 \wp\right) + \\ +\mathbf{v} \cdot \dfrac{\partial}{\partial \mathbf{v}}\left(\rho_{0q}\mathbf{A} + \rho_0 \mathbf{D} + \rho_0 \mathbf{U} + \rho_0 \mathbf{\Pi}\right) \end{bmatrix} u^0 \sqrt{-g}\, dx^1 dx^2 dx^3 + \sum_{n=1}^{N} \left( \frac{\partial L_f}{\partial \mathbf{v}_n} \right).$$

(48)

In a particular case, when the special theory of relativity is valid, the expression of four-momentum $P_\mu$ (45) can be given a more visual meaning. In this case, the system's momentum $\mathbf{P}$ will be directed along the velocity $\mathbf{V}$ of motion of the system's center of momentum, and the four-momentum $P_\mu$ is directed along the four-velocity $u_\mu$ of motion of the system's center of momentum

$$P_\mu = \left( \frac{E}{c}, -\mathbf{P} \right) = \frac{E'}{c^2} u_\mu = \left( u_0 \frac{E'}{c^2}, -u_0 \frac{E'}{c^3} \mathbf{V} \right). \tag{49}$$

In (49), $E'$ denotes the system's energy, calculated using (36) in the center-of-momentum reference frame $O'$; $u_0$ is the time component of four-velocity $u_\mu$ of the center of momentum in reference frame $O$, taken with a covariant index. Representation in the form (49) is possible because, by definition, the momentum of a physical system is zeroed in the reference frame $O'$, the four-momentum has the form $P'_\mu = \left( \frac{E'}{c}, 0 \right)$, and the Lorentz transformation of four-momentum $P'$ into an arbitrary reference frame $O$ leads to (49).

In (49), the following definitions of four-position and four-velocity with covariant indices, valid in the special theory of relativity, were used:

$$x_\mu = (ct, -\mathbf{r}), \qquad u_\mu = \frac{dx_\mu}{d\tau} = \left(c\frac{dt}{d\tau}, -\frac{d\mathbf{r}}{d\tau}\right) = \left(u_0, -\frac{u_0}{c}\frac{d\mathbf{r}}{dt}\right) = \left(u_0, -\frac{u_0}{c}\mathbf{V}\right). \qquad (50)$$

Similar expressions with a contravariant index have the following form:

$$x^\mu = (ct, \mathbf{r}),$$

$$u^\nu = \frac{dx^\nu}{d\tau} = \left(c\frac{dt}{d\tau}, \frac{d\mathbf{r}}{d\tau}\right) = \left(u^0, \frac{u^0}{c}\frac{d\mathbf{r}}{dt}\right) = \left(u^0, \frac{u^0}{c}\mathbf{V}\right) = \left(u^0, \frac{u^0}{c}V^1, \frac{u^0}{c}V^2, \frac{u^0}{c}V^3\right).$$

(51)

In (51), the velocity $\mathbf{V}$ of motion of the center of momentum is expressed in terms of contravariant components in the form $\mathbf{V} = (V^1, V^2, V^3)$. Note that expressions (51) are considered primary in the sense that they are valid even in curved spacetime.

Let us transform the four-velocity (51) of the system's center of momentum into an expression with a covariant index using the metric tensor at the center of momentum:

$$u_\mu = (u_0, u_1, u_2, u_3) = g_{\mu\nu} u^\nu = \left(g_{0\nu} u^\nu, g_{1\nu} u^\nu, g_{2\nu} u^\nu, g_{3\nu} u^\nu\right). \qquad (52)$$

The four-velocity components (52) are as follows:

$$u_0 = g_{0\nu} u^\nu = g_{00} u^0 + g_{01}\frac{u^0}{c}V^1 + g_{02}\frac{u^0}{c}V^2 + g_{03}\frac{u^0}{c}V^3.$$

$$u_1 = g_{1\nu} u^\nu = g_{10} u^0 + g_{11}\frac{u^0}{c}V^1 + g_{12}\frac{u^0}{c}V^2 + g_{13}\frac{u^0}{c}V^3.$$

$$u_2 = g_{2\nu} u^\nu = g_{20} u^0 + g_{21}\frac{u^0}{c}V^1 + g_{22}\frac{u^0}{c}V^2 + g_{23}\frac{u^0}{c}V^3.$$

$$u_3 = g_{3\nu} u^\nu = g_{30} u^0 + g_{31} \frac{u^0}{c} V^1 + g_{32} \frac{u^0}{c} V^2 + g_{33} \frac{u^0}{c} V^3.$$

(53)

From (52-53) it is clear that even in the case when $\mathbf{V} = (V^1, V^2, V^3) = (0,0,0) = 0$ and the center of momentum is stationary in the reference frame $O$, the spatial components $u_1$, $u_2$ and $u_3$, of four-velocity $u_\mu$ may not be equal to zero. A comparison of the components of four-velocity $u_\mu$ (52-53) with the components of $u^\nu$ (51) shows that the spatial components of $u_\mu$ in the general case change asymmetrically with respect to the spatial components of $u^\nu$. This means that the relativistic momentum $\mathbf{P}$ of the system in (45) may not be directed along the velocity $\mathbf{V}$, and then the equality on the right side of (49) does not hold.

From the above it follows that the four-momentum $P_\mu$ is represented by the sum of two integral vectors, the generalized four-vector $p_\mu$ (38) with components in (39-40), and four-vector $K_\mu$ (46) with components in (47-48).

### 3.5. Components of four-momentum with contravariant index

The generalized four-momentum with a contravariant index can be represented in terms of components as follows: $p^\mu = (p^0, p^j) = (p^0, p^1, p^2, p^3)$. Then, the expressions for $p^0$ and $p^j$ follow from (41):

$$p^0 = \frac{1}{c} \int_{V_m} \left( \rho_{0q} A^0 + \rho_0 D^0 + \rho_0 U^0 + \rho_0 \pi^0 \right) u^0 \sqrt{-g}\, dx^1 dx^2 dx^3 =$$
$$= \frac{1}{c} \int_{V_m} \left( \rho_{0q} g^{0\nu} A_\nu + \rho_0 g^{0\nu} D_\nu + \rho_0 g^{0\nu} U_\nu + \rho_0 g^{0\nu} \pi_\nu \right) u^0 \sqrt{-g}\, dx^1 dx^2 dx^3.$$

(54)

$$p^j = \frac{1}{c} \int_{V_m} \left( \rho_{0q} A^j + \rho_0 D^j + \rho_0 U^j + \rho_0 \pi^j \right) u^0 \sqrt{-g}\, dx^1 dx^2 dx^3 =$$
$$= \frac{1}{c} \int_{V_m} \left( \rho_{0q} g^{j\nu} A_\nu + \rho_0 g^{j\nu} D_\nu + \rho_0 g^{j\nu} U_\nu + \rho_0 g^{j\nu} \pi_\nu \right) u^0 \sqrt{-g}\, dx^1 dx^2 dx^3.$$

(55)

In (55), the index $j = 1, 2, 3$ defines space components of the generalized four-momentum with a contravariant index. We can substitute into (54) the time components of fields' four-potentials $A_\nu = \left(\dfrac{\varphi}{c}, -\mathbf{A}\right)$, $D_\nu = \left(\dfrac{\psi}{c}, -\mathbf{D}\right)$, $U_\nu = \left(\dfrac{\vartheta}{c}, -\mathbf{U}\right)$ and $\pi_\nu = \left(\dfrac{\wp}{c}, -\mathbf{\Pi}\right)$. In addition, only in Minkowski spacetime, where metric tensor $g^{\mu\nu}$ has constant diagonal components $(1, -1, -1, -1)$ and other components are equal to zero, does the time component $p^0$ (54) become equal to the time component $p_0$ (39). In this case, the time component $p^0$ up to a factor in the form of the speed of light can be part of the energy $\mathbb{E}$ (36), defining the particles' energy in scalar field potentials. In this regard and in order to simplify the results, all the subsequent arguments apply only to Minkowski spacetime.

Let us determine a four-vector with a contravariant index $K^\mu = (K^0, K^j) = (K^0, K^1, K^2, K^3)$. By analogy with (46), it should be

$$P^\mu = p^\mu + K^\mu, \qquad \mathbb{E} = cp^0 + cK^0, \qquad P^j = p^j + K^j, \tag{56}$$

where the index $j = 1, 2, 3$.

The quantity $K^0$ (56) coincides with $K_0$ (47) because we are now writing the formulas in Minkowski spacetime.

Like in (1), the system's four-momentum with a contravariant index is written as follows:

$$P^\mu = \left(\dfrac{\mathbb{E}}{c}, P^j\right) = (P^0, P^1, P^2, P^3). \tag{57}$$

In Minkowski spacetime, the center of momentum of a physical system moves at a certain constant velocity $\mathbf{V}$, which is part of four-velocity (51). As in (49), we will again assume that the components of system's momentum $P^j$ (57) are directed along the components of velocity $V^j$ of motion of the system's center of momentum, and the four-momentum $P^\mu$ is directed along the four-velocity $u^\mu$ of motion of the system's center of momentum:

$$P^\mu = \left(\dfrac{\mathbb{E}}{c}, P^j\right) = \dfrac{\mathbb{E}'}{c^2} u^\mu = \left(u^0 \dfrac{\mathbb{E}'}{c^2}, u^0 \dfrac{\mathbb{E}'}{c^3} V^j\right). \tag{58}$$

In (58) $\mathbb{E}'$ denotes the system's energy, calculated using (36) in the center-of-momentum frame $O'$; $u^0$ is the time component of four-velocity of the center of momentum, taken with a contravariant index.

From (49) and (58), we can see that different expressions for the same energy $\mathbb{E}$ in the form $u_0 \frac{\mathbb{E}'}{c}$ and $u^0 \frac{\mathbb{E}'}{c}$ are possible because, only in the special theory of relativity for time components of four-velocity with covariant and contravariant indices, the following relation holds: $u_0 = u^0 = \gamma c$, where $\gamma$ is the Lorentz factor of the center of momentum. Moreover, four-momenta (49) and (58) are related by the formula $P^\mu = \eta^{\mu\nu} P_\nu$, where $\eta^{\mu\nu}$ is the metric tensor of Minkowski spacetime.

For a moving material point, the standard expression for four-momentum is $P^\mu = g^{\mu\nu} P_\nu$, where the metric tensor $g^{\mu\nu}$ is taken at the location of the material point. Obviously, for a system with many particles, such a local expression of the four-momentum $P^\mu$ through the metric tensor $g^{\mu\nu}$ at any one point turns out to be unacceptable. For a system of particles, the expression $P^\mu = p^\mu + K^\mu$ in (56), valid in the special theory of relativity, should be used instead of $P^\mu = g^{\mu\nu} P_\nu$. In curved spacetime, defining the four-momentum $P^\mu$ with the contravariant index requires additional assumptions.

### 3.6. Relativistic uniform system at rest

Let us apply the formulas obtained above to calculate the four-momentum of a physical system, which is a relativistic uniform system. To simplify, we perform calculations in Minkowski spacetime, that is, within the framework of the special theory of relativity.

The relativistic uniform system was investigated in a number of papers [11], [21-22] and it has been well studied. It is a physical system of spherical shape consisting of charged particles and fields that is held in equilibrium by its own gravitational field and is counteracted by electromagnetic field, acceleration field and pressure field. All the mentioned fields are considered vector fields, and gravitation is represented in the framework of covariant theory of gravitation [8], [23-25]. It is assumed that the particles are moving randomly and that the global vector potentials $\mathbf{A}$, $\mathbf{D}$, $\mathbf{U}$, and $\mathbf{\Pi}$ of all the fields in the center-of-momentum frame $O'$ are equal to zero. As a result, in the sphere at rest, all solenoidal vectors, such as magnetic field and torsion field (which is called gravitomagnetic field in theory of gravitoelectromagnetism), are also equal to zero.

Since the vector potentials in $O'$ are equal to zero, then, according to (40), $\mathbf{p}' = 0$. For the time component of generalized four-momentum (39), then in $O'$ it was calculated in [17] in the following form:

$$p'_0 = \frac{m_b}{c}\left[\frac{c^2 \gamma_c}{\eta}(\eta + \sigma)\cos\left(\frac{a}{c}\sqrt{4\pi\eta\rho_0}\right) + \wp_c - \frac{\sigma c^2 \gamma_c}{\eta}\right] \approx \frac{m_b}{c}\left(c^2 \gamma_c + \wp_c\right), \qquad (59)$$

where $\gamma_c$ is the Lorentz factor of particles at the center of the sphere, $\eta$ is acceleration field coefficient, $\sigma$ is pressure field coefficient, $a$ is radius of the sphere densely filled with particles, and $\wp_c$ is scalar potential of pressure field at the center of the sphere. The mass $m_b$ is sum of invariant masses of all the system's particles. This mass is equal to gravitational mass $m_g$ of the system and is found with the help of Lorentz factor $\gamma'$ of particles, depending on the current radius. The mass $m_b$ is determined by the following formula:

$$m_b = m_g = \int dm = \int \rho_0 \gamma' dV_m = \frac{c^2 \gamma_c}{\eta}\left[\frac{c}{\sqrt{4\pi\eta\rho_0}}\sin\left(\frac{a}{c}\sqrt{4\pi\eta\rho_0}\right) - a\cos\left(\frac{a}{c}\sqrt{4\pi\eta\rho_0}\right)\right].$$

(60)

The total charge of the sphere is calculated in a similar way as the sum of the invariant charges of all the particles, which are found in the particles' comoving reference frames:

$$q_b = \int dq = \int \rho_{0q} \gamma' dV_m = \frac{c^2 \gamma_c \rho_{0q}}{\eta \rho_0}\left[\frac{c}{\sqrt{4\pi\eta\rho_0}}\sin\left(\frac{a}{c}\sqrt{4\pi\eta\rho_0}\right) - a\cos\left(\frac{a}{c}\sqrt{4\pi\eta\rho_0}\right)\right].$$

(61)

To calculate in the center-of-momentum frame $O'$ the fields' energy $K'_0$, we use the results from [24], [26]. For volume, occupied by matter, we obtain the following:

$$\frac{1}{4\mu_0 c}\int_{r=0}^{a} F'_{\mu\nu} F'^{\mu\nu} dV_m =$$

$$= -\frac{\rho_{0q}^2 c^3 \gamma_c^2}{8\pi\varepsilon_0 \eta^2 \rho_0^2}\left[\frac{a}{2} + \frac{c}{4\sqrt{4\pi\eta\rho_0}}\sin\left(\frac{2a}{c}\sqrt{4\pi\eta\rho_0}\right) - \frac{c^2}{4\pi\eta\rho_0 a}\sin^2\left(\frac{a}{c}\sqrt{4\pi\eta\rho_0}\right)\right].$$

$$-\frac{c}{16\pi G} \int_{r=0}^{a} \Phi'_{\mu\nu} \Phi'^{\mu\nu} dV_m =$$

$$= \frac{Gc^3 \gamma_c^2}{2\eta^2} \left[ \frac{a}{2} + \frac{c}{4\sqrt{4\pi\eta\rho_0}} \sin\left( \frac{2a}{c} \sqrt{4\pi\eta\rho_0} \right) - \frac{c^2}{4\pi\eta\rho_0 a} \sin^2\left( \frac{a}{c} \sqrt{4\pi\eta\rho_0} \right) \right].$$

$$\frac{c}{16\pi\eta} \int_{r=0}^{a} u'_{\mu\nu} u'^{\mu\nu} dV_m =$$

$$= -\frac{c^3 \gamma_c^2}{2\eta} \left[ \frac{a}{2} + \frac{c}{4\sqrt{4\pi\eta\rho_0}} \sin\left( \frac{2a}{c} \sqrt{4\pi\eta\rho_0} \right) - \frac{c^2}{4\pi\eta\rho_0 a} \sin^2\left( \frac{a}{c} \sqrt{4\pi\eta\rho_0} \right) \right].$$

$$\frac{c}{16\pi\sigma} \int_{r=0}^{a} f'_{\mu\nu} f'^{\mu\nu} dV_m =$$

$$= -\frac{\sigma c^3 \gamma_c^2}{2\eta^2} \left[ \frac{a}{2} + \frac{c}{4\sqrt{4\pi\eta\rho_0}} \sin\left( \frac{2a}{c} \sqrt{4\pi\eta\rho_0} \right) - \frac{c^2}{4\pi\eta\rho_0 a} \sin^2\left( \frac{a}{c} \sqrt{4\pi\eta\rho_0} \right) \right].$$

(62)

According to [15], [21], in the system under consideration, the relation between the field coefficients follows from the equation of particle motion:

$$\eta + \sigma = G - \frac{\rho_{0q}^2}{4\pi\varepsilon_0 \rho_0^2}, \qquad (63)$$

where $\varepsilon_0$ is the electric constant.

If we sum the integrals of all tensor invariants in (62) and take into account (63), we obtain zero:

$$\frac{1}{c} \int_{r=0}^{a} \left( \frac{1}{4\mu_0} F_{\mu\nu} F^{\mu\nu} - \frac{c^2}{16\pi G} \Phi_{\mu\nu} \Phi^{\mu\nu} + \frac{c^2}{16\pi\eta} u_{\mu\nu} u^{\mu\nu} + \frac{c^2}{16\pi\sigma} f_{\mu\nu} f^{\mu\nu} \right) \sqrt{-g}\, dx^1 dx^2 dx^3 = 0.$$

(64)

The equation (64) corresponds to the fact that the energy $Z$ in (30) becomes equal to zero. Therefore, in the system under consideration, fields inside the matter will not contribute to the component $K'_0$ according to (47).

Outside matter there are only electromagnetic and gravitational fields, for which instead of (62) taking into account (60-61) we can write:

$$\frac{1}{4\mu_0 c} \int_{r=a}^{\infty} F'_{\mu\nu} F'^{\mu\nu} dV =$$
$$= -\frac{\rho_{0q}^2 c^3 \gamma_c^2}{8\pi\varepsilon_0 \eta^2 \rho_0^2 a} \left[ \frac{c}{\sqrt{4\pi\eta\rho_0}} \sin\left(\frac{a}{c}\sqrt{4\pi\eta\rho_0}\right) - a\cos\left(\frac{a}{c}\sqrt{4\pi\eta\rho_0}\right) \right]^2 = -\frac{q_b^2}{8\pi\varepsilon_0 ca}.$$

(65)

$$-\frac{c}{16\pi G} \int_{r=a}^{\infty} \Phi'_{\mu\nu} \Phi'^{\mu\nu} dV =$$
$$= \frac{Gc^3 \gamma_c^2}{2\eta^2 a} \left[ \frac{c}{\sqrt{4\pi\eta\rho_0}} \sin\left(\frac{a}{c}\sqrt{4\pi\eta\rho_0}\right) - a\cos\left(\frac{a}{c}\sqrt{4\pi\eta\rho_0}\right) \right]^2 = \frac{Gm_b^2}{2ca}.$$

(66)

The sum of (64), (65) and (66) gives the integral of the sum of tensor invariants in (47), taking into account the fields inside and outside the matter:

$$\frac{1}{c} \int \left( \frac{1}{4\mu_0} F_{\mu\nu} F^{\mu\nu} - \frac{c^2}{16\pi G} \Phi_{\mu\nu} \Phi^{\mu\nu} + \frac{c^2}{16\pi\eta} u_{\mu\nu} u^{\mu\nu} + \frac{c^2}{16\pi\sigma} f_{\mu\nu} f^{\mu\nu} \right) \sqrt{-g}\, dx^1 dx^2 dx^3 =$$
$$= \frac{Gm_b^2}{2ca} - \frac{q_b^2}{8\pi\varepsilon_0 ca}.$$

(67)

Taking into account (67) from (47) we find:

$$K'_0 = \frac{1}{c^2} \int \left[ \begin{array}{l} -\mathbf{v}' \cdot \dfrac{\partial}{\partial \mathbf{v}'} \left( \rho_{0q} \varphi' + \rho_0 \psi' + \rho_0 \vartheta' + \rho_0 \wp' \right) + \\ + v'^2 \dfrac{\partial}{\partial \mathbf{v}'} \left( \rho_{0q} \mathbf{A}' + \rho_0 \mathbf{D}' + \rho_0 \mathbf{U}' + \rho_0 \mathbf{\Pi}' \right) \end{array} \right] u'^0 \sqrt{-g}\, dx^1 dx^2 dx^3 +$$

$$+ \frac{Gm_b^2}{2ca} - \frac{q_b^2}{8\pi\varepsilon_0 ca} + \frac{1}{c} \sum_{n=1}^{N} \left( \mathbf{v}'_n \cdot \frac{\partial L_f}{\partial \mathbf{v}'_n} \right).$$

(68)

All the primed quantities are calculated in the center-of-momentum frame $O'$ associated with the center of the sphere.

Within the framework of the special theory of relativity, the global scalar and vector field potentials inside a sphere with chaotically moving particles obey the equations [16]:

$$\partial_\beta \partial^\beta \varphi' = \frac{\gamma' \rho_{0q}}{\varepsilon_0}, \qquad\qquad \partial_\beta \partial^\beta \mathbf{A}' = \mu_0 \mathbf{j}',$$

$$\partial_\beta \partial^\beta \psi' = -4\pi G \gamma' \rho_0, \qquad\qquad \partial_\beta \partial^\beta \mathbf{D}' = -\frac{4\pi G}{c^2} \mathbf{J}',$$

$$\partial_\beta \partial^\beta \vartheta' = 4\pi \eta \gamma' \rho_0, \qquad\qquad \partial_\beta \partial^\beta \mathbf{U}' = \frac{4\pi \eta}{c^2} \mathbf{J}',$$

$$\partial_\beta \partial^\beta \wp' = 4\pi \sigma \gamma' \rho_0, \qquad\qquad \partial_\beta \partial^\beta \mathbf{\Pi}' = \frac{4\pi \sigma}{c^2} \mathbf{J}'. \tag{69}$$

In a stationary and non-rotating sphere under equilibrium conditions, the charge current density $\mathbf{j}'$ and mass current density $\mathbf{J}'$ are equal to zero, since it is assumed that all physical quantities are independent of time, and the directed flows of charge and mass necessary for the emergence of $\mathbf{j}'$ and $\mathbf{J}'$ are absent. As a consequence, the vector potentials $\mathbf{A}'$, $\mathbf{D}'$, $\mathbf{U}'$ and $\mathbf{\Pi}'$ of fields in (69) are equal to zero. The scalar potentials $\varphi'$, $\psi'$, $\vartheta'$ and $\wp'$ of fields in (69) depend on the square $v'^2 = \mathbf{v}'^2$ of the velocity of typical particles at the observation point, since $v'^2$ is included in the Lorentz factor $\gamma' = \dfrac{1}{\sqrt{1 - v'^2/c^2}}$. As a result, at $O'$ in the limit of continuous medium, the global scalar field potentials inside the sphere with randomly moving particles depend on the velocity $\mathbf{v}'$ of typical particles up to terms containing the square $c^2$ of the speed of light in the denominator.

According to (28), the Lagrangian density $\mathcal{L}_f$ depends on the field tensors, each of which is found by calculating the four-rotor from the corresponding four-potential containing scalar and vector potentials. Therefore, in the reference frame $O'$, part $\mathcal{L}_f$ of the Lagrangian density and the corresponding part of the Lagrangian function $L_f = \int \mathcal{L}_f \sqrt{-g}\, dx^1 dx^2 dx^3$ have some weak dependence on velocity $\mathbf{v}'$. In (68) it is required to find the derivatives $\dfrac{\partial}{\partial \mathbf{v}'}$ with respect to the velocities of particles from the field potentials, and when calculating the sum $\dfrac{1}{c}\sum\limits_{n=1}^{N}\left(\mathbf{v}'_n \cdot \dfrac{\partial L_f}{\partial \mathbf{v}'_n}\right)$ it is necessary to find the derivatives $\dfrac{\partial L_f}{\partial \mathbf{v}'_n}$ with respect to the velocities $\mathbf{v}'_n$ of typical particles. This leads to the fact that the time component $K'_0$ (68) acquires small additional terms containing the square of the speed of light in the denominator. In order to simplify calculations, we will not consider such terms, leaving only the main terms.

As a result, the time component $K'_0$ (68) in $O'$ will be approximately equal to

$$K'_0 \approx \frac{Gm_b^2}{2ca} - \frac{q_b^2}{8\pi\varepsilon_0 ca}. \qquad (70)$$

In $O'$ relation (46) must hold for energies: $\mathbb{E}' = c p'_0 + c K'_0$. Hence, taking into account (59) and (70), the energy of the sphere at rest will be equal to:

$$\mathbb{E}' \approx m_b\left[\frac{c^2\gamma_c}{\eta}(\eta+\sigma)\cos\left(\frac{a}{c}\sqrt{4\pi\eta\rho_0}\right) + \wp_c - \frac{\sigma c^2 \gamma_c}{\eta}\right] + \frac{Gm_b^2}{2a} - \frac{q_b^2}{8\pi\varepsilon_0 a}. \qquad (71)$$

According to (71), the relativistic energy $\mathbb{E}'$ of the system at rest is expressed in terms of the total energy of particles in field potentials minus the energy of gravitational and electromagnetic fields outside the matter.

Since in $O'$ both the total momentum, and the generalized momentum are equal to zero, $\mathbf{P}' = 0$, and $\mathbf{p}' = 0$, then according to (46) the field momentum will be equal to zero: $\mathbf{K}' = 0$.

For a fixed sphere $\mathbf{V}=0$, the Lorentz factor $\gamma=1$, the time component of four-velocity of sphere $u_0 = c\gamma = c$, and four-momentum (49) in $O'$ are written as follows:

$$P'_\mu = \left(\frac{E'}{c}, -\mathbf{P}'\right) = \left(\frac{E'}{c}, 0\right). \tag{72}$$

### 3.7. Moving relativistic uniform system

In [17], transformation of the four-velocity of an arbitrary particle from $O'$ to inertial reference frame $O$ was carried out using Lorentz transformations for the case, when the sphere with particles was moving at constant velocity $\mathbf{V}$ along the axis $OX$:

$$u^\mu = \left(\gamma_p c, \gamma_p v_x, \gamma_p v_y, \gamma_p v_z\right) = \left[c\gamma\gamma'\left(1+Vv'_x/c^2\right), \gamma\gamma'\left(v'_x+V\right), \gamma' v'_y, \gamma' v'_z\right]. \tag{73}$$

$$v_x = \frac{v'_x + V}{1+Vv'_x/c^2}, \quad v_y = \frac{v'_y}{\gamma\left(1+Vv'_x/c^2\right)}, \quad v_z = \frac{v'_z}{\gamma\left(1+Vv'_x/c^2\right)}, \quad \gamma_p = \gamma\gamma'\left(1+Vv'_x/c^2\right). \tag{74}$$

Here, $\gamma_p$ denotes the Lorentz factor of the particle in $O$; $v_x$, $v_y$ and $v_z$ are the components of the particle's velocity in $O$; $\gamma$ is the Lorentz factor of the center of momentum, which moves together with the physical system at velocity $\mathbf{V}$; $\gamma'$ is the Lorentz factor of a particle in the center-of-the momentum frame $O'$; $v'_x$, $v'_y$ and $v'_z$ are the components of the particle's velocity in $O'$.

In (73), the time component of the four-velocity of particle is $u^0 = c\gamma\gamma'\left(1+Vv'_x/c^2\right)$. Using this in (38-40), after transformation of fields' four-potentials from the reference frame $O'$ into $O$ and then averaging over the velocities $\mathbf{v}'$ of randomly and multidirectional moving particles, the following was found in [17]:

$$p_0 = \frac{\gamma m_b}{c}\left[\frac{c^2\gamma_c}{\eta}(\eta+\sigma)\cos\left(\frac{a}{c}\sqrt{4\pi\eta\rho_0}\right) + \wp_c - \frac{\sigma c^2\gamma_c}{\eta}\right] \approx \frac{\gamma m_b}{c}\left(c^2\gamma_c + \wp_c\right). \tag{75}$$

In comparison with (59), the component $p_0$ (75) is increased by a factor of $\gamma$ due to the motion of the physical system as a whole at velocity $\mathbf{V}$. Moreover, the relation $p_x = \dfrac{p_0 V}{c}$ is satisfied. Thus, in $O$ we find all the components of the generalized four-momentum for a sphere with particles moving at constant velocity $\mathbf{V}$ along the axis $OX$:

$p_\mu = (p_0, -\mathbf{p}) = (p_0, -p_x, 0, 0) = (p_0, p_1, p_2, p_3)$.

Now we need to calculate the components of the four-vector $K_\mu = (K_0, -\mathbf{K}) = (K_0, -K_x, -K_y, -K_z) = (K_0, K_1, K_2, K_3)$, associated with the energy and momentum of the fields, both in the matter and beyond. According to (47), the time component $K_0$ is found using field four-potentials and field tensors; moreover, to calculate the field tensors the strengths and solenoidal vectors in $O$ are needed.

There are two equivalent methods for determining strengths and solenoidal vectors in $O$. In the first of them, we can take these quantities in $O'$ and then apply the transformation of tensor components from $O'$ into $O$. The other method involves first transforming the fields' four-potentials from $O'$ to $O$ using Lorentz transformations, and then calculating the strengths and solenoidal vectors in $O$ using these four-potentials with the help of four-curl.

For clarity, we use the first method and find the components of electromagnetic tensor in $O$.

In Cartesian coordinates, even in curved spacetime, the following relations are valid for the components of electric field strength $\mathbf{E}$, magnetic field induction $\mathbf{B}$ and electromagnetic field tensor $F_{\mu\nu}$ with covariant indices:

$$\mathbf{E} = -\nabla \varphi - \frac{\partial \mathbf{A}}{\partial t}, \qquad \mathbf{B} = \nabla \times \mathbf{A}. \qquad (76)$$

$$F_{\mu\nu} = \begin{pmatrix} 0 & \dfrac{E_x}{c} & \dfrac{E_y}{c} & \dfrac{E_z}{c} \\ -\dfrac{E_x}{c} & 0 & -B_z & B_y \\ -\dfrac{E_y}{c} & B_z & 0 & -B_x \\ -\dfrac{E_z}{c} & -B_y & B_x & 0 \end{pmatrix}. \qquad (77)$$

The electromagnetic field tensor with contravariant indices can be found, knowing the components of $F_{\mu\nu}$ in (77) and the metric tensor $g^{\eta\mu}$, using the formula:

$$F^{\eta\lambda} = g^{\eta\mu} g^{\lambda\iota} F_{\mu\nu}. \tag{78}$$

In the special theory of relativity, the metric tensor $g^{\eta\mu}$ becomes equal to the tensor $\eta^{\eta\mu}$ of the following form:

$$\eta^{\eta\mu} = \begin{pmatrix} 1 & 0 & 0 & 0 \\ 0 & -1 & 0 & 0 \\ 0 & 0 & -1 & 0 \\ 0 & 0 & 0 & -1 \end{pmatrix}. \tag{79}$$

Substitution of (77) and (79) into (78) gives the tensor expression:

$$F^{\eta\lambda} = \begin{pmatrix} 0 & -\dfrac{E_x}{c} & -\dfrac{E_y}{c} & -\dfrac{E_z}{c} \\ \dfrac{E_x}{c} & 0 & -B_z & B_y \\ \dfrac{E_y}{c} & B_z & 0 & -B_x \\ \dfrac{E_z}{c} & -B_y & B_x & 0 \end{pmatrix}. \tag{80}$$

Let us consider relations (76) in the reference frame $O'$. Since the global vector potentials $\mathbf{A}'$, $\mathbf{D}'$, $\mathbf{U}'$ and $\mathbf{\Pi}'$ of fields in $O'$ are equal to zero, as follows from (69), the vector of electric field strength $\mathbf{E}'_i$ inside the sphere at rest is expressed in (76) in terms of the scalar electric potential $\varphi'_i$, found in [26], according to standard formula for electrostatics:

$$\mathbf{E}'_i = -\nabla \varphi'_i = \frac{\rho_{0q} c^2 \gamma_c \mathbf{r}'}{4\pi \varepsilon_0 \eta \rho_0 r'^3} \left[ \frac{c}{\sqrt{4\pi \eta \rho_0}} \sin\left(\frac{r'}{c}\sqrt{4\pi \eta \rho_0}\right) - r'\cos\left(\frac{r'}{c}\sqrt{4\pi \eta \rho_0}\right) \right] \approx$$

$$\approx \frac{\rho_{0q} \gamma_c \mathbf{r}'}{3\varepsilon_0} \left(1 - \frac{2\pi \eta \rho_0 r'^2}{5c^2}\right).$$

(81)

In (81) $r'$ is the current radius inside the sphere, and the index $i$ indicates that the strength $\mathbf{E}'_i$ and scalar potential $\varphi'_i$ refer to the internal field of the sphere. Since $\mathbf{A}' = 0$, then according to (76) in $O'$ the magnetic field is equal to zero everywhere, $\mathbf{B}'_i = 0$.

The components of the antisymmetric electromagnetic tensor $F'^{\mu\nu}$ in $O'$ in the special theory of relativity are expressed according to (80) in terms of components of vectors $\mathbf{E}'_i$ and $\mathbf{B}'_i$ as follows:

$$F'^{\mu\nu} = \begin{pmatrix} 0 & -\dfrac{E'_{ix}}{c} & -\dfrac{E'_{iy}}{c} & -\dfrac{E'_{iz}}{c} \\ \dfrac{E'_{ix}}{c} & 0 & -B'_{iz} & B'_{iy} \\ \dfrac{E'_{iy}}{c} & B'_{iz} & 0 & -B'_{ix} \\ \dfrac{E'_{iz}}{c} & -B'_{iy} & B'_{ix} & 0 \end{pmatrix}. \qquad (82)$$

$$F'^{01} = -\frac{E'_{ix}}{c}, \quad F'^{02} = -\frac{E'_{iy}}{c}, \quad F'^{03} = -\frac{E'_{iz}}{c}, \quad F'^{12} = -B'_{iz}, \quad F'^{23} = -B'_{ix}, \quad F'^{31} = -B'_{iy}.$$

$$(83)$$

The Lorentz transformation of tensor components from $O'$ to $O$ is carried out according to standard formulas (§ 24. Lorentz transformation of the field, in [2]):

$$F^{01} = F'^{01}, \quad F^{02} = \gamma F'^{02} + \gamma V F'^{12}/c, \quad F^{03} = \gamma F'^{03} + \gamma V F'^{13}/c,$$

$$F^{12} = \gamma F'^{12} + \gamma V F'^{02}/c, \quad F^{23} = F'^{23}, \quad F^{31} = \gamma F'^{31} + \gamma V F'^{30}/c. \qquad (84)$$

Substituting (83) into (84), in view of relations $F'^{13} = -F'^{31}$, $F'^{30} = -F'^{03}$, and $\mathbf{B}'_i = 0$, gives the following:

$$E_{ix} = E'_{ix}, \quad E_{iy} = \gamma E'_{iy}, \quad E_{iz} = \gamma E'_{iz},$$

$$B_{ix} = B'_{ix} = 0, \qquad B_{iy} = -\frac{\gamma V E'_{iz}}{c^2}, \qquad B_{iz} = \frac{\gamma V E'_{iy}}{c^2}. \qquad (85)$$

According to (85), due to the motion of the sphere with an internal electric field, a magnetic field appears in the reference frame $O$, although in the reference frame $O'$ associated with the sphere there is no magnetic field. This is a consequence of the principle of relativity in relation to the components of the electromagnetic field, when the own electric field of a moving object generates an additional magnetic field in another reference frame, and the own magnetic field of a moving object generates an additional electric field in another reference frame. In this case, the additional fields turn out to be proportional to the velocity $\mathbf{V}$ of the object.

The contribution to $K_0$ (47) from the tensor invariant of electromagnetic field in the reference frame $O$ is as follows:

$$\frac{1}{4\mu_0 c} \int_{r=0}^{a} F_{\mu\nu} F^{\mu\nu} dV_m = -\frac{\varepsilon_0}{2c} \int_{r=0}^{a} (E_i^2 - c^2 B_i^2) dV_m. \qquad (86)$$

In (86), expressions $F_{\mu\nu}$ (77) and $F^{\mu\nu}$ (80) were taken into account, for which we obtain $F_{\mu\nu} F^{\mu\nu} = 2B^2 - \frac{2E^2}{c^2}$. The subscript $i$ in $E_i$ and in $B_i$ (86) indicates that the electric field strength and magnetic field induction are taken inside the moving sphere. The magnetic constant $\mu_0$ and the electric constant $\varepsilon_0$ are related to each other and to the square of the speed of light by the relation $c^2 \mu_0 \varepsilon_0 = 1$.

Let us calculate the quantities $E_i^2 = E_{ix}^2 + E_{iy}^2 + E_{iz}^2$ and $B_i^2 = B_{ix}^2 + B_{iy}^2 + B_{iz}^2$ using the components $\mathbf{E}_i$ and $\mathbf{B}_i$ (85), and substitute $E_i^2$ and $B_i^2$ in (86), taking into account the expression for the Lorentz factor $\gamma = \frac{1}{\sqrt{1 - V^2/c^2}}$:

$$\frac{1}{4\mu_0 c} \int_{r=0}^{a} F_{\mu\nu} F^{\mu\nu} dV_m = -\frac{\varepsilon_0}{2c} \int_{r=0}^{a} (E_{ix}'^2 + E_{iy}'^2 + E_{iz}'^2) dV_m = -\frac{\varepsilon_0}{2c} \int_{r=0}^{a} E_i'^2 dV_m. \qquad (87)$$

In contrast to (62), in (87) the integration is carried out over the moving volume of the sphere. A moving sphere in the special theory of relativity is considered a Heaviside ellipsoid.

Like in [27-28], we introduce in $O$ new coordinates $r, \theta, \phi$, associated with Cartesian coordinates:

$$x - Vt = \frac{1}{\gamma} r \cos\theta, \qquad y = r \sin\theta \cos\phi, \qquad z = r \sin\theta \sin\phi. \qquad (88)$$

The volume element in these coordinates is defined by the formula $dV_m = \frac{1}{\gamma} r^2 \sin\theta \, dr \, d\theta \, d\phi$. The equation of the Heaviside ellipsoid surface, in view of (88), is as follows:

$$\gamma^2 (x - Vt)^2 + y^2 + z^2 = a^2, \qquad r = a. \qquad (89)$$

Thus, the limits of integration over the sphere's volume in new coordinates will be as follows: radius $r$ should vary from 0 to $a$, and angles $\theta$ and $\phi$ vary the same as in spherical coordinates (from 0 to $\pi$ for the angle $\theta$ and from 0 to $2\pi$ for the angle $\phi$).

If in $O'$ we denote the current radius by $r' = \sqrt{x'^2 + y'^2 + z'^2}$, and express the coordinates $x', y', z'$ in terms of the coordinates $x, y, z$ in $O$ with the help of Lorentz transformations and use (88), we obtain the following:

$$r' = \sqrt{\gamma^2 (x - Vt)^2 + y^2 + z^2} = r. \qquad (90)$$

According to (90), instead of $r'$ we can use the coordinate $r$ in (81), after which we can substitute the vector $\mathbf{E}'_i$ into (87). Considering the relation $dV_m = \frac{1}{\gamma} r^2 \sin\theta \, dr \, d\theta \, d\phi$, we obtain:

$$\frac{1}{4\mu_0 c} \int_{r=0}^{a} F_{\mu\nu} F^{\mu\nu} dV_m =$$

$$= -\frac{\rho_{0q}^2 c^3 \gamma_c^2}{8\pi \varepsilon_0 \eta^2 \rho_0^2 \gamma} \int_{r=0}^{a} \frac{1}{r^2} \left[ \frac{c}{\sqrt{4\pi \eta \rho_0}} \sin\left(\frac{r}{c}\sqrt{4\pi \eta \rho_0}\right) - r\cos\left(\frac{r}{c}\sqrt{4\pi \eta \rho_0}\right) \right]^2 dr. \qquad (91)$$

Similarly, we can repeat the same steps for remaining fields. In $O'$ gravitational field strength inside the sphere, strengths of acceleration field and pressure field are expressed in terms of scalar potentials, as found in [24], [26], [29]:

$$\boldsymbol{\Gamma}'_i = -\nabla \psi'_i = -\frac{Gc^2 \gamma_c \mathbf{r}'}{\eta r'^3}\left[\frac{c}{\sqrt{4\pi\eta\rho_0}}\sin\left(\frac{r'}{c}\sqrt{4\pi\eta\rho_0}\right) - r'\cos\left(\frac{r'}{c}\sqrt{4\pi\eta\rho_0}\right)\right] \approx$$

$$\approx -\frac{4\pi G \rho_0 \gamma_c \mathbf{r}'}{3}\left(1 - \frac{2\pi\eta\rho_0 r'^2}{5c^2}\right).$$

$$\mathbf{S}' = -\nabla \vartheta' = \frac{c^2 \gamma_c \mathbf{r}'}{r'^3}\left[\frac{c}{\sqrt{4\pi\eta\rho_0}}\sin\left(\frac{r'}{c}\sqrt{4\pi\eta\rho_0}\right) - r'\cos\left(\frac{r'}{c}\sqrt{4\pi\eta\rho_0}\right)\right] \approx$$

$$\approx \frac{4\pi\eta\rho_0 \gamma_c \mathbf{r}'}{3}\left(1 - \frac{2\pi\eta\rho_0 r'^2}{5c^2}\right).$$

$$\mathbf{C}' = -\nabla \wp' = \frac{\sigma c^2 \gamma_c \mathbf{r}'}{\eta r'^3}\left[\frac{c}{\sqrt{4\pi\eta\rho_0}}\sin\left(\frac{r'}{c}\sqrt{4\pi\eta\rho_0}\right) - r'\cos\left(\frac{r'}{c}\sqrt{4\pi\eta\rho_0}\right)\right] \approx$$

$$\approx \frac{4\pi\sigma\rho_0 \gamma_c \mathbf{r}'}{3}\left(1 - \frac{2\pi\eta\rho_0 r'^2}{5c^2}\right).$$

(92)

Like in (84-85), the field's strengths and solenoidal vectors in the reference frame $O$ are equal to:

$$\Gamma_{ix} = \Gamma'_{ix}, \qquad \Gamma_{iy} = \gamma \Gamma'_{iy}, \qquad \Gamma_{iz} = \gamma \Gamma'_{iz},$$

$$\Omega_{ix} = \Omega'_{ix} = 0, \qquad \Omega_{iy} = -\frac{\gamma V \Gamma'_{iz}}{c^2}, \qquad \Omega_{iz} = \frac{\gamma V \Gamma'_{iy}}{c^2}.$$

$$S_x = S'_x, \qquad S_y = \gamma S'_y, \qquad S_z = \gamma S'_z,$$

$$N_x = N'_x = 0, \qquad N_y = -\frac{\gamma V S'_z}{c^2}, \qquad N_z = \frac{\gamma V S'_y}{c^2}.$$

$$C_x = C'_x, \qquad C_y = \gamma C'_y, \qquad C_z = \gamma C'_z,$$

$$I_x = I'_x = 0, \qquad I_y = -\frac{\gamma V C'_z}{c^2}, \qquad I_z = \frac{\gamma V C'_y}{c^2}.$$

(93)

Taking into account (86-91), as well as (92-93), the results of integrating tensor invariants over the moving sphere's volume for three remaining fields are as follows:

$$-\frac{c}{16\pi G}\int_{r=0}^{a}\Phi_{\mu\nu}\Phi^{\mu\nu}dV_m = \frac{1}{8\pi Gc}\int_{r=0}^{a}(\Gamma_i^2 - c^2\Omega_i^2)dV_m = \frac{1}{8\pi Gc}\int_{r=0}^{a}\Gamma_i'^2 dV_m =$$

$$= \frac{Gc^3\gamma_c^2}{2\eta^2\gamma}\int_{r=0}^{a}\frac{1}{r^2}\left[\frac{c}{\sqrt{4\pi\eta\rho_0}}\sin\left(\frac{r}{c}\sqrt{4\pi\eta\rho_0}\right) - r\cos\left(\frac{r}{c}\sqrt{4\pi\eta\rho_0}\right)\right]^2 dr.$$

$$\frac{c}{16\pi\eta}\int_{r=0}^{a}u_{\mu\nu}u^{\mu\nu}dV_m = -\frac{1}{8\pi\eta c}\int_{r=0}^{a}(S^2 - c^2 N^2)dV_m = -\frac{1}{8\pi\eta c}\int_{r=0}^{a}S'^2 dV_m =$$

$$= -\frac{c^3\gamma_c^2}{2\eta\gamma}\int_{r=0}^{a}\frac{1}{r^2}\left[\frac{c}{\sqrt{4\pi\eta\rho_0}}\sin\left(\frac{r}{c}\sqrt{4\pi\eta\rho_0}\right) - r\cos\left(\frac{r}{c}\sqrt{4\pi\eta\rho_0}\right)\right]^2 dr.$$

$$\frac{c}{16\pi\sigma}\int_{r=0}^{a}f_{\mu\nu}f^{\mu\nu}dV_m = -\frac{1}{8\pi\sigma c}\int_{r=0}^{a}(C^2 - c^2 I^2)dV_m = -\frac{1}{8\pi\sigma c}\int_{r=0}^{a}C'^2 dV_m =$$

$$= -\frac{\sigma c^3\gamma_c^2}{2\eta^2\gamma}\int_{r=0}^{a}\frac{1}{r^2}\left[\frac{c}{\sqrt{4\pi\eta\rho_0}}\sin\left(\frac{r}{c}\sqrt{4\pi\eta\rho_0}\right) - r\cos\left(\frac{r}{c}\sqrt{4\pi\eta\rho_0}\right)\right]^2 dr.$$

(94)

Let us sum the terms in (91) and (94) and take into account (63):

$$\frac{1}{c}\int_{r=0}^{a}\left(\frac{1}{4\mu_0}F_{\mu\nu}F^{\mu\nu} - \frac{c^2}{16\pi G}\Phi_{\mu\nu}\Phi^{\mu\nu} + \frac{c^2}{16\pi\eta}u_{\mu\nu}u^{\mu\nu} + \frac{c^2}{16\pi\sigma}f_{\mu\nu}f^{\mu\nu}\right)dV_m =$$

$$= \frac{c^3\gamma_c^2}{2\eta^2\gamma}\left(G - \frac{\rho_{0q}^2}{4\pi\varepsilon_0\rho_0^2} - \eta - \sigma\right)\int_{r=0}^{a}\frac{1}{r^2}\left[\frac{c}{\sqrt{4\pi\eta\rho_0}}\sin\left(\frac{r}{c}\sqrt{4\pi\eta\rho_0}\right) - r\cos\left(\frac{r}{c}\sqrt{4\pi\eta\rho_0}\right)\right]^2 dr = 0.$$

Sum (95) is included as an integral part in $K_0$ (47). Thus, in matter inside the moving sphere the sum of contributions to $K_0$ from all the fields becomes equal to zero, as in the case of the sphere at rest when summing integrals of all tensor invariants in (64).

Now it is necessary to consider in $K_0$ electromagnetic and gravitational fields outside the sphere. In the center-of-momentum frame $O'$, there are both an external electric field strength and an external gravitational field strength:

$$\mathbf{E}'_o = -\nabla \varphi'_o = \frac{\rho_{0q} c^2 \gamma_c \mathbf{r}'}{4\pi \varepsilon_0 \rho_0 \eta\, r'^3} \left[ \frac{c}{\sqrt{4\pi \eta \rho_0}} \sin\left(\frac{a}{c}\sqrt{4\pi \eta \rho_0}\right) - a\cos\left(\frac{a}{c}\sqrt{4\pi \eta \rho_0}\right) \right] = \frac{q_b \mathbf{r}'}{4\pi \varepsilon_0 r'^3}.$$

$$\mathbf{\Gamma}'_o = -\nabla \psi'_o = -\frac{G c^2 \gamma_c \mathbf{r}'}{\eta\, r'^3} \left[ \frac{c}{\sqrt{4\pi \eta \rho_0}} \sin\left(\frac{a}{c}\sqrt{4\pi \eta \rho_0}\right) - a\cos\left(\frac{a}{c}\sqrt{4\pi \eta \rho_0}\right) \right] = -\frac{G m_b \mathbf{r}'}{r'^3}.$$

(96)

In (96), the index $o$ indicates that a quantity refers to space outside the matter. In the reference frame $O$, in which the sphere is moving at constant velocity $\mathbf{V}$ along the axis $OX$, similar to (85) and (93) a magnetic field $\mathbf{B}_o$ appears, as does a torsion field $\mathbf{\Omega}_o$:

$$E_{ox} = E'_{ox}, \qquad E_{0y} = \gamma E'_{oy}, \qquad E_{oz} = \gamma E'_{oz},$$

$$B_{ox} = B'_{ox} = 0, \qquad B_{oy} = -\frac{\gamma V E'_{oz}}{c^2}, \qquad B_{oz} = \frac{\gamma V E'_{oy}}{c^2}.$$

$$\Gamma_{ox} = \Gamma'_{ox}, \qquad \Gamma_{0y} = \gamma \Gamma'_{oy}, \qquad \Gamma_{oz} = \gamma \Gamma'_{oz},$$

$$\Omega_{ox} = \Omega'_{ox} = 0, \qquad \Omega_{oy} = -\frac{\gamma V \Gamma'_{oz}}{c^2}, \qquad \Omega_{oz} = \frac{\gamma V \Gamma'_{oy}}{c^2}. \tag{97}$$

Taking into account relations (87-90), as well as (97) and the relation $dV_o = \frac{1}{\gamma} r^2 \sin\theta \, dr \, d\theta \, d\phi$, for integrals of tensor invariants of external fields of moving sphere we find the following:

$$\frac{1}{4\mu_0 c} \int_{r=a}^{\infty} F_{\mu\nu} F^{\mu\nu} dV_o = -\frac{\varepsilon_0}{2c} \int_{r=a}^{\infty} (E_o^2 - c^2 B_o^2) dV_o = -\frac{\varepsilon_0}{2c} \int_{r=a}^{\infty} E_o'^2 dV_o = -\frac{q_b^2}{8\pi\varepsilon_0 ca\gamma}.$$

$$-\frac{c}{16\pi G} \int_{r=a}^{\infty} \Phi_{\mu\nu} \Phi^{\mu\nu} dV_o = \frac{1}{8\pi G c} \int_{r=a}^{\infty} (\Gamma_o^2 - c^2 \Omega_o^2) dV_o = \frac{1}{8\pi G c} \int_{r=a}^{\infty} \Gamma_0'^2 dV_o = \frac{Gm_b^2}{2ca\gamma}.$$

(98)

Substituting (98) into (47) and taking into account (95) for the fields inside the sphere, in $O$ we determine the time component $K_0$:

$$K_0 = \frac{1}{c^2} \int \left[ \begin{array}{l} -\mathbf{v} \cdot \dfrac{\partial}{\partial \mathbf{v}} \left( \rho_{0q} \varphi_i + \rho_0 \psi_i + \rho_0 \vartheta + \rho_0 \wp \right) + \\ +v^2 \dfrac{\partial}{\partial \mathbf{v}} \left( \rho_{0q} \mathbf{A}_i + \rho_0 \mathbf{D}_i + \rho_0 \mathbf{U} + \rho_0 \mathbf{\Pi} \right) \end{array} \right] u^0 \sqrt{-g} \, dx^1 dx^2 dx^3 +$$

$$+ \frac{Gm_b^2}{2ca\gamma} - \frac{q_b^2}{8\pi\varepsilon_0 ca\gamma} + \frac{1}{c} \sum_{n=1}^{N} \left( \mathbf{v}_n \cdot \frac{\partial L_f}{\partial \mathbf{v}_n} \right).$$

(99)

In the system under consideration $\rho_{0q} = const$, $\rho_0 = const$, and the continuous medium approximation is used, while in the center-of-momentum frame $O'$, in accordance with (69), the vector potentials are considered equal to zero. At the same time, the scalar field potentials at the location of a particle weakly depend on the speed $\mathbf{v}'$ of this particle, and are determined up to terms containing the square of the speed of light in the denominator. The situation does not change in the reference frame $O$; although potentials become dependent on the velocity $\mathbf{V}$ of the sphere's motion, they still weakly depend on the components of particles' velocity $\mathbf{v} = (v_x, v_y, v_z)$ presented in (74). Neglecting small terms, we find an approximate expression for $K_0$ (99):

$$K_0 \approx \frac{Gm_b^2}{2ca\gamma} - \frac{q_b^2}{8\pi\varepsilon_0 ca\gamma} + \frac{1}{c}\sum_{n=1}^{N}\left(\mathbf{v}_n \cdot \frac{\partial L_f}{\partial \mathbf{v}_n}\right). \qquad (100)$$

Let us represent $L_f$ in the form $L_f = L_{fi} + L_{fo}$. Inside the sphere, taking into account (2) and (28), as well as conditions for energy gauging and metrics [11], [20] in the form $ckR = 2ck\Lambda$, we obtain:

$$L_{fi} = \int_{V_m} \mathcal{L}_{fi} \sqrt{-g}\, dx^1 dx^2 dx^3 =$$
$$= -\int_{V_m}\left(\frac{1}{4\mu_0}F_{\mu\nu}F^{\mu\nu} - \frac{c^2}{16\pi G}\Phi_{\mu\nu}\Phi^{\mu\nu} + \frac{c^2}{16\pi\eta}u_{\mu\nu}u^{\mu\nu} + \frac{c^2}{16\pi\sigma}f_{\mu\nu}f^{\mu\nu}\right)\sqrt{-g}\, dx^1 dx^2 dx^3.$$

(101)

If we take into account (95), then inside the moving sphere we obtain $L_{fi} = 0$, so that in this case $L_{fi}$ (101) does not contribute to $L_f$ in (100). In view of (98), we find $L_{fo}$ in space outside the sphere, where there are only electromagnetic and gravitational fields:

$$L_{fo} = -\int\left(\frac{1}{4\mu_0}F_{\mu\nu}F^{\mu\nu} - \frac{c^2}{16\pi G}\Phi_{\mu\nu}\Phi^{\mu\nu}\right)\sqrt{-g}\, dx^1 dx^2 dx^3 = \frac{q_b^2}{8\pi\varepsilon_0 a\gamma} - \frac{Gm_b^2}{2a\gamma}. \qquad (102)$$

In (102) the Lorentz factor $\gamma = \dfrac{1}{\sqrt{1 - V^2/c^2}}$ is present, which is a function of the velocity $\mathbf{V}$ of the sphere motion along the axis $OX$ of reference frame $O$. According to (74), the particle velocity in $O$ is equal to:

$$\mathbf{v} = (v_x, v_y, v_z) = \left[\frac{v'_x + V}{1 + Vv'_x/c^2}, \frac{v'_y}{\gamma(1 + Vv'_x/c^2)}, \frac{v'_z}{\gamma(1 + Vv'_x/c^2)}\right]. \qquad (103)$$

If we average velocities $\mathbf{v}$ of neighboring particles directed in all directions, in (103) we obtain $\overline{\mathbf{v}} = (\overline{v}_x, \overline{v}_y, \overline{v}_z) \approx (V, 0, 0)$. This can be explained by the following calculations:

$$\bar{v}_x = \left\langle \frac{v'_x + V}{1 + Vv'_x/c^2} \right\rangle \approx \left\langle (v'_x + V)\left(1 - \frac{Vv'_x}{c^2}\right) \right\rangle \approx \left\langle v'_x + V - \frac{Vv'_x(v'_x + V)}{c^2} \right\rangle \approx$$

$$\approx \langle v'_x \rangle + \langle V \rangle - \left\langle \frac{Vv'^2_x}{c^2} \right\rangle - \frac{V^2}{c^2}\langle v'_x \rangle \approx V - \frac{V\langle v'^2_x \rangle}{c^2} \approx V.$$

(104)

$$\bar{v}_y = \left\langle \frac{v'_y}{\gamma(1 + Vv'_x/c^2)} \right\rangle \approx \left\langle \frac{v'_y\left(1 - \frac{Vv'_x}{c^2}\right)}{\gamma} \right\rangle \approx \frac{1}{\gamma}\langle v'_y \rangle - \frac{V}{c^2\gamma}\langle v'_x v'_y \rangle \approx$$

$$\approx -\frac{V}{c^2\gamma}\langle v'_x \rangle \langle v'_y \rangle \approx 0.$$

$$\bar{v}_z = \left\langle \frac{v'_z}{\gamma(1 + Vv'_x/c^2)} \right\rangle \approx \left\langle \frac{v'_z\left(1 - \frac{Vv'_x}{c^2}\right)}{\gamma} \right\rangle \approx \frac{1}{\gamma}\langle v'_z \rangle - \frac{V}{c^2\gamma}\langle v'_x v'_z \rangle \approx$$

$$\approx -\frac{V}{c^2\gamma}\langle v'_x \rangle \langle v'_z \rangle \approx 0.$$

(105)

In (104) it is assumed that after averaging over all directions $\langle v'_x \rangle = 0$, and that the small term $\frac{V\langle v'^2_x \rangle}{c^2}$ can be neglected. In (105), it is similarly assumed that the average values of $\langle v'_y \rangle = 0$ and $\langle v'_z \rangle = 0$, , and can be neglected by small terms containing the square of the speed of light in the denominator.

Since $L_f = L_{fi} + L_{fo}$, $L_{fi} = 0$, in (100) it is necessary to calculate the sum $\frac{1}{c}\sum_{n=1}^{N}\left(\mathbf{v}_n \cdot \frac{\partial L_{fo}}{\partial \mathbf{v}_n}\right)$, including the derivative $\frac{\partial L_{fo}}{\partial \mathbf{v}_n}$ with respect to the particles' velocities $\mathbf{v}_n$. However, $L_{fo}$ is a result of averaging over the velocities of individual particles, and to a first approximation depends only on $V$. Fields outside the sphere look as if they are created by one body without internal motion of particles, and this body moves at velocity $\mathbf{V} = (V, 0, 0) \approx \bar{\mathbf{v}}$ and has a mass and charge equal to the sum of the masses and charges of individual particles of the system. This allows us to replace this sum with its average value by replacing $\mathbf{v}_n$ with $\bar{\mathbf{v}}$:

$$\left\langle \frac{1}{c} \sum_{n=1}^{N} \left( \mathbf{v}_n \cdot \frac{\partial L_{fo}}{\partial \mathbf{v}_n} \right) \right\rangle \approx \frac{1}{c} \overline{\mathbf{v}} \cdot \frac{\partial L_{fo}}{\partial \overline{\mathbf{v}}} \approx \frac{V}{c} \frac{\partial L_{fo}}{\partial V}. \tag{106}$$

To better understand (106), we can consider the electromagnetic field outside a moving sphere with radius $a$, uniformly and symmetrically filled with a large number $N$ of charged particles, each of which has a charge $q$. Due to symmetry taking into account Gauss's theorem, the field outside the sphere will be the same as if all the charges of the sphere were placed at the center of the sphere and there would be the charge $Q = Nq$. So the sphere with $N$ charges in relation to the field outside the sphere becomes equivalent to one charge $Q$, which has a point particle radius much smaller than the radius $a$. In this case, we can assume that the sum in (106) contains only one term for one particle with charge $Q = Nq$, and $L_{fo}$ is a Lagrange function for a point charged particle moving with the velocity $\mathbf{V}$. The field of such a particle with the charge $Q$ is equivalent to the field of the moving sphere, which allows us to replace the sum of terms in (106) by one term.

Substituting into (106) $L_{fo}$ from (102), taking into account the expression for the Lorentz factor $\gamma = \dfrac{1}{\sqrt{1 - V^2/c^2}}$, we find:

$$\frac{\partial L_{fo}}{\partial V} = \left( \frac{q_b^2}{8\pi\varepsilon_0 a} - \frac{Gm_b^2}{2a} \right) \frac{\partial}{\partial V} \left( \frac{1}{\gamma} \right) = \frac{\gamma V}{c^2} \left( \frac{Gm_b^2}{2a} - \frac{q_b^2}{8\pi\varepsilon_0 a} \right).$$

$$\frac{1}{c} \sum_{n=1}^{N} \left( \mathbf{v}_n \cdot \frac{\partial L_f}{\partial \mathbf{v}_n} \right) = \left\langle \frac{1}{c} \sum_{n=1}^{N} \left( \mathbf{v}_n \cdot \frac{\partial L_{fo}}{\partial \mathbf{v}_n} \right) \right\rangle \approx \frac{\gamma V^2}{c^3} \left( \frac{Gm_b^2}{2a} - \frac{q_b^2}{8\pi\varepsilon_0 a} \right). \tag{107}$$

This sum (107) is presented in (100), which allows us to clarify the form $K_0$:

$$K_0 \approx \frac{\gamma}{c} \left( \frac{Gm_b^2}{2a} - \frac{q_b^2}{8\pi\varepsilon_0 a} \right). \tag{108}$$

Taking into account (75) and (108), which should be substituted into (46), the energy of the moving sphere will be equal to:

$$\mathbb{E} = c p_0 + c K_0 \approx \gamma m_b \left[ \frac{c^2 \gamma_c}{\eta}(\eta+\sigma)\cos\left(\frac{a}{c}\sqrt{4\pi\eta\rho_0}\right) + \wp_c - \frac{\sigma c^2 \gamma_c}{\eta} \right] + \frac{\gamma G m_b^2}{2a} - \frac{\gamma q_b^2}{8\pi\varepsilon_0 a}. \tag{109}$$

By comparing (109) and (71) for the case of the sphere at rest we can see that the energy of moving sphere increases by a factor of $\gamma$.

Now we use (48) to calculate the vector $\mathbf{K}$, again assuming that the charge density, mass density, and field potentials inside the moving sphere in the first approximation do not depend on velocities of individual particles at the integration point. Then in (48) the integral vanishes and the following equation holds:

$$\mathbf{K} \approx \sum_{n=1}^{N} \left( \frac{\partial L_f}{\partial \mathbf{v}_n} \right). \tag{110}$$

In (110) $L_f = L_{fi} + L_{fo}$, and from (95) it follows that inside the sphere $L_{fi}$ in (101) is equal to zero. Considering the average value of the sum in (110), taking into account the expression for velocity $\mathbf{V} = (V,0,0) \approx \bar{\mathbf{v}}$, the value $L_{fo}$ (102) and $\gamma = \frac{1}{\sqrt{1-V^2/c^2}}$, we find:

$$\mathbf{K} \approx \left\langle \sum_{n=1}^{N} \left( \frac{\partial L_{fo}}{\partial \mathbf{v}_n} \right) \right\rangle \approx \frac{\partial L_{fo}}{\partial \bar{\mathbf{v}}} \approx \frac{\partial L_{fo}}{\partial \mathbf{V}},$$

$$K_x \approx \frac{\gamma V}{c^2}\left( \frac{G m_b^2}{2a} - \frac{q_b^2}{8\pi\varepsilon_0 a} \right), \qquad K_y = 0, \qquad K_z = 0. \tag{111}$$

Comparison of (108) and (111) gives the relations:

$$K_x = \frac{K_0 V}{c}, \qquad K_\mu = (K_0, -\mathbf{K}) = (K_0, -K_x, 0, 0) = (K_0, K_1, K_2, K_3). \tag{112}$$

In (75), the component $p_0$ was calculated, which allows determination of the component $p_x$ for generalized four-momentum $p_\mu$ of moving sphere:

$$p_x = \frac{p_0 V}{c}, \qquad p_\mu = (p_0, -\mathbf{p}) = (p_0, -p_x, 0, 0) = (p_0, p_1, p_2, p_3). \qquad (113)$$

Relations (112) for $K_\mu$ have the same form as relations for $p_\mu$ in (113).

Adding vectors $\mathbf{p}$ and $\mathbf{K}$ according to (46), we find the momentum of the system:

$$\mathbf{P} = \frac{\gamma m_b}{c^2}\left[\frac{c^2 \gamma_c}{\eta}(\eta + \sigma)\cos\left(\frac{a}{c}\sqrt{4\pi\eta\rho_0}\right) + \wp_c - \frac{\sigma c^2 \gamma_c}{\eta}\right]\mathbf{V} + \frac{\gamma}{c^2}\left(\frac{G m_b^2}{2a} - \frac{q_b^2}{8\pi\varepsilon_0 a}\right)\mathbf{V}. \qquad (114)$$

According to the method of its calculation in (13), the generalized four-momentum $p_\mu$ (113) is a nonlocal four-vector of integral type with a covariant index associated with the interaction of particles with each other and with fields. Nonlocality of $p_\mu$ is a consequence of its definition through the integral over the volume occupied by all typical particles of the system. The four-vector $K_\mu$ (112), which specifies the four-momentum of the fields of the system, has the same properties. The sum of four-vectors $p_\mu$ and $K_\mu$ в (46) gives the total four-momentum of the physical system $P_\mu$ with the covariant index.

### 3.8. Integral vector inside moving sphere

In [7] we calculated the integral vector $\mathcal{J}_\alpha$ (9) for a fixed sphere within the framework of the special theory of relativity. Now we determine this vector for the case of motion of the sphere with particles at constant velocity $\mathbf{V}$ along the $OX$ axis in the reference frame $O$. The stress-energy tensor of a physical system consists of the sum of the stress-energy tensors of electromagnetic and gravitational fields, acceleration field and pressure field. Taking into account the metric signature $(+---)$ we use, we can write:

$$T_\alpha^{\ \beta} = W_\alpha^{\ \beta} + U_\alpha^{\ \beta} + B_\alpha^{\ \beta} + P_\alpha^{\ \beta}. \qquad (115)$$

$$W_\alpha^{\ \beta} = \varepsilon_0 c^2\, g^{\mu\kappa}\left(-\delta_\alpha^\lambda g^{\sigma\beta} + \frac{1}{4}\delta_\alpha^\beta g^{\sigma\lambda}\right)F_{\mu\lambda} F_{\kappa\sigma},$$

$$U_\alpha{}^\beta = -\frac{c^2}{4\pi G} g^{\mu\kappa} \left( -\delta_\alpha^\lambda g^{\sigma\beta} + \frac{1}{4} \delta_\alpha^\beta g^{\sigma\lambda} \right) \Phi_{\mu\lambda} \Phi_{\kappa\sigma},$$

$$B_\alpha{}^\beta = \frac{c^2}{4\pi\eta} g^{\mu\kappa} \left( -\delta_\alpha^\lambda g^{\sigma\beta} + \frac{1}{4} \delta_\alpha^\beta g^{\sigma\lambda} \right) u_{\mu\lambda} u_{\kappa\sigma},$$

$$P_\alpha{}^\beta = \frac{c^2}{4\pi\sigma} g^{\mu\kappa} \left( -\delta_\alpha^\lambda g^{\sigma\beta} + \frac{1}{4} \delta_\alpha^\beta g^{\sigma\lambda} \right) f_{\mu\lambda} f_{\kappa\sigma}. \tag{116}$$

The expression for the energy-momentum tensor $T_\alpha{}^\beta$ of a physical system in (115) follows from the principle of least action [11], and the tensor $T_\alpha{}^\beta$ is included in the equation for calculating the metric tensor $g^{\sigma\beta}$. In (116) $W_\alpha{}^\beta$, $U_\alpha{}^\beta$, $B_\alpha{}^\beta$ and $P_\alpha{}^\beta$ are, respectively, the energy-momentum tensors of the electromagnetic and gravitational fields, the acceleration field and the pressure field.. In this case, the tensor $W_\alpha{}^\beta$ is expressed through the electromagnetic field tensor $F_{\mu\lambda}$, the tensor $U_\alpha{}^\beta$ is expressed through the gravitational field tensor $\Phi_{\mu\lambda}$, the tensor $B_\alpha{}^\beta$ is expressed through the acceleration field tensor $u_{\mu\lambda}$, and the tensor $P_\alpha{}^\beta$ is expressed through the pressure field tensor $f_{\mu\lambda}$. In (116) $c$ is the speed of light, $\varepsilon_0$ is electrical constant, $G$ is gravitational constant, $\eta$ and $\sigma$ are constant of acceleration field and pressure field, respectively.

Let us first consider the situation inside the sphere, that is, in its continuously distributed matter. We calculate components $W_\alpha{}^0$, $U_\alpha{}^0$, $B_\alpha{}^0$ and $P_\alpha{}^0$ of fields' stress-energy tensors and consider that in Minkowski spacetime metric tensor $g^{\sigma\beta}$ becomes the metric tensor $\eta^{\sigma\beta}$ (79), which does not depend on time or coordinates.

In Cartesian coordinates, the time components of the stress-energy tensor of electromagnetic field in (116) can be written in terms of electric field **E** and magnetic field **B**:

$$W_0{}^0 = \frac{\varepsilon_0}{2}(E^2 + c^2 B^2), \qquad W_1{}^0 = -\varepsilon_0 c\, [\mathbf{E} \times \mathbf{B}]_x,$$

$$W_2{}^0 = -\varepsilon_0 c\, [\mathbf{E} \times \mathbf{B}]_y, \qquad W_3{}^0 = -\varepsilon_0 c\, [\mathbf{E} \times \mathbf{B}]_z. \tag{117}$$

In (117), the vector $[\mathbf{E}\times\mathbf{B}]$ is the cross product of the vectors $\mathbf{E}$ and $\mathbf{B}$; $[\mathbf{E}\times\mathbf{B}]_x$ denotes the projection of the vector $[\mathbf{E}\times\mathbf{B}]$ onto the $OX$ axis of the Cartesian coordinate system. Similarly, $[\mathbf{E}\times\mathbf{B}]_y$ is a projection of the vector $[\mathbf{E}\times\mathbf{B}]$ onto the $OY$ axis, and $[\mathbf{E}\times\mathbf{B}]_z$ is a projection of the vector $[\mathbf{E}\times\mathbf{B}]$ onto the $OZ$ axis of the Cartesian coordinate system.

If we take into account the definition of the Poynting vector in the form $\mathbf{S}_p = \frac{1}{\mu_0}[\mathbf{E}\times\mathbf{B}]$ and relation $c^2 \mu_0 \varepsilon_0 = 1$, then (117) can be represented in standard form as follows:

$$W_0^{\ 0} = \frac{1}{2}\left(\varepsilon_0 E^2 + \frac{1}{\mu_0} B^2\right), \qquad W_1^{\ 0} = -\frac{S_{px}}{c}, \qquad W_2^{\ 0} = -\frac{S_{py}}{c}, \qquad W_3^{\ 0} = -\frac{S_{pz}}{c}.$$

(118)

Substitution of components $\mathbf{E}$ and $\mathbf{B}$ (85) into (117) taking into account (81) and the relation $\gamma^2 = \frac{1}{1-V^2/c^2}$ gives inside a moving sphere:

$$W_0^{\ 0} = \frac{\varepsilon_0}{2}(E^2 + c^2 B^2) = \frac{\varepsilon_0}{2}\left[E_{ix}'^2 + \gamma^2\left(1+\frac{V^2}{c^2}\right)\left(E_{iy}'^2 + E_{iz}'^2\right)\right] = \frac{\varepsilon_0}{2}\left[E_i'^2 + \frac{2\gamma^2 V^2}{c^2}\left(E_{iy}'^2 + E_{iz}'^2\right)\right] =$$

$$= \frac{\rho_{0q}^2 c^4 \gamma_c^2}{32\pi^2 \varepsilon_0 \eta^2 \rho_0^2 r'^4}\left[1+\frac{2\gamma^2 V^2\left(y'^2 + z'^2\right)}{c^2 r'^2}\right]\left[\frac{c}{\sqrt{4\pi\eta\rho_0}}\sin\left(\frac{r'}{c}\sqrt{4\pi\eta\rho_0}\right) - r'\cos\left(\frac{r'}{c}\sqrt{4\pi\eta\rho_0}\right)\right]^2.$$

$$W_1^{\ 0} = -\varepsilon_0 c\,[\mathbf{E}\times\mathbf{B}]_x = -\varepsilon_0 c\left(E_y B_z - E_z B_y\right) = -\frac{\varepsilon_0 \gamma^2 V}{c}\left(E_{iy}'^2 + E_{iz}'^2\right) =$$

$$= -\frac{\rho_{0q}^2 c^3 \gamma_c^2 \gamma^2 V\left(y'^2 + z'^2\right)}{16\pi^2 \varepsilon_0 \eta^2 \rho_0^2 r'^6}\left[\frac{c}{\sqrt{4\pi\eta\rho_0}}\sin\left(\frac{r'}{c}\sqrt{4\pi\eta\rho_0}\right) - r'\cos\left(\frac{r'}{c}\sqrt{4\pi\eta\rho_0}\right)\right]^2.$$

$$W_2^{\ 0} = -\varepsilon_0 c\,[\mathbf{E}\times\mathbf{B}]_y = -\varepsilon_0 c\left(E_z B_x - E_x B_z\right) = \frac{\gamma V \varepsilon_0 E_{ix}' E_{iy}'}{c} =$$

$$= \frac{\rho_{0q}^2 c^3 \gamma_c^2 \gamma V x' y'}{16\pi^2 \varepsilon_0 \eta^2 \rho_0^2 r'^6}\left[\frac{c}{\sqrt{4\pi\eta\rho_0}}\sin\left(\frac{r'}{c}\sqrt{4\pi\eta\rho_0}\right) - r'\cos\left(\frac{r'}{c}\sqrt{4\pi\eta\rho_0}\right)\right]^2.$$

$$W_3^{\ 0} = -\varepsilon_0 c\, [\mathbf{E}\times\mathbf{B}]_z = -\varepsilon_0 c\left(E_x B_y - E_y B_x\right) = \frac{\gamma V \varepsilon_0 E'_{ix} E'_{iz}}{c} =$$

$$= \frac{\rho_{0q}^2 c^3 \gamma_c^2 \gamma V x' z'}{16\pi^2 \varepsilon_0 \eta^2 \rho_0^2 r'^6}\left[\frac{c}{\sqrt{4\pi\eta\rho_0}}\sin\left(\frac{r'}{c}\sqrt{4\pi\eta\rho_0}\right) - r'\cos\left(\frac{r'}{c}\sqrt{4\pi\eta\rho_0}\right)\right]^2. \quad (119)$$

Similarly to (117), the time components of the stress-energy tensors of gravitational field, acceleration field and pressure field in (116) can be written in terms of strengths and solenoidal vectors of corresponding fields [11]:

$$U_0^{\ 0} = -\frac{1}{8\pi G}(\Gamma^2 + c^2\Omega^2), \qquad U_1^{\ 0} = \frac{c}{4\pi G}[\mathbf{\Gamma}\times\mathbf{\Omega}]_x,$$

$$U_2^{\ 0} = \frac{c}{4\pi G}[\mathbf{\Gamma}\times\mathbf{\Omega}]_y, \qquad U_3^{\ 0} = \frac{c}{4\pi G}[\mathbf{\Gamma}\times\mathbf{\Omega}]_z.$$

$$B_0^{\ 0} = \frac{1}{8\pi\eta}(S^2 + c^2 N^2), \qquad B_1^{\ 0} = -\frac{c}{4\pi\eta}[\mathbf{S}\times\mathbf{N}]_x,$$

$$B_2^{\ 0} = -\frac{c}{4\pi\eta}[\mathbf{S}\times\mathbf{N}]_y, \qquad B_3^{\ 0} = -\frac{c}{4\pi\eta}[\mathbf{S}\times\mathbf{N}]_z.$$

$$P_0^{\ 0} = \frac{1}{8\pi\sigma}(C^2 + c^2 I^2), \qquad P_1^{\ 0} = -\frac{c}{4\pi\sigma}[\mathbf{C}\times\mathbf{I}]_x,$$

$$P_2^{\ 0} = -\frac{c}{4\pi\sigma}[\mathbf{C}\times\mathbf{I}]_y, \qquad P_3^{\ 0} = -\frac{c}{4\pi\sigma}[\mathbf{C}\times\mathbf{I}]_z. \quad (120)$$

In (120) $\mathbf{\Gamma}$ and $\mathbf{\Omega}$ are the strength and torsion field of gravitational field; $\mathbf{S}$ and $\mathbf{N}$ are the strength and solenoidal vector of acceleration field; $\mathbf{C}$ and $\mathbf{I}$ are the strength and solenoidal vector of pressure field.

Let us substitute (93) into (120) and take into account (92):

$$U_0^{\ 0} = -\frac{Gc^4\gamma_c^2}{8\pi\eta^2 r'^4}\left[1+\frac{2\gamma^2 V^2\left(y'^2+z'^2\right)}{c^2 r'^2}\right]\left[\frac{c}{\sqrt{4\pi\eta\rho_0}}\sin\left(\frac{r'}{c}\sqrt{4\pi\eta\rho_0}\right)-r'\cos\left(\frac{r'}{c}\sqrt{4\pi\eta\rho_0}\right)\right]^2.$$

$$U_1^{\ 0} = \frac{Gc^3\gamma_c^2\gamma^2 V\left(y'^2+z'^2\right)}{4\pi\eta^2 r'^6}\left[\frac{c}{\sqrt{4\pi\eta\rho_0}}\sin\left(\frac{r'}{c}\sqrt{4\pi\eta\rho_0}\right)-r'\cos\left(\frac{r'}{c}\sqrt{4\pi\eta\rho_0}\right)\right]^2.$$

$$U_2^{\ 0} = -\frac{Gc^3\gamma_c^2\gamma V x'y'}{4\pi\eta^2 r'^6}\left[\frac{c}{\sqrt{4\pi\eta\rho_0}}\sin\left(\frac{r'}{c}\sqrt{4\pi\eta\rho_0}\right)-r'\cos\left(\frac{r'}{c}\sqrt{4\pi\eta\rho_0}\right)\right]^2.$$

$$U_3^{\ 0} = -\frac{Gc^3\gamma_c^2\gamma V x'z'}{4\pi\eta^2 r'^6}\left[\frac{c}{\sqrt{4\pi\eta\rho_0}}\sin\left(\frac{r'}{c}\sqrt{4\pi\eta\rho_0}\right)-r'\cos\left(\frac{r'}{c}\sqrt{4\pi\eta\rho_0}\right)\right]^2.$$

$$B_0^{\ 0} = \frac{c^4\gamma_c^2}{8\pi\eta r'^4}\left[1+\frac{2\gamma^2 V^2\left(y'^2+z'^2\right)}{c^2 r'^2}\right]\left[\frac{c}{\sqrt{4\pi\eta\rho_0}}\sin\left(\frac{r'}{c}\sqrt{4\pi\eta\rho_0}\right)-r'\cos\left(\frac{r'}{c}\sqrt{4\pi\eta\rho_0}\right)\right]^2.$$

$$B_1^{\ 0} = -\frac{c^3\gamma_c^2\gamma^2 V\left(y'^2+z'^2\right)}{4\pi\eta r'^6}\left[\frac{c}{\sqrt{4\pi\eta\rho_0}}\sin\left(\frac{r'}{c}\sqrt{4\pi\eta\rho_0}\right)-r'\cos\left(\frac{r'}{c}\sqrt{4\pi\eta\rho_0}\right)\right]^2.$$

$$B_2^{\ 0} = \frac{c^3\gamma_c^2\gamma V x'y'}{4\pi\eta r'^6}\left[\frac{c}{\sqrt{4\pi\eta\rho_0}}\sin\left(\frac{r'}{c}\sqrt{4\pi\eta\rho_0}\right)-r'\cos\left(\frac{r'}{c}\sqrt{4\pi\eta\rho_0}\right)\right]^2.$$

$$B_3^{\ 0} = \frac{c^3\gamma_c^2\gamma V x'z'}{4\pi\eta r'^6}\left[\frac{c}{\sqrt{4\pi\eta\rho_0}}\sin\left(\frac{r'}{c}\sqrt{4\pi\eta\rho_0}\right)-r'\cos\left(\frac{r'}{c}\sqrt{4\pi\eta\rho_0}\right)\right]^2.$$

$$P_0^{\ 0} = \frac{\sigma c^4\gamma_c^2}{8\pi\eta^2 r'^4}\left[1+\frac{2\gamma^2 V^2\left(y'^2+z'^2\right)}{c^2 r'^2}\right]\left[\frac{c}{\sqrt{4\pi\eta\rho_0}}\sin\left(\frac{r'}{c}\sqrt{4\pi\eta\rho_0}\right)-r'\cos\left(\frac{r'}{c}\sqrt{4\pi\eta\rho_0}\right)\right]^2.$$

$$P_1^{\ 0} = -\frac{\sigma c^3\gamma_c^2\gamma^2 V\left(y'^2+z'^2\right)}{4\pi\eta^2 r'^6}\left[\frac{c}{\sqrt{4\pi\eta\rho_0}}\sin\left(\frac{r'}{c}\sqrt{4\pi\eta\rho_0}\right)-r'\cos\left(\frac{r'}{c}\sqrt{4\pi\eta\rho_0}\right)\right]^2.$$

$$P_2^{\ 0} = \frac{\sigma c^3 \gamma_c^2 \gamma V x' y'}{4\pi \eta^2 r'^6}\left[\frac{c}{\sqrt{4\pi \eta \rho_0}}\sin\left(\frac{r'}{c}\sqrt{4\pi \eta \rho_0}\right) - r'\cos\left(\frac{r'}{c}\sqrt{4\pi \eta \rho_0}\right)\right]^2.$$

$$P_3^{\ 0} = \frac{\sigma c^3 \gamma_c^2 \gamma V x' z'}{4\pi \eta^2 r'^6}\left[\frac{c}{\sqrt{4\pi \eta \rho_0}}\sin\left(\frac{r'}{c}\sqrt{4\pi \eta \rho_0}\right) - r'\cos\left(\frac{r'}{c}\sqrt{4\pi \eta \rho_0}\right)\right]^2. \quad (121)$$

If we substitute (119) and (121) into (115) and find the time components of the total stress-energy tensor $T_\alpha^{\ 0}$, then by integrating $T_\alpha^{\ 0}$ over volume of moving sphere, according to (9), we can find the components of the integral vector $\mathcal{J}_\alpha$ inside the sphere:

$$\mathcal{J}_0 = \int (W_0^{\ 0} + U_0^{\ 0} + B_0^{\ 0} + P_0^{\ 0}) dx^1 dx^2 dx^3 = \frac{c^4 \gamma_c^2}{8\pi\eta^2}\left(\frac{\rho_{0q}^2}{4\pi\varepsilon_0 \rho_0^2} - G + \eta + \sigma\right) \times$$

$$\times \int \frac{1}{r'^4}\left[1 + \frac{2\gamma^2 V^2 (y'^2 + z'^2)}{c^2 r'^2}\right]\left[\frac{c}{\sqrt{4\pi\eta\rho_0}}\sin\left(\frac{r'}{c}\sqrt{4\pi\eta\rho_0}\right) - r'\cos\left(\frac{r'}{c}\sqrt{4\pi\eta\rho_0}\right)\right]^2 dx^1 dx^2 dx^3 = 0.$$

$$\mathcal{J}_1 = \int (W_1^{\ 0} + U_1^{\ 0} + B_1^{\ 0} + P_1^{\ 0}) dx^1 dx^2 dx^3 = -\frac{c^3 \gamma_c^2 \gamma^2 V}{4\pi\eta^2}\left(\frac{\rho_{0q}^2}{4\pi\varepsilon_0 \rho_0^2} - G + \eta + \sigma\right) \times$$

$$\times \int \frac{(y'^2 + z'^2)}{r'^6}\left[\frac{c}{\sqrt{4\pi\eta\rho_0}}\sin\left(\frac{r'}{c}\sqrt{4\pi\eta\rho_0}\right) - r'\cos\left(\frac{r'}{c}\sqrt{4\pi\eta\rho_0}\right)\right]^2 dx^1 dx^2 dx^3 = 0.$$

$$\mathcal{J}_2 = \int (W_2^{\ 0} + U_2^{\ 0} + B_2^{\ 0} + P_2^{\ 0}) dx^1 dx^2 dx^3 = \frac{c^3 \gamma_c^2 \gamma V}{4\pi\eta^2}\left(\frac{\rho_{0q}^2}{4\pi\varepsilon_0 \rho_0^2} - G + \eta + \sigma\right) \times$$

$$\times \int \frac{x' y'}{r'^6}\left[\frac{c}{\sqrt{4\pi\eta\rho_0}}\sin\left(\frac{r'}{c}\sqrt{4\pi\eta\rho_0}\right) - r'\cos\left(\frac{r'}{c}\sqrt{4\pi\eta\rho_0}\right)\right]^2 dx^1 dx^2 dx^3 = 0.$$

$$\mathcal{J}_3 = \int (W_3^{\ 0} + U_3^{\ 0} + B_3^{\ 0} + P_3^{\ 0}) dx^1 dx^2 dx^3 = \frac{c^3 \gamma_c^2 \gamma V}{4\pi\eta^2}\left(\frac{\rho_{0q}^2}{4\pi\varepsilon_0 \rho_0^2} - G + \eta + \sigma\right) \times$$

$$\times \int \frac{x' z'}{r'^6}\left[\frac{c}{\sqrt{4\pi\eta\rho_0}}\sin\left(\frac{r'}{c}\sqrt{4\pi\eta\rho_0}\right) - r'\cos\left(\frac{r'}{c}\sqrt{4\pi\eta\rho_0}\right)\right]^2 dx^1 dx^2 dx^3 = 0.$$

(122)

Before each integral in (122), in the corresponding bracket, there is a sum which, according to (63), is equal to zero: $\frac{\rho_{0q}^2}{4\pi\varepsilon_0\rho_0^2} - G + \eta + \sigma = 0$. This equality, containing field coefficients, was found in [15], [21], as a consequence of the balance of forces from all fields acting on typical particles.

Thus, inside moving sphere the integral vector becomes equal to zero, $\mathcal{J}_\alpha = 0$.

### 3.9. Integral vector outside moving sphere

To calculate the total integral vector $\mathcal{J}_\alpha$, it is necessary to integrate $T_\alpha^{\ 0}$ over entire volume, both inside and outside the sphere. Since acceleration field and pressure field are present only in matter, only electromagnetic and gravitational fields remain outside the sphere. Let us find the time components of the stress-energy tensors of these fields, taking into account (117), (120), (96-97):

$$W_0^{\ 0} = \frac{\varepsilon_0}{2}\left[E_{ox}'^2 + \gamma^2\left(1+\frac{V^2}{c^2}\right)\left(E_{oy}'^2 + E_{oz}'^2\right)\right] = \frac{\varepsilon_0}{2}\left[E_o'^2 + \frac{2\gamma^2 V^2}{c^2}\left(E_{oy}'^2 + E_{oz}'^2\right)\right] =$$

$$= \frac{q_b^2}{32\pi^2\varepsilon_0 r'^4}\left[1 + \frac{2\gamma^2 V^2\left(y'^2 + z'^2\right)}{c^2 r'^2}\right].$$

$$W_1^{\ 0} = -\varepsilon_0 c\left(E_y B_z - E_z B_y\right) = -\frac{\varepsilon_0 \gamma^2 V}{c}\left(E_{oy}'^2 + E_{oz}'^2\right) = -\frac{q_b^2 \gamma^2 V\left(y'^2 + z'^2\right)}{16\pi^2\varepsilon_0 c r'^6}.$$

$$W_2^{\ 0} = -\varepsilon_0 c\left(E_z B_x - E_x B_z\right) = \frac{\gamma V \varepsilon_0 E_{ox}' E_{oy}'}{c} = \frac{q_b^2 \gamma V x' y'}{16\pi^2\varepsilon_0 c r'^6}.$$

$$W_3^{\ 0} = -\varepsilon_0 c\left(E_x B_y - E_y B_x\right) = \frac{\gamma V \varepsilon_0 E_{ox}' E_{oz}'}{c} = \frac{q_b^2 \gamma V x' z'}{16\pi^2\varepsilon_0 c r'^6}.$$

$$U_0^{\ 0} = -\frac{Gm_b^2}{8\pi r'^4}\left[1 + \frac{2\gamma^2 V^2\left(y'^2 + z'^2\right)}{c^2 r'^2}\right], \qquad U_1^{\ 0} = \frac{Gm_b^2 \gamma^2 V\left(y'^2 + z'^2\right)}{4\pi c r'^6},$$

$$U_2{}^0 = -\frac{Gm_b^2 \gamma V x' y'}{4\pi c r'^6}, \qquad U_3{}^0 = -\frac{Gm_b^2 \gamma V x' z'}{4\pi c r'^6}. \qquad (123)$$

By substituting $W_0{}^0$ and $U_0{}^0$ from (123) into (115), we find $T_0{}^0$ and the component $\mathcal{J}_0$ in (9):

$$\mathcal{J}_0 = \int \left(W_0{}^0 + U_0{}^0\right) dx^1 dx^2 dx^3 = \left(\frac{q_b^2}{32\pi^2 \varepsilon_0} - \frac{Gm_b^2}{8\pi}\right) \int \frac{1}{r'^4}\left[1 + \frac{2\gamma^2 V^2 \left(y'^2 + z'^2\right)}{c^2 r'^2}\right] dx^1 dx^2 dx^3.$$

(124)

The integral in (124) is taken over the volume outside the moving sphere, and the sphere is considered a Heaviside ellipsoid. According to (90), we suppose that $r' = r$, where $r'$ is current radius in the reference frame $O'$, associated with the center of sphere, and $r$ is radial coordinate in (88). From Lorentz transformations, in the case of the sphere's motion at constant velocity **V** along the axis $OX$ in reference frame $O$, it follows that $x' = \gamma(x - Vt)$, $y' = y$, and $z' = z$. Then, in view of (88), we have:

$$x' = r\cos\theta, \quad y' = r\sin\theta\cos\phi, \quad z' = r\sin\theta\sin\phi, \quad y'^2 + z'^2 = r^2 \sin^2\theta. \qquad (125)$$

As in (91), the volume element will equal $dx^1 dx^2 dx^3 = \frac{1}{\gamma} r^2 \sin\theta \, dr \, d\theta \, d\phi$. Using all this in (124), we find:

$$\mathcal{J}_0 = \frac{1}{\gamma}\left(\frac{q_b^2}{32\pi^2 \varepsilon_0} - \frac{Gm_b^2}{8\pi}\right) \int_{\phi=0}^{2\pi} \int_{\theta=0}^{\pi} \int_{r=a}^{\infty} \frac{1}{r^2}\left(1 + \frac{2\gamma^2 V^2 \sin^2\theta}{c^2}\right) \sin\theta \, dr \, d\theta \, d\phi =$$
$$= \frac{1}{\gamma}\left(1 + \frac{4\gamma^2 V^2}{3c^2}\right)\left(\frac{q_b^2}{8\pi\varepsilon_0 a} - \frac{Gm_b^2}{2a}\right).$$

(126)

For space components of the integral vector $\mathcal{J}_\alpha$ in (9), using (123) and (125) we find the following in a similar way:

$$\mathcal{J}_1 = \int \left(W_1{}^0 + U_1{}^0\right) dx^1 dx^2 dx^3 = \left(-\frac{q_b^2 \gamma^2 V}{16\pi^2 \varepsilon_0 c} + \frac{Gm_b^2 \gamma^2 V}{4\pi c}\right) \int \frac{y'^2 + z'^2}{r'^6} dx^1 dx^2 dx^3 =$$

$$= -\frac{1}{\gamma}\left(\frac{q_b^2 \gamma^2 V}{16\pi^2 \varepsilon_0 c} - \frac{Gm_b^2 \gamma^2 V}{4\pi c}\right) \int_{\phi=0}^{2\pi}\int_{\theta=0}^{\pi}\int_{r=a}^{\infty} \frac{1}{r^2} \sin^3\theta\, dr\, d\theta\, d\phi = -\frac{4\gamma}{3}\left(\frac{q_b^2}{8\pi\varepsilon_0 a} - \frac{Gm_b^2}{2a}\right)\frac{V}{c}.$$

$$\mathcal{J}_2 = \int \left(W_2{}^0 + U_2{}^0\right) dx^1 dx^2 dx^3 = \left(\frac{q_b^2 \gamma V}{16\pi^2 \varepsilon_0 c} - \frac{Gm_b^2 \gamma V}{4\pi c}\right) \int \frac{x'y'}{r'^6} dx^1 dx^2 dx^3 =$$

$$= \left(\frac{q_b^2 V}{16\pi^2 \varepsilon_0 c} - \frac{Gm_b^2 V}{4\pi c}\right) \int_{\phi=0}^{2\pi}\int_{\theta=0}^{\pi}\int_{r=a}^{\infty} \frac{\sin^2\theta \cos\theta \cos\phi}{r^2}\, dr\, d\theta\, d\phi = 0.$$

$$\mathcal{J}_3 = \int \left(W_3{}^0 + U_3{}^0\right) dx^1 dx^2 dx^3 = \left(\frac{q_b^2 \gamma V}{16\pi^2 \varepsilon_0 c} - \frac{Gm_b^2 \gamma V}{4\pi c}\right) \int \frac{x'z'}{r'^6} dx^1 dx^2 dx^3 =$$

$$= \left(\frac{q_b^2 V}{16\pi^2 \varepsilon_0 c} - \frac{Gm_b^2 V}{4\pi c}\right) \int_{\phi=0}^{2\pi}\int_{\theta=0}^{\pi}\int_{r=a}^{\infty} \frac{\sin^2\theta \cos\theta \sin\phi}{r^2}\, dr\, d\theta\, d\phi = 0. \tag{127}$$

### 4. Discussion of results

In (122) we found that in the matter inside a sphere all the components of the integral vector $\mathcal{J}_\alpha$ become equal to zero. Therefore, it suffices to consider only the components of the integral vector outside the matter.

According to (111-112) and (127), we have the following:

$$K_1 = -K_x = -\frac{\gamma V}{c^2}\left(\frac{Gm_b^2}{2a} - \frac{q_b^2}{8\pi\varepsilon_0 a}\right) = -\frac{3\mathcal{J}_1}{4c}. \tag{128}$$

Comparison of (108) and (126) gives:

$$K_0 = -\frac{3c\gamma^2}{3c^2 + 4\gamma^2 V^2} \mathcal{J}_0. \tag{129}$$

From (128) it follows that the space component $\mathcal{J}_1$ of the integral vector $\mathcal{J}_\alpha$ is proportional to the space component $K_1$ of the four-vector $K_\mu$, which defines contribution of fields to energy-momentum of the system under consideration. According to (112), the relation $K_1 = -\dfrac{K_0 V}{c}$ holds for components of four-vector $K_\mu$. However, for the components of the integral vector, according to (126-127), we obtain a different relation:

$$\mathcal{J}_1 = -\frac{4c\gamma^2 \mathcal{J}_0 V}{3c^2 + 4\gamma^2 V^2} \approx -\frac{4\gamma^2 \mathcal{J}_0 V}{3c}. \tag{130}$$

From (130) it follows that the component $\mathcal{J}_1$ depends in a complex manner on the velocity $V$ of the sphere's motion. Even in the limit of low velocities in (130), we obtain a coefficient $\dfrac{4\gamma^2}{3}$ that is not equal to unity, in contrast to the relation $K_1 = -\dfrac{K_0 V}{c}$ for components of the four-vector $K_\mu$, where similar coefficient is equal to 1. Thus, the integral vector $\mathcal{J}_\alpha$ is not a four-vector; therefore, the integral vector cannot specify either the four-momentum of the system or the four-momentum of the fields.

The same can be said in other words. According to (9), the integral vector is obtained by volume integration of the time components $T_\alpha^{\ 0}$ of the total stress-energy tensor of the system. Hence, it is not enough to know the stress-energy tensor of a system to calculate the four-momentum.

On the other hand, the 4/3 factor in (130) is associated with the so-called 4/3 problem, according to which the mass-energy contained in $\mathcal{J}_1$ is approximately 4/3 times greater than the mass-energy contained in component $\mathcal{J}_0$. Obviously, such behavior of mass-energies is inconsistent with the role of mass in the four-momentum of a single point particle, where the mass is part of both energy and momentum to the same extent.

In this case, what does the integral vector $\mathcal{J}_\alpha$ represent? According to its meaning, it is a volume integral of the equation of motion, and it was shown in [15] that components $T_\alpha^{\ 0}$ of the stress-energy tensor correspond to the generalized Poynting theorem, according to which a change in the fields' energy in any given volume is exactly compensated by the fields' energy flux through the surface, surrounding the given volume. If we try to find the integral vector $\mathcal{J}_\alpha$ using $T_\alpha^{\ 0}$, then it turns out to be equal neither to the system's four-momentum $P_\mu$ nor to the four-vector $K_\mu$.

However, in a closed system moving at a constant velocity, the integral vector $\mathcal{J}_\alpha$ is conserved. This can be seen from (129-130), since the quantity $K_0$ (108), proportional to fields' energy, which is calculated using tensor invariants, is conserved.

## 5. Conclusion

Instead of four-momentum $P^\mu$ with a contravariant index (1), we proceed from definition of four-momentum $P_\mu$ with a covariant index in (45). The transition between these forms of four-momentum in the form $P^\mu = g^{\mu\nu} P_\nu$, $P_\mu = g_{\mu\nu} P^\nu$, with the participation of the metric tensor $g^{\mu\nu}$, is possible only in the special theory of relativity. In the more general case, in curved spacetime, additional assumptions are required for the definition of $P^\mu$.

The main reason for the primary definition of four-momentum $P_\mu$ as a four-vector with covariant index is the need to take into account in $P_\mu$ contributions from particles in the form of generalized four-momentum $p_\mu$, as well as contributions from fields, present in the system, by means of four-vector $K_\mu$. As a result, we obtain the relation $P_\mu = p_\mu + K_\mu$, where $p_\mu$ is expressed in a covariant way in terms of the Lagrangian density in (13) and (38).

For Lagrangian density (27), in which four vector fields are presented, the energy and momentum of a system are determined in (36) and (37), and the components $K_\mu = (K_0, -\mathbf{K})$ are given in (47) and (48). As a consequence, our four-momentum $P_\mu$ (46) does not coincide with any of the expressions (1-9) presented in literature for characteristic of four-momentum.

In the case of four-dimensional variation in the action function, in (25-26) we obtain a covariant expression for function $Z$ that is conserved in a closed system, and for Lagrangian density (27) defines the energy (30) of fields in matter associated with tensor invariants. From (62-64) it follows that the energy $Z$ in the volume occupied by matter becomes equal to zero, and the same follows from (95) for the matter inside the moving sphere. According to (30) and (101), $Z = -L_{fi}$, where $L_{fi}$ denotes that part of the Lagrangian that is associated with tensor invariants in the matter.

From comparison of (38) and (41), it follows that the generalized four-momenta $p_\mu$ and $p^\mu$, which have an integral form, differ from the standard four-vectors by their nonlocality, and the expressions of the type $p_\mu = g_{\mu\nu} p^\nu$ will be incorrect. The same is true for four-vectors $K_\mu$ and $K^\mu$. In the special theory of relativity, the four-momentum $P_\mu$ is expressed according to (49) through the four-velocity $u_\mu$ of the center of the system's momentum. It is due to the use of four-dimensional Lagrangian formalism and nonlocality of $p_\mu$ it becomes clear that the four-momentum must be defined as a covariant four-vector $P_\mu$.

To apply the obtained formulas for four-vectors, the components of these four-vectors were calculated for a relativistic uniform system in the form of a sphere with particles and fields within the framework of the special theory of relativity. For a fixed sphere, the three-dimensional relativistic momentum is equal to zero, and the energy of the system is defined in (71). In this case, the energy of fields inside the matter becomes equal to zero, and the system's energy consists of the particles' energy in scalar field potentials, taking into account the contribution from the energy of fields outside the matter.

For a moving sphere, from comparison of (59) and (75) it follows that due to motion time component $p_0$ of the generalized four-momentum increases by a factor of $\gamma$, where $\gamma$ denotes the Lorentz factor of center of momentum of the moving sphere. According to (70) and (108), the same is true for the time component $K_0$, which is associated with the energy from the fields' tensor invariants. These changes in $p_0$ and $K_0$ are clearly observed in the formula for the system's energy (109), which can be compared with (71) for a fixed sphere.

In addition to the four-vectors $P_\mu$, $p_\mu$ and $K_\mu$, we also calculate the components of integral vector $\mathcal{J}_\alpha$ (9). According to (122), the integral vector inside the moving sphere becomes equal to zero, so that in the case under consideration, it suffices to calculate its components outside the sphere. If we compare the relations for $K_\mu$ in (112) and for $p_\mu$ in (113) with the relation for $\mathcal{J}_\alpha$ in (130), we see that $\mathcal{J}_\alpha$ is not a real four-vector.

The results obtained can be summarized as follows: for continuously distributed matter, to uniquely determine four-momentum, it should be defined as a sum of two nonlocal four-vectors of the integral type, that is, as a four-vector with a covariant index in the form $P_\mu = p_\mu + K_\mu$, taking into account contributions from the energy and momentum of all the system's particles and fields.

For the integral vector $\mathcal{J}_\alpha$, obtained by volume integration of the time components of the system's stress-energy tensor, such a vector is not a four-vector or four-momentum, although it is conserved in a closed system.

Another conclusion that follows from the Lagrangian formalism for vector fields is that the most natural representation of some physically meaningful four-vectors is their form with a covariant index. These four-vectors include the generalized four-momentum $p_\mu$, relativistic four-momentum $P_\mu$, fields' four-momentum $K_\mu$, four-force $F_\mu = \dfrac{DP_\mu}{D\tau}$, four-potentials $A_\mu, D_\mu, U_\mu, \pi_\mu$ of electromagnetic and gravitational fields, acceleration field and pressure field, respectively.

A consequence of the fact that four-potentials of fields are defined as four-vectors with a covariant index is that field tensors are expressed most simply as tensors with covariant indices. An example here is the electromagnetic field tensor $F_{\mu\nu}$, calculated using a four-dimensional rotor from the four-potential $A_\nu$ according to the formula: $F_{\mu\nu} = \nabla_\mu A_\nu - \nabla_\nu A_\mu = \partial_\mu A_\nu - \partial_\nu A_\mu = \partial_\mu \times A_\nu$.

The fact that the four-momentum $P_\mu$ of a physical system can be determined in covariant form through the sum of two nonlocal four-vectors $p_\mu$ and $K_\mu$, significantly changes our understanding of the energy and momentum of the system. Unlike the cases of one point particle or many non-interacting point particles, in systems with a continuous distribution of matter there is an active exchange of energy and momentum among all interacting particles and fields. In addition, even in stationary systems, the metric tensor present in the formulas is a function of time and coordinates. All this leads to the nonlocality of four-vectors $p_\mu$ and $K_\mu$. In addition to the generalized four-momentum $p_\mu$ associated with the particles of the system, the definition of $P_\mu$ must take into account another four-momentum $K_\mu$ associated with the fields of the system.

Another feature of the considered approach is that when taking into account the metric, the energy calibration procedure is necessary [11], [20], due to which the expressions for the energy and momentum of the system cease to depend on the cosmological constant $\Lambda$ and become uniquely defined in a covariant form.

From the practical point of view, the derived formulas for the relativistic four-momentum allow one to find the energy and momentum of any physical system. This is especially true in systems in which the role of fields acting on particles is significant, or when it is necessary to study the effects associated with the energy and momentum of the fields themselves.

The results obtained also show that neither the integral vector $\mathcal{J}_\alpha$ (9), the components of which were calculated in (126-127), nor the integral vector $\mathcal{J}^\mu$ (8), proposed in the general theory of relativity [2], can be considered as a four-momentum of a physical system. Indeed, for the calculation $\mathcal{J}_\alpha$ and $\mathcal{J}^\mu$ it is necessary to know the time components of the energy-momentum tensor, taking into account all the fields of the system, including the gravitational field. However, the time components of the energy-momentum tensor of a system, by definition, include the energy densities of all fields and the energy fluxes of these fields. After integrating the time components of the energy-momentum tensor over the moving volume, the corresponding components of the integral vectors $\mathcal{J}_\alpha$ and $\mathcal{J}^\mu$ appear, which are proportional to the energy and momentum of the fields.

Although the components $\mathcal{J}_\alpha$ and $\mathcal{J}^\mu$ are conserved in a closed system, they are related not to the four-momentum of the physical system, but to Poynting's four-dimensional theorem, as was shown in [15] in relation to $\mathcal{J}_\alpha$. According to Poynting's theorem, in each volume of the system the loss of energy is associated with the corresponding energy flow from this volume. The components of $\mathcal{J}_\alpha$ satisfy Poynting's theorem exactly, but do not form a four-vector. Instead, in accordance with (130), the so-called 4/3 problem arises for the components of $\mathcal{J}_\alpha$, so that in every even small volume of the system, the mass-energy in this volume differs by approximately 4/3 times from the mass-energy contained in the field momentum in this volume.

From the above it follows that to calculate the four-momentum of a physical system, it is necessary to know not the energy-momentum tensor of the system, but the four-potentials and tensors of all fields acting in the system.

## References

1. Landau L.D., Lifshitz E.M. Mechanics. Vol. 1 (3rd ed.). (1976). Butterworth-Heinemann. ISBN 978-0-7506-2896-9.

2. Landau L.D., Lifshitz E.M. The Classical Theory of Fields. (1951). Pergamon Press. ISBN 7-5062-4256-7.

3. Goldstein H., Poole C.P. and Safko J.L. Classical Mechanics (3rd ed.). (2001). Addison-Wesley. p. 680. ISBN 9780201657029.

4. M. Sharif, Tasnim Fatima. Energy-Momentum Distribution: A Crucial Problem in General Relativity. Int. J. Mod. Phys. A, Vol. 20, p. 4309 (2005). https://doi.org/10.1142/S0217751X05020793.

5. Denisov V. I., Logunov A. A. The inertial mass defined in the general theory of relativity has no physical meaning. Theoretical and Mathematical Physics, Vol. 51, Issue 2, pp. 421-426 (1982). http://dx.doi.org/10.1007/BF01036205.

6. Denisov V. I., Logunov A. A. Does the general theory of relativity have a classical Newtonian limit? Theoretical and Mathematical Physics, Vol. 45, Issue 3, pp. 1035-1041 (1980). https://doi.org/10.1007%2FBF01016702.

7. Fedosin S.G. The covariant additive integrals of motion in the theory of relativistic vector fields. Bulletin of Pure and Applied Sciences, Vol. 37 D (Physics), No. 2, pp. 64-87 (2018). http://dx.doi.org/10.5958/2320-3218.2018.00013.1.

8. Fedosin S.G. The Principle of Least Action in Covariant Theory of Gravitation. Hadronic Journal, Vol. 35, No. 1, pp. 35-70 (2012). http://dx.doi.org/10.5281/zenodo.889804.

9. Dirac P. A. M. General Theory of Relativity. (1975). Princeton University Press, quick presentation of the bare essentials of GTR. ISBN 0-691-01146-X.

10. Khrapko, R. I. The Truth about the Energy-Momentum Tensor and Pseudotensor. Gravitation and Cosmology, Vol. 20, No. 4, pp. 264-273 (2014). http://dx.doi.org/10.1134/S0202289314040082.

11. Fedosin S.G. About the cosmological constant, acceleration field, pressure field and energy. Jordan Journal of Physics, Vol. 9, No. 1, pp. 1-30 (2016). http://dx.doi.org/10.5281/zenodo.889304.

12. Fedosin S.G. The Procedure of Finding the Stress-Energy Tensor and Equations of Vector Field of Any Form. Advanced Studies in Theoretical Physics, Vol. 8, No. 18, pp. 771-779 (2014). http://dx.doi.org/10.12988/astp.2014.47101.

13. Fedosin S.G. The concept of the general force vector field. OALib Journal, Vol. 3, pp. 1-15 (2016). http://dx.doi.org/10.4236/oalib.1102459.

14. Fedosin S.G. Two components of the macroscopic general field. Reports in Advances of Physical Sciences, Vol. 1, No. 2, 1750002 (2017). http://dx.doi.org/10.1142/S2424942417500025.

15. Fedosin S.G. The generalized Poynting theorem for the general field and solution of the 4/3 problem. International Frontier Science Letters, Vol. 14, pp. 19-40 (2019). https://doi.org/10.18052/www.scipress.com/IFSL.14.19.

16. Fedosin S.G. Equations of Motion in the Theory of Relativistic Vector Fields. International Letters of Chemistry, Physics and Astronomy, Vol. 83, pp. 12-30 (2019). https://doi.org/10.18052/www.scipress.com/ILCPA.83.12..

17. Fedosin S.G. Generalized Four-momentum for Continuously Distributed Materials. Gazi University Journal of Science, (2024). https://doi.org/10.35378/gujs.1231793.

18. Fock V.A. The Theory of Space, Time and Gravitation. (1959). Pergamon Press, London..

19. Fedosin S.G. Four-Dimensional Equation of Motion for Viscous Compressible and Charged Fluid with Regard to the Acceleration Field, Pressure Field and Dissipation Field. International Journal of Thermodynamics, Vol. 18, No. 1, pp. 13-24 (2015). https://doi.org/10.5541/ijot.5000034003.

20. Fedosin S.G. Energy and metric gauging in the covariant theory of gravitation. Aksaray University Journal of Science and Engineering, Vol. 2, Issue 2, pp. 127-143 (2018). http://dx.doi.org/10.29002/asujse.433947.

21. Fedosin S.G. Estimation of the physical parameters of planets and stars in the gravitational equilibrium model. Canadian Journal of Physics, Vol. 94, No. 4, pp. 370-379 (2016). http://dx.doi.org/10.1139/cjp-2015-0593.

22. Fedosin S.G. The virial theorem and the kinetic energy of particles of a macroscopic system in the general field concept. Continuum Mechanics and Thermodynamics, Vol. 29, Issue 2, pp. 361-371 (2016). https://dx.doi.org/10.1007/s00161-016-0536-8.

23. Fedosin S.G. The Hamiltonian in Covariant Theory of Gravitation. Advances in Natural Science, Vol. 5, No. 4, pp. 55-75 (2012). http://dx.doi.org/10.3968%2Fj.ans.1715787020120504.2023.

24. Fedosin S.G. The Gravitational Field in the Relativistic Uniform Model within the Framework of the Covariant Theory of Gravitation. International Letters of Chemistry, Physics and Astronomy, Vol. 78, pp. 39-50 (2018). http://dx.doi.org/10.18052/www.scipress.com/ILCPA.78.39.

25. Fedosin S.G. The Pioneer Anomaly in Covariant Theory of Gravitation. Canadian Journal of Physics, Vol. 93, No. 11, pp. 1335-1342 (2015). http://dx.doi.org/10.1139/cjp-2015-0134.

26. Fedosin S.G. Relativistic Energy and Mass in the Weak Field Limit. Jordan Journal of Physics. Vol. 8, No. 1, pp. 1-16 (2015). http://dx.doi.org/10.5281/zenodo.889210.

27. Searle G.F.C. On the steady motion of an electrified ellipsoid. The Philosophical Magazine Series 5, 44 (269), 329-341 (1897). http://dx.doi.org/10.1088/1478-7814/15/1/323.

28. Fedosin S.G. 4/3 Problem for the Gravitational Field. Advances in Physics Theories and Applications, Vol. 23, pp. 19-25 (2013). http://dx.doi.org/10.5281/zenodo.889383.

29. Fedosin S.G. The Integral Energy-Momentum 4-Vector and Analysis of 4/3 Problem Based on the Pressure Field and Acceleration Field. American Journal of Modern Physics. Vol. 3, No. 4, pp. 152-167 (2014). http://dx.doi.org/10.11648/j.ajmp.20140304.12.

**Appendix A.**

**Q.1 How can the four-momentum of a physical system to four-dimensional space toroidal geometrical topology (or other topology) possibly be general, with this approach?**

To answer this question, we use double numbering of formulas, in which the last digits indicate the number of the formula in the text of the article.

The relativistic four-momentum of a physical system located in four-dimensional spacetime with arbitrary geometry and topology can be determined in covariant form by the formula

$$P_\mu = \left(\frac{E}{c}, -\mathbf{P}\right) = \left(\frac{E}{c}, -P_x, -P_y, -P_z\right) = (P_0, P_1, P_2, P_3). \tag{1-45}$$

where the energy $E$ of the system is expressed by the formula

$$E = c\int \mathbf{v} \cdot \frac{\partial}{\partial \mathbf{v}}\left(\frac{\mathcal{L}_p}{u^0}\right) dV_0 + \sum_{n=1}^{N}\left(\mathbf{v}_n \cdot \frac{\partial L_f}{\partial \mathbf{v}_n}\right) - \int (\mathcal{L}_p + \mathcal{L}_f)\sqrt{-g}\, dx^1 dx^2 dx^3 =$$

$$= \int \left[\mathbf{v} \cdot \frac{\partial}{\partial \mathbf{v}}\left(\frac{\mathcal{L}_p}{u^0}\right) u^0 - \mathcal{L}_p\right] \sqrt{-g}\, dx^1 dx^2 dx^3 - \int \mathcal{L}_f \sqrt{-g}\, dx^1 dx^2 dx^3 + \sum_{n=1}^{N}\left(\mathbf{v}_n \cdot \frac{\partial L_f}{\partial \mathbf{v}_n}\right).$$

$$\tag{2-31}$$

In (2-31), the energy $E$ is determined by the formula in which the Lagrangian density $\mathcal{L} = \mathcal{L}_p + \mathcal{L}_f$ of the system is determined by the sum of two terms, and the term $\mathcal{L}_p$ directly depends on the four-currents $j^\mu$ and $J^\mu$, and therefore $\mathcal{L}_p$ depends on the particle's velocity $\mathbf{v}$.

Using for the components of Lagrangian density expressions corresponding to vector fields in the form

$$\mathcal{L}_p = -A_\mu j^\mu - D_\mu J^\mu - U_\mu J^\mu - \pi_\mu J^\mu, \tag{3-27}$$

$$\mathcal{L}_f = -\frac{1}{4\mu_0} F_{\mu\nu} F^{\mu\nu} + \frac{c^2}{16\pi G}\Phi_{\mu\nu}\Phi^{\mu\nu} - \frac{c^2}{16\pi\eta}u_{\mu\nu}u^{\mu\nu} - \frac{c^2}{16\pi\sigma}f_{\mu\nu}f^{\mu\nu} + ckR - 2ck\Lambda,$$

$$\tag{4-28}$$

we can simplify expression (2-31) for the energy of the system:

$$\mathbb{E} = \frac{1}{c}\int \left[ \begin{array}{l} \rho_{0q}\varphi + \rho_0\psi + \rho_0\vartheta + \rho_0\wp - \mathbf{v}\cdot\dfrac{\partial}{\partial \mathbf{v}}\left(\rho_{0q}\varphi + \rho_0\psi + \rho_0\vartheta + \rho_0\wp\right) + \\ +v^2\dfrac{\partial}{\partial \mathbf{v}}\left(\rho_{0q}\mathbf{A} + \rho_0\mathbf{D} + \rho_0\mathbf{U} + \rho_0\mathbf{\Pi}\right) \end{array} \right] u^0\sqrt{-g}\,dx^1dx^2dx^3 +$$

$$+\int\left(\frac{1}{4\mu_0}F_{\mu\nu}F^{\mu\nu} - \frac{c^2}{16\pi G}\Phi_{\mu\nu}\Phi^{\mu\nu} + \frac{c^2}{16\pi\eta}u_{\mu\nu}u^{\mu\nu} + \frac{c^2}{16\pi\sigma}f_{\mu\nu}f^{\mu\nu}\right)\sqrt{-g}\,dx^1dx^2dx^3 +$$

$$+\sum_{n=1}^{N}\left(\mathbf{v}_n\cdot\frac{\partial L_f}{\partial \mathbf{v}_n}\right).$$

(5-36)

The momentum $\mathbf{P}$ of the system in (1-45) is expressed by the formula

$$\mathbf{P} = \sum_{n=1}^{N}\left(\frac{\partial L}{\partial \mathbf{v}_n}\right) = c\int\frac{\partial}{\partial \mathbf{v}}\left(\frac{\mathcal{L}_p}{u^0}\right)dV_0 + \sum_{n=1}^{N}\left(\frac{\partial L_f}{\partial \mathbf{v}_n}\right) =$$

$$= \frac{1}{c}\int \left[ \begin{array}{l} \rho_{0q}\mathbf{A} + \rho_0\mathbf{D} + \rho_0\mathbf{U} + \rho_0\mathbf{\Pi} - \dfrac{\partial}{\partial \mathbf{v}}\left(\rho_{0q}\varphi + \rho_0\psi + \rho_0\vartheta + \rho_0\wp\right) + \\ +\mathbf{v}\cdot\dfrac{\partial}{\partial \mathbf{v}}\left(\rho_{0q}\mathbf{A} + \rho_0\mathbf{D} + \rho_0\mathbf{U} + \rho_0\mathbf{\Pi}\right) \end{array} \right] u^0\sqrt{-g}\,dx^1dx^2dx^3 +$$

$$+\sum_{n=1}^{N}\left(\frac{\partial L_f}{\partial \mathbf{v}_n}\right).$$

(6-37)

In this case, expressions (3-27) and (4-28) are taken into account in (6-37).

The second method of calculating the relativistic four-momentum $P_\mu$ of an arbitrary physical system involves splitting $P_\mu$ into two nonlocal four-vectors:

$$P_\mu = p_\mu + K_\mu.$$

(7-46)

To calculate the generalized four-momentum $p_\mu$ in (7-46), it is necessary, under given initial conditions, to determine the dependence of the metric tensor $g_{\mu\nu}$ and its determinant $g$ on time and coordinates, and to find the time component $u^0$ of the four-velocity $u^\mu$ of the particles of the system. In addition, it is required to know the Lagrangian density $\mathcal{L} = \mathcal{L}_p + \mathcal{L}_f$ of the system so that the derivatives $\dfrac{\partial \mathcal{L}_p}{\partial u^\mu}$ can be calculated and then $p_\mu$ determined by integration over the moving volume of the system:

$$p_\mu = -\frac{1}{c}\int_{V_m} \frac{\partial \mathcal{L}_p}{\partial u^\mu} u^0 \sqrt{-g}\, dx^1 dx^2 dx^3 = \frac{1}{c}\int_{V_m} \mathcal{P}_\mu u^0 \sqrt{-g}\, dx^1 dx^2 dx^3 = \int_{V_{m0}} \mathcal{P}_\mu dV_0 . \qquad (8\text{-}13)$$

Using $\mathcal{L}_p$ (3-27), for $p_\mu$ and its components we have:

$$p_\mu = \frac{1}{c}\int_{V_m} \left( \rho_{0q} A_\mu + \rho_0 D_\mu + \rho_0 U_\mu + \rho_0 \pi_\mu \right) u^0 \sqrt{-g}\, dx^1 dx^2 dx^3 . \qquad (9\text{-}38)$$

$$p_0 = \frac{1}{c^2}\int_{V_m} \left( \rho_{0q}\varphi + \rho_0 \psi + \rho_0 \vartheta + \rho_0 \wp \right) u^0 \sqrt{-g}\, dx^1 dx^2 dx^3 . \qquad (10\text{-}39)$$

$$\mathbf{p} = \frac{1}{c}\int_{V_m} \left( \rho_{0q}\mathbf{A} + \rho_0 \mathbf{D} + \rho_0 \mathbf{U} + \rho_0 \mathbf{\Pi} \right) u^0 \sqrt{-g}\, dx^1 dx^2 dx^3 . \qquad (11\text{-}40)$$

Components of four-vector

$$K_\mu = (K_0, -\mathbf{K}) = (K_0, -K_x, -K_y, -K_z) = (K_0, K_1, K_2, K_3)$$

in (7-46) with the use of (3-27) and (4-28) are expressed by the formulas:

$$K_0 = \frac{1}{c^2}\int \begin{bmatrix} -\mathbf{v}\cdot\dfrac{\partial}{\partial \mathbf{v}}\left(\rho_{0q}\varphi + \rho_0\psi + \rho_0\vartheta + \rho_0\wp\right) + \\ +v^2 \dfrac{\partial}{\partial \mathbf{v}}\left(\rho_{0q}\mathbf{A} + \rho_0 \mathbf{D} + \rho_0 \mathbf{U} + \rho_0 \mathbf{\Pi}\right) \end{bmatrix} u^0 \sqrt{-g}\, dx^1 dx^2 dx^3 +$$

$$+\frac{1}{c}\int \left( \frac{1}{4\mu_0} F_{\mu\nu} F^{\mu\nu} - \frac{c^2}{16\pi G}\Phi_{\mu\nu}\Phi^{\mu\nu} + \frac{c^2}{16\pi\eta} u_{\mu\nu} u^{\mu\nu} + \frac{c^2}{16\pi\sigma} f_{\mu\nu} f^{\mu\nu} \right) \sqrt{-g}\, dx^1 dx^2 dx^3 +$$

$$+\frac{1}{c}\sum_{n=1}^{N}\left( \mathbf{v}_n \cdot \frac{\partial L_f}{\partial \mathbf{v}_n} \right).$$

$$(12\text{-}47)$$

$$\mathbf{K} = \frac{1}{c} \int \left[ \begin{array}{l} -\dfrac{\partial}{\partial \mathbf{v}} \left( \rho_{0q} \varphi + \rho_0 \psi + \rho_0 \vartheta + \rho_0 \wp \right) + \\ + \mathbf{v} \cdot \dfrac{\partial}{\partial \mathbf{v}} \left( \rho_{0q} \mathbf{A} + \rho_0 \mathbf{D} + \rho_0 \mathbf{U} + \rho_0 \mathbf{\Pi} \right) \end{array} \right] u^0 \sqrt{-g}\, dx^1 dx^2 dx^3 + \sum_{n=1}^{N} \left( \frac{\partial L_f}{\partial \mathbf{v}_n} \right). \qquad (13\text{-}48)$$

Both of the methods presented above for determining the relativistic four-momentum $P_\mu$ imply that first the equations for each field acting in the system are solved, and the solutions to the equation for the metric and the equation of motion of matter particles are also found. After this, it becomes possible to determine the components of the four-currents, metric tensor, four-potentials and field tensors required to find $P_\mu$.

### Appendix B. Equations (119-122)

Next, we use double numbering of formulas, in which the last digits indicate the number of the formula in the text of the article.

We consider the stress-energy tensor of the electromagnetic field with mixed indices:

$$W_\alpha^{\ \beta} = \varepsilon_0 c^2\, g^{\mu\kappa} \left( -\delta_\alpha^\lambda g^{\sigma\beta} + \frac{1}{4} \delta_\alpha^\beta g^{\sigma\lambda} \right) F_{\mu\lambda} F_{\kappa\sigma}. \qquad (1\text{-}116)$$

The time components of the tensor $W_\alpha^{\ \beta}$ (1-116) are expressed in terms of the electric field strength $\mathbf{E}$ and magnetic field induction $\mathbf{B}$ in Cartesian coordinates as follows:

$$W_0^{\ 0} = \frac{\varepsilon_0}{2}(E^2 + c^2 B^2), \qquad W_1^{\ 0} = -\varepsilon_0 c\, [\mathbf{E} \times \mathbf{B}]_x,$$

$$W_2^{\ 0} = -\varepsilon_0 c\, [\mathbf{E} \times \mathbf{B}]_y, \qquad W_3^{\ 0} = -\varepsilon_0 c\, [\mathbf{E} \times \mathbf{B}]_z. \qquad (2\text{-}117)$$

Inside the moving sphere in the reference system $O$, the components of $\mathbf{E}_i$ and $\mathbf{B}_i$ are determined by the expressions

$$E_{ix} = E'_{ix}, \qquad E_{iy} = \gamma E'_{iy}, \qquad E_{iz} = \gamma E'_{iz},$$

$$B_{ix} = B'_{ix} = 0, \qquad B_{iy} = -\frac{\gamma V E'_{iz}}{c^2}, \qquad B_{iz} = \frac{\gamma V E'_{iy}}{c^2}. \tag{3-85}$$

In the reference frame $O'$ associated with the center of the sphere, there is an electric field strength $\mathbf{E}'_i$ inside the sphere, and the magnetic field $\mathbf{B}'_i$ is zero due to the absence of internal currents. Expressions (3-85) are obtained by transforming the electromagnetic field tensor, containing components $\mathbf{E}_i$ and $\mathbf{B}_i$, from the reference system $O'$ to the reference system $O$ using Lorentz transformations.

The electric field $\mathbf{E}'_i$ inside a sphere with charged particles in case of the relativistic uniform system is expressed by the formula:

$$\mathbf{E}'_i = -\nabla \varphi'_i = \frac{\rho_{0q} c^2 \gamma_c \mathbf{r}'}{4\pi \varepsilon_0 \eta \rho_0 r'^3} \left[ \frac{c}{\sqrt{4\pi \eta \rho_0}} \sin\left(\frac{r'}{c}\sqrt{4\pi \eta \rho_0}\right) - r'\cos\left(\frac{r'}{c}\sqrt{4\pi \eta \rho_0}\right) \right] \approx$$

$$\approx \frac{\rho_{0q} \gamma_c \mathbf{r}'}{3\varepsilon_0}\left(1 - \frac{2\pi \eta \rho_0 r'^2}{5c^2}\right).$$

$$\tag{4-81}$$

Substituting the components $\mathbf{E}'_i$ from (4-81) into (3-85), taking into account the fact that in (4-81) there is the radius-vector $\mathbf{r}' = (x', y', z')$, gives the following:

$$E_{ix} = \frac{\rho_{0q} c^2 \gamma_c x'}{4\pi \varepsilon_0 \eta \rho_0 r'^3} \left[ \frac{c}{\sqrt{4\pi \eta \rho_0}} \sin\left(\frac{r'}{c}\sqrt{4\pi \eta \rho_0}\right) - r'\cos\left(\frac{r'}{c}\sqrt{4\pi \eta \rho_0}\right) \right].$$

$$E_{iy} = \frac{\rho_{0q} c^2 \gamma \gamma_c y'}{4\pi \varepsilon_0 \eta \rho_0 r'^3} \left[ \frac{c}{\sqrt{4\pi \eta \rho_0}} \sin\left(\frac{r'}{c}\sqrt{4\pi \eta \rho_0}\right) - r'\cos\left(\frac{r'}{c}\sqrt{4\pi \eta \rho_0}\right) \right].$$

$$E_{iz} = \frac{\rho_{0q} c^2 \gamma \gamma_c z'}{4\pi \varepsilon_0 \eta \rho_0 r'^3} \left[ \frac{c}{\sqrt{4\pi \eta \rho_0}} \sin\left(\frac{r'}{c}\sqrt{4\pi \eta \rho_0}\right) - r'\cos\left(\frac{r'}{c}\sqrt{4\pi \eta \rho_0}\right) \right].$$

$$B_{ix} = 0, \qquad B_{iy} = -\frac{\rho_{0q} V \gamma \gamma_c z'}{4\pi \varepsilon_0 \eta \rho_0 r'^3} \left[ \frac{c}{\sqrt{4\pi \eta \rho_0}} \sin\left(\frac{r'}{c}\sqrt{4\pi \eta \rho_0}\right) - r'\cos\left(\frac{r'}{c}\sqrt{4\pi \eta \rho_0}\right) \right].$$

$$B_{iz} = \frac{\rho_{0q} V \gamma \gamma_c y'}{4\pi\varepsilon_0 \eta \rho_0 r'^3} \left[ \frac{c}{\sqrt{4\pi\eta\rho_0}} \sin\left(\frac{r'}{c}\sqrt{4\pi\eta\rho_0}\right) - r'\cos\left(\frac{r'}{c}\sqrt{4\pi\eta\rho_0}\right) \right]. \tag{5}$$

Taking into account the field components (5), we find:

$$E_i^2 = E_{ix}^2 + E_{iy}^2 + E_{iz}^2 =$$

$$= \left(\frac{\rho_{0q} c^2 \gamma_c}{4\pi\varepsilon_0 \eta \rho_0 r'^3}\right)^2 \left[x'^2 + \gamma^2\left(y'^2 + z'^2\right)\right] \begin{bmatrix} \frac{c}{\sqrt{4\pi\eta\rho_0}} \sin\left(\frac{r'}{c}\sqrt{4\pi\eta\rho_0}\right) - \\ -r'\cos\left(\frac{r'}{c}\sqrt{4\pi\eta\rho_0}\right) \end{bmatrix}^2.$$

$$B_i^2 = B_{ix}^2 + B_{iy}^2 + B_{iz}^2 =$$

$$= \left(y'^2 + z'^2\right)\left(\frac{\rho_{0q} V \gamma \gamma_c}{4\pi\varepsilon_0 \eta \rho_0 r'^3}\right)^2 \left[\frac{c}{\sqrt{4\pi\eta\rho_0}} \sin\left(\frac{r'}{c}\sqrt{4\pi\eta\rho_0}\right) - r'\cos\left(\frac{r'}{c}\sqrt{4\pi\eta\rho_0}\right)\right]^2.$$

$$[\mathbf{E}_i \times \mathbf{B}_i]_x = \left(E_{iy} B_{iz} - E_{iz} B_{iy}\right) =$$

$$= V\left(y'^2 + z'^2\right)\left(\frac{c\rho_{0q} \gamma \gamma_c}{4\pi\varepsilon_0 \eta \rho_0 r'^3}\right)^2 \left[\frac{c}{\sqrt{4\pi\eta\rho_0}} \sin\left(\frac{r'}{c}\sqrt{4\pi\eta\rho_0}\right) - r'\cos\left(\frac{r'}{c}\sqrt{4\pi\eta\rho_0}\right)\right]^2.$$

$$[\mathbf{E}_i \times \mathbf{B}_i]_y = \left(E_{iz} B_{ix} - E_{ix} B_{iz}\right) =$$

$$= -\gamma V x' y' \left(\frac{c\rho_{0q} \gamma_c}{4\pi\varepsilon_0 \eta \rho_0 r'^3}\right)^2 \left[\frac{c}{\sqrt{4\pi\eta\rho_0}} \sin\left(\frac{r'}{c}\sqrt{4\pi\eta\rho_0}\right) - r'\cos\left(\frac{r'}{c}\sqrt{4\pi\eta\rho_0}\right)\right]^2.$$

$$[\mathbf{E}_i \times \mathbf{B}_i]_z = \left(E_{ix} B_{iy} - E_{iy} B_{ix}\right) =$$

$$= -\gamma V x' z' \left(\frac{c\rho_{0q} \gamma_c}{4\pi\varepsilon_0 \eta \rho_0 r'^3}\right)^2 \left[\frac{c}{\sqrt{4\pi\eta\rho_0}} \sin\left(\frac{r'}{c}\sqrt{4\pi\eta\rho_0}\right) - r'\cos\left(\frac{r'}{c}\sqrt{4\pi\eta\rho_0}\right)\right]^2.$$

$$\tag{6}$$

Substituting in (2-117) $\mathbf{E}_i$ instead of $\mathbf{E}$ and $\mathbf{B}_i$ instead of $\mathbf{B}$, taking into account (6) and the relation $\gamma^2 = \frac{1}{1 - V^2/c^2}$, we find:

$$W_0^{\ 0} =$$
$$= \frac{\rho_{0q}^2 c^4 \gamma_c^2}{32\pi^2 \varepsilon_0 \eta^2 \rho_0^2 r'^4}\left[1+\frac{2\gamma^2 V^2\left(y'^2+z'^2\right)}{c^2 r'^2}\right]\left[\frac{c}{\sqrt{4\pi\eta\rho_0}}\sin\left(\frac{r'}{c}\sqrt{4\pi\eta\rho_0}\right)-r'\cos\left(\frac{r'}{c}\sqrt{4\pi\eta\rho_0}\right)\right]^2.$$

$$W_1^{\ 0} = -\frac{\rho_{0q}^2 c^3 \gamma_c^2 \gamma^2 V\left(y'^2+z'^2\right)}{16\pi^2 \varepsilon_0 \eta^2 \rho_0^2 r'^6}\left[\frac{c}{\sqrt{4\pi\eta\rho_0}}\sin\left(\frac{r'}{c}\sqrt{4\pi\eta\rho_0}\right)-r'\cos\left(\frac{r'}{c}\sqrt{4\pi\eta\rho_0}\right)\right]^2,$$

$$W_2^{\ 0} = \frac{\rho_{0q}^2 c^3 \gamma_c^2 \gamma V x' y'}{16\pi^2 \varepsilon_0 \eta^2 \rho_0^2 r'^6}\left[\frac{c}{\sqrt{4\pi\eta\rho_0}}\sin\left(\frac{r'}{c}\sqrt{4\pi\eta\rho_0}\right)-r'\cos\left(\frac{r'}{c}\sqrt{4\pi\eta\rho_0}\right)\right]^2.$$

$$W_3^{\ 0} = \frac{\rho_{0q}^2 c^3 \gamma_c^2 \gamma V x' z'}{16\pi^2 \varepsilon_0 \eta^2 \rho_0^2 r'^6}\left[\frac{c}{\sqrt{4\pi\eta\rho_0}}\sin\left(\frac{r'}{c}\sqrt{4\pi\eta\rho_0}\right)-r'\cos\left(\frac{r'}{c}\sqrt{4\pi\eta\rho_0}\right)\right]^2.$$

(7-119)

From the principle of least action follow the equations of electromagnetic and gravitational fields, acceleration fields and pressure fields. All these equations have the same form, similar to Maxwell's equations. This is a consequence of the fact that these fields are vector fields and have the same representation through the four-potentials and tensors of these fields. As a result, the expression for the field tensors, as well as for the stress-energy tensors for the electromagnetic field $W_\alpha^{\ \beta}$, for the gravitational field $U_\alpha^{\ \beta}$, the acceleration field $B_\alpha^{\ \beta}$ and the pressure field $P_\alpha^{\ \beta}$ turn out to be similar to each other and have the same dependence on time and coordinates. To obtain the time components of the stress-energy tensor $U_\alpha^{\ \beta}$ of the gravitational field, it is enough in (7-119) to replace the charge density $\rho_{0q}$ with the mass density $\rho_0$, and replace the electric constant $\varepsilon_0$ with $-\frac{1}{4\pi G}$, as can be seen in (116):

$$U_0^{\ 0} =$$
$$= -\frac{G c^4 \gamma_c^2}{8\pi \eta^2 r'^4}\left[1+\frac{2\gamma^2 V^2\left(y'^2+z'^2\right)}{c^2 r'^2}\right]\left[\frac{c}{\sqrt{4\pi\eta\rho_0}}\sin\left(\frac{r'}{c}\sqrt{4\pi\eta\rho_0}\right)-r'\cos\left(\frac{r'}{c}\sqrt{4\pi\eta\rho_0}\right)\right]^2.$$

$$U_1^0 = \frac{Gc^3\gamma_c^2\gamma^2 V(y'^2+z'^2)}{4\pi\eta^2 r'^6}\left[\frac{c}{\sqrt{4\pi\eta\rho_0}}\sin\left(\frac{r'}{c}\sqrt{4\pi\eta\rho_0}\right)-r'\cos\left(\frac{r'}{c}\sqrt{4\pi\eta\rho_0}\right)\right]^2,$$

$$U_2^0 = -\frac{Gc^3\gamma_c^2\gamma V x'y'}{4\pi\eta^2 r'^6}\left[\frac{c}{\sqrt{4\pi\eta\rho_0}}\sin\left(\frac{r'}{c}\sqrt{4\pi\eta\rho_0}\right)-r'\cos\left(\frac{r'}{c}\sqrt{4\pi\eta\rho_0}\right)\right]^2.$$

$$U_3^0 = -\frac{Gc^3\gamma_c^2\gamma V x'z'}{4\pi\eta^2 r'^6}\left[\frac{c}{\sqrt{4\pi\eta\rho_0}}\sin\left(\frac{r'}{c}\sqrt{4\pi\eta\rho_0}\right)-r'\cos\left(\frac{r'}{c}\sqrt{4\pi\eta\rho_0}\right)\right]^2.$$

(8)

In the same way, we can find the time components of the stress-energy tensors of the acceleration field and the pressure field by replacing the charge density $\rho_{0q}$ in (7-119) with the mass density $\rho_0$, and replacing the electrical constant $\varepsilon_0$ with $\dfrac{1}{4\pi\eta}$ and with $\dfrac{1}{4\pi\sigma}$, respectively:

$$B_0^0 =$$
$$= \frac{c^4\gamma_c^2}{8\pi\eta r'^4}\left[1+\frac{2\gamma^2 V^2(y'^2+z'^2)}{c^2 r'^2}\right]\left[\frac{c}{\sqrt{4\pi\eta\rho_0}}\sin\left(\frac{r'}{c}\sqrt{4\pi\eta\rho_0}\right)-r'\cos\left(\frac{r'}{c}\sqrt{4\pi\eta\rho_0}\right)\right]^2.$$

$$B_1^0 = -\frac{c^3\gamma_c^2\gamma^2 V(y'^2+z'^2)}{4\pi\eta r'^6}\left[\frac{c}{\sqrt{4\pi\eta\rho_0}}\sin\left(\frac{r'}{c}\sqrt{4\pi\eta\rho_0}\right)-r'\cos\left(\frac{r'}{c}\sqrt{4\pi\eta\rho_0}\right)\right]^2,$$

$$B_2^0 = \frac{c^3\gamma_c^2\gamma V x'y'}{4\pi\eta r'^6}\left[\frac{c}{\sqrt{4\pi\eta\rho_0}}\sin\left(\frac{r'}{c}\sqrt{4\pi\eta\rho_0}\right)-r'\cos\left(\frac{r'}{c}\sqrt{4\pi\eta\rho_0}\right)\right]^2.$$

$$B_3^0 = \frac{c^3\gamma_c^2\gamma V x'z'}{4\pi\eta r'^6}\left[\frac{c}{\sqrt{4\pi\eta\rho_0}}\sin\left(\frac{r'}{c}\sqrt{4\pi\eta\rho_0}\right)-r'\cos\left(\frac{r'}{c}\sqrt{4\pi\eta\rho_0}\right)\right]^2.$$

$$P_0^{\ 0} =$$
$$= \frac{\sigma c^4 \gamma_c^2}{8\pi \eta^2 r'^4} \left[1 + \frac{2\gamma^2 V^2 (y'^2 + z'^2)}{c^2 r'^2}\right] \left[\frac{c}{\sqrt{4\pi\eta\rho_0}} \sin\left(\frac{r'}{c}\sqrt{4\pi\eta\rho_0}\right) - r'\cos\left(\frac{r'}{c}\sqrt{4\pi\eta\rho_0}\right)\right]^2.$$

$$P_1^{\ 0} = -\frac{\sigma c^3 \gamma_c^2 \gamma^2 V (y'^2 + z'^2)}{4\pi \eta^2 r'^6} \left[\frac{c}{\sqrt{4\pi\eta\rho_0}} \sin\left(\frac{r'}{c}\sqrt{4\pi\eta\rho_0}\right) - r'\cos\left(\frac{r'}{c}\sqrt{4\pi\eta\rho_0}\right)\right]^2,$$

$$P_2^{\ 0} = \frac{\sigma c^3 \gamma_c^2 \gamma V x' y'}{4\pi \eta^2 r'^6} \left[\frac{c}{\sqrt{4\pi\eta\rho_0}} \sin\left(\frac{r'}{c}\sqrt{4\pi\eta\rho_0}\right) - r'\cos\left(\frac{r'}{c}\sqrt{4\pi\eta\rho_0}\right)\right]^2.$$

$$P_3^{\ 0} = \frac{\sigma c^3 \gamma_c^2 \gamma V x' z'}{4\pi \eta^2 r'^6} \left[\frac{c}{\sqrt{4\pi\eta\rho_0}} \sin\left(\frac{r'}{c}\sqrt{4\pi\eta\rho_0}\right) - r'\cos\left(\frac{r'}{c}\sqrt{4\pi\eta\rho_0}\right)\right]^2.$$

(9)

The stress-energy tensors in (8-9) correspond to the expressions for the stress-energy tensors in Eq. (121) of article.

The stress-energy tensor of a physical system is obtained by summing the stress-energy tensors of all fields:

$$T_\alpha^{\ \beta} = W_\alpha^{\ \beta} + U_\alpha^{\ \beta} + B_\alpha^{\ \beta} + P_\alpha^{\ \beta}. \qquad (10\text{-}115)$$

Substituting (7-119), (8) and (9) into (10-115) makes it possible to find the time components of the stress-energy tensor:

$$T_0^{\ 0} = W_0^{\ 0} + U_0^{\ 0} + B_0^{\ 0} + P_0^{\ 0} =$$
$$= \frac{c^4 \gamma_c^2}{8\pi \eta^2 r'^4} \left(\frac{\rho_{0q}^2}{4\pi^2 \varepsilon_0 \rho_0^2} - G + \eta + \sigma\right)\left[1 + \frac{2\gamma^2 V^2 (y'^2 + z'^2)}{c^2 r'^2}\right] \times$$
$$\times \left[\frac{c}{\sqrt{4\pi\eta\rho_0}} \sin\left(\frac{r'}{c}\sqrt{4\pi\eta\rho_0}\right) - r'\cos\left(\frac{r'}{c}\sqrt{4\pi\eta\rho_0}\right)\right]^2.$$

$$T_1^{\ 0} = W_1^{\ 0} + U_1^{\ 0} + B_1^{\ 0} + P_1^{\ 0} =$$

$$= -\frac{c^3 \gamma_c^2 \gamma^2 V (y'^2 + z'^2)}{4\pi \eta^2 r'^6} \left( \frac{\rho_{0q}^2}{4\pi \varepsilon_0 \rho_0^2} - G + \eta + \sigma \right) \times$$

$$\times \left[ \frac{c}{\sqrt{4\pi \eta \rho_0}} \sin\left( \frac{r'}{c} \sqrt{4\pi \eta \rho_0} \right) - r' \cos\left( \frac{r'}{c} \sqrt{4\pi \eta \rho_0} \right) \right]^2 .$$

$$T_2^{\ 0} = W_2^{\ 0} + U_2^{\ 0} + B_2^{\ 0} + P_2^{\ 0} =$$

$$= \frac{c^3 \gamma_c^2 \gamma V x' y'}{4\pi \eta^2 r'^6} \left( \frac{\rho_{0q}^2}{4\pi \varepsilon_0 \rho_0^2} - G + \eta + \sigma \right) \left[ \frac{c}{\sqrt{4\pi \eta \rho_0}} \sin\left( \frac{r'}{c} \sqrt{4\pi \eta \rho_0} \right) - r' \cos\left( \frac{r'}{c} \sqrt{4\pi \eta \rho_0} \right) \right]^2 .$$

$$T_3^{\ 0} = W_3^{\ 0} + U_3^{\ 0} + B_3^{\ 0} + P_3^{\ 0} =$$

$$= \frac{c^3 \gamma_c^2 \gamma V x' z'}{4\pi \eta^2 r'^6} \left( \frac{\rho_{0q}^2}{4\pi \varepsilon_0 \rho_0^2} - G + \eta + \sigma \right) \left[ \frac{c}{\sqrt{4\pi \eta \rho_0}} \sin\left( \frac{r'}{c} \sqrt{4\pi \eta \rho_0} \right) - r' \cos\left( \frac{r'}{c} \sqrt{4\pi \eta \rho_0} \right) \right]^2 .$$

(11)

The equation of motion of matter particles under the influence of fields is obtained from the principle of least action. In the case of a relativistic uniform system, the equation of motion implies the following relation for the field coefficients:

$$\eta + \sigma = G - \frac{\rho_{0q}^2}{4\pi \varepsilon_0 \rho_0^2} . \qquad (12\text{-}63)$$

If we substitute (12-63) into (11), it becomes clear that in the matter of the moving sphere, which is a relativistic uniform system, the time components $T_0^{\ 0}$, $T_1^{\ 0}$, $T_2^{\ 0}$ and $T_3^{\ 0}$ of stress-energy tensor of the system become equal to zero.

The integral vector is determined by the expression:

$$\mathcal{J}_\alpha = \int T_\alpha^{\ 0} dx^1 dx^2 dx^3 , \qquad (13\text{-}9)$$

where index $\alpha = 0, 1, 2, 3$.

Since all time components $T_\alpha{}^0$ in the matter inside the moving sphere are equal to zero according to (11) and (12-63), then the components of the integral vector $\mathcal{J}_\alpha$ in (13-9) become zero.

### Appendix C. List of symbols

$A_\mu$ is four-potential of electromagnetic field

**A** is vector potential of electromagnetic field

$a$ is radius of sphere

**B** is magnetic field induction

$B_\alpha{}^\beta$ is stress-energy tensor of acceleration field with mixed indexes

**C** is strength of pressure field

$c$ is speed of light

$dx^1 dx^2 dx^3$ is volume element in the form of product of differentials of space coordinates

$D_\mu$ is four-potential of gravitational field within the framework of covariant theory of gravitation

**D** is vector potential of gravitational field

**E** is electric field strength

$\mathbb{E}$ is energy of a system

$\mathbb{E}'$ is energy of a system in reference frame $O'$

$\varepsilon_0$ is electric constant

$F_\mu$ is four-force

$(f_\mu)_n$ is density of the four-force acting in a unit element of the volume in which a typical particle with the number $n$ is located

$F_{\mu\nu}$ is tensor of electromagnetic field

$\Phi_{\mu\nu}$ is tensor of gravitational field

$f_{\mu\nu}$ is tensor of pressure field

$\varphi$ is scalar potential of electromagnetic field

$\phi$ is an angle coordinate of spherical coordinate system

$G$ is gravitational constant

$g_{\mu\nu}$ or $g^{\mu\nu}$ is metric tensor

$g$ is determinant of metric tensor

$\eta^{\mu\nu}$ is metric tensor of Minkowski spacetime

$\eta$ is acceleration field coefficient

$\mathbf{\Gamma}$ is strength of gravitational field

$\gamma$ denotes the Lorentz factor of center of momentum of the moving sphere

$\gamma'$ is the Lorentz factor of particles in reference frame $O'$

$\gamma_c$ is the Lorentz factor of particles at the center of the sphere in reference frame $O'$

$\gamma_p$ denotes the Lorentz factor of a particle in reference frame $O$

$\mathbf{I}$ is solenoidal vector of pressure field

$\mathcal{J}^\mu$ is integral vector in general theory of relativity

$\mathcal{J}_\alpha$ is integral vector in covariant theory of gravitation

$\mathfrak{I}_\mu$ is auxiliary four-dimensional quantity

$j^\mu$ is charge four-current

$J^\mu$ is mass four-current

$\mathbf{J'}$ is mass current density in reference frame $O'$

$\mathbf{j'}$ is charge current density in reference frame $O'$

$K^\mu$ or $K_\mu$ is field four-momentum

$K_0$ is time component of $K_\mu$, which is associated with the energy in the term with the fields' tensor invariants

$\mathbf{K}$ is momentum of fields

$L$ is Lagrangian

$L_f$ represents the part of the Lagrangian $L$, that is associated with tensor invariants

$L_{fi}$ denotes that part of the Lagrangian that is associated with tensor invariants in the matter

$L_{fo}$ denotes that part of the Lagrangian that is associated with tensor invariants outside the matter

$L_p$ represents the part of the Lagrangian $L$, which contains mass four-current $J^\mu$ and charge four-current $j^\mu$

$L_m$ is Lagrangian inside the matter

$\mathcal{L}$ is Lagrangian density

$\mathcal{L}_f$ represents the part of the Lagrangian density $\mathcal{L}$, which contains tensor invariants

$\mathcal{L}_p$ represents the part of the Lagrangian density $\mathcal{L}$, which contains mass four-current $J^\mu$ and charge four-current $j^\mu$

$\mathcal{L}_m$ is Lagrangian density inside the matter

$\Lambda$ is cosmological constant

$m_b$ is total mass of particles of relativistic uniform system

$m_g$ is gravitational mass of relativistic uniform system

$\mu_0$ is magnetic constant

$N$ is total number of particles of a physical system

$\mathbf{N}$ is solenoidal vector of acceleration field

$n$ is current number of particle or volume element

$O$ is reference frame of coordinate observer

$O'$ is reference frame, associated with the center-of-momentum of moving sphere

$\mathbf{\Omega}$ is torsion field vector of gravitational field

$\mathbf{P}$ is three-dimensional relativistic momentum of a system

$\mathbf{P}_n$ is three-dimensional momentum of one volume element or one particle

$\mathbf{P}_p$ is relativistic momentum of particles

$P^\mu$ or $P_\mu$ is four-momentum of a system

$P_\alpha^{\ \beta}$ is stress-energy tensor of pressure field with mixed indexes

$\pi_\mu$ is four-potential of pressure field

$\mathcal{P}_\mu$ is density of the generalized four-momentum

$p^\mu$ or $p_\mu$ is generalized four-momentum

$p_0$ is time component of the generalized four-momentum

$\mathbf{p}$ is generalized momentum

$\mathbf{\Pi}$ is vector potential of pressure field

$q$ is charge of a particle

$q_b$ is total charge of relativistic uniform system

$Q = Nq$ is charge of moving sphere with particles

$\theta$ is an angle coordinate of spherical coordinate system

$\vartheta$ is scalar potential of acceleration field

$R$ is scalar curvature

**r** is three-dimensional vector of position

$r$ is radial coordinate of spherical coordinate system

$r'$ is current radius inside the sphere

$\rho_0$ is mass density in particle's comoving reference frame

$\rho_{0q}$ is charge density in particle's comoving reference frame

$\wp$ is scalar potential of pressure field

$\wp_c$ is scalar potential of pressure field at the center of the sphere in reference frame $O'$

$S$ is action function

**S** is strength of acceleration field

$\mathbf{S}_p$ is Poynting vector

$\sigma$ is pressure field coefficient

$t$ is coordinate time

$\tau$ is proper time

$T^{\mu\nu}$ is stress-energy tensor

$t^{\mu\nu}$ is gravitational field pseudotensor

$T^{\mu 0}$ are time components of the stress-energy tensor

$t^{\mu 0}$ are time components of the gravitational field pseudotensor

**U** is vector potential of acceleration field

$U_\mu$ is four-potential of acceleration field

$U_\alpha^{\ \beta}$ is stress-energy tensor of gravitational field with mixed indexes

$u^\mu$ or $u_\mu$ is four-velocity

$u_n^0$ is time component of four-velocity of particle with the number $n$

$u_{\mu\nu}$ is tensor of acceleration field

$dV_0$ is differential of volume element in comoving reference frame

$V_m$ is volume occupied by matter

**V** is velocity of motion of the system's center of momentum

**v** is velocity of a particle or velocity of a volume element of matter in reference frame $O$

$\mathbf{v}_n$ is velocity of matter particle with the current number $n$

$\mathbf{v}'$ is velocity of typical particles in reference frame $O'$

$W_\alpha^{\ \beta}$ is stress-energy tensor of electromagnetic field with mixed indexes

$x^{\mu}$ or $x_{\mu}$ is four-position

$\psi$ is scalar potential of gravitational field

$Z$ is total field energy inside the matter, associated with tensor invariants

$x, y, z$ are Cartesian coordinates in reference frame $O$

$x', y', z'$ are Cartesian coordinates in reference frame $O'$